\documentclass[12pt]{article}

\pdfoutput=1
\usepackage{graphicx} 
\usepackage{color}                                                                                                                                                                                                                                                                                                                                                                                                                                                                                                                                                                                                                                                                                                                                                                                                                                                                                                                                                                                                                        
\usepackage{cancel,tabularx,moreverb,fancybox,amsmath,float,bm,braket,slashbox,txfonts,amssymb,bm,accents}
\usepackage[top=30truemm,bottom=30truemm,left=25truemm,right=25truemm]{geometry}
\usepackage{latexsym}
\usepackage{here}
\usepackage{cite}
\usepackage{hyperref}

\newcommand{\ex}[1]{\mathrm{e}^{#1}}

\newcommand{\pa}[1]{\left(#1 \right)}

\newcommand{\br}[1]{\left[#1 \right]}
\newcommand{\BR}[1]{\Biggl[#1 \Biggr]}

\newcommand{\kagi}[1]{\lbrack#1 \rbrack}
\newcommand{\bb}[1]{\mathbb{#1}}

\newcommand{\ca}[1]{\mathcal{#1}}

\newcommand{\abs}[1]{\left|#1\right|}
\newcommand{\ave}[1]{\langle #1\rangle}
\newcommand{\ar}[1]{\xrightarrow[#1]{}}
\newcommand{\pd}[1]{\frac{\partial}{\partial #1}}

\newcommand{\ti}[1]{\tilde{#1}}
\newcommand{\dg}[1]{#1^{\dagger}}
\newcommand{\fr}{\frac}
\newcommand{\s}[1]{\sqrt{#1}}

\def\be{\begin{equation}}
\def\ee{\end{equation}}
\def\ba{\begin{eqnarray}}
\def\ea{\end{eqnarray}}

\def\del{{\partial}}

\def\la{{\lambda}}
 \def\w{{\omega}}
 
 \def\ep{{\epsilon}}
 \def\d{{\delta}}

 \def\a{{\alpha}}
 \def\ba{{\bar{\alpha}}}
 
 \def\l{{\lambda}}
 \def\G{{\Gamma}}
 \def\D{{\Delta}}
 \def\g{{\gamma}}
 
 \def\b{{\beta}}
 \def\e{{\epsilon}}

 \def\p{\partial}
\def\ii{{\mathrm{i}}}
\def\tr{{\text{tr}}}
\def\bz{{\bar{z}}}

\def\dd{{\mathrm{d}}}
\def\sgn{{\text{sgn}}}

  \makeatletter
    
    \@addtoreset{equation}{section}
  \makeatother

\begin{document}

\begin{titlepage}
\thispagestyle{empty}

\begin{flushright}
YITP-19-30 \\
UT-19-08
\\

\end{flushright}

\bigskip

\begin{center}
\noindent{{\large \textbf{
Entanglement Entropy, OTOC and Bootstrap in 2D CFTs\\
from Regge and Light Cone Limits of Multi-point Conformal Block
}}}\\
\vspace{2cm}
Yuya Kusuki${}^a$, Masamichi Miyaji${}^{b}$
\vspace{1cm}

{\it
${}^{a}$Center for Gravitational Physics, \\
Yukawa Institute for Theoretical Physics (YITP), Kyoto University, \\
Kitashirakawa Oiwakecho, Sakyo-ku, Kyoto 606-8502, Japan.\\

${}^{b}$Department of Physics, Faculty of Science, The University of Tokyo,\\ Bunkyo-ku, Tokyo 113-0033, Japan}
\vskip 2em
\end{center}

\begin{abstract}
We explore the structures of light cone and Regge limit singularities of $n$-point Virasoro conformal blocks in $c>1$ two-dimensional conformal field theories with no chiral primaries, using fusion matrix approach. These CFTs include not only holographic CFTs dual to classical gravity, but also their full quantum corrections, since this approach allows us to explore full $1/c$ corrections. As the important applications, we study time dependence of Renyi entropy after a local quench and out-of-time ordered correlator (OTOC) at late time.

 We first show that, the $n$-th $\pa{n>2}$ Renyi entropy after a local quench in our CFT grows logarithmically at late time, for any $c$ and any conformal dimensions of excited primary. In particular, we find that this behavior is independent of $c$, contrary to the expectation that the finite $c$ correction fixes the late time Renyi entropy to be constant. We also show that the constant part of the late time Renyi entropy is given by a monodromy matrix.

We also investigate OTOCs by using the monodromy matrix. We first rewrite the monodromy matrix in terms of fusion matrix explicitly. By this expression, we find that the OTOC decays exponentially in time, and the decay rates are divided into three patterns, depending on the dimensions of external operators. We note that our result is valid for any $c>1$ and any external operator dimensions. Our monodromy matrix approach can be generalized to the Liouville theory and we show that the Liouville OTOC approaches constant in the late time regime.

We emphasize that, there is a number of other applications of the fusion and the monodromy matrix approaches, such as solving the conformal bootstrap equation. Therefore, it is tempting to believe that the fusion and monodromy matrix approaches provide a key to understanding the AdS/CFT correspondence.
 \end{abstract}

\end{titlepage}

\restoregeometry

\tableofcontents

\section{Introduction \& Summary}
The AdS/CFT correspondence is a useful tool to investigate quantum gravity in terms of CFT. In particular, two dimensional CFTs have rich enough structures that can accommodate non trivial dual gravity dynamics, yet they are simple enough to allow us to investigate them analytically. There is a particular class of CFTs, which are believed to have gravity duals. The most simplest CFTs in this class are unitary, compact CFTs with central charge $c>1$ and without chiral primaries (hence, no extra currents apart from the Virasoro current), which we call {\it pure CFTs}.
\footnote{There are exceptions in holographic CFTs of interest. For example, we sometimes consider the $D1$-$D5$ CFT in the holographic context, but in such a theory, there are many chiral primaries. Therefore, it is not classified  as pure CFTs. }
\footnote{
Strictly speaking, we also need to require ``non-zero'' gap between the vacuum and the first-excited state.
For example, we exclude the possibility that there are an infinite sequence of primary operators such that their twist accumulate to zero.
}

 These CFTs are often considered in the context of holography (for e.g., \cite{Collier2016,CollierKravchukLinYin2017}); however, we do not know the explicit construction of such CFTs yet, that is, we have no example of a pure CFT. This is the reason why we have only a poor understanding of a pure CFT despite a number of developments in 2D CFTs.  Moreover, there are few known tools to access the dynamics of such theories. In this article, we will show how the fusion matrix approach enables us to explore the Regge and light cone singularities of the conformal blocks, for any value of central charge $c$ and conformal dimensions of operators. As a result, we determine the generic behavior of the Renyi entanglement entropy of locally excited state and late time out-of-time ordered correlator.

\subsection{Virasoro Block from Fusion and Monodromy Approaches}

Many works studied the relation between the Virasoro blocks and the dual gravity in AdS${}_3$\cite{Fitzpatrick2015, Alkalaev2015,Hijano2015,Asplund2015, Alkalaev2016a, Alkalaev2016b, Alkalaev2017, Maxfield2017,Alkalaev2018, Hikida2018c, Hikida2018a, Alkalaev2018a, Chen2018}. Most of the works relied on the HHLL approximation, 
which is very useful when we explore dynamics of light particles in black hole background, and treating back reactions as perturbations. However, this treatment is not applicable to more interesting gravity dynamics in which we cannot treat back reactions as perturbations, such as black hole merger and other quantum gravity effects. These phenomena require us to have knowledge of finite c correlators with heavy external and intermediate operators. In order to study such correlators, it is natural to look at singularity structure of correlators or conformal blocks, where we may have general results. This is one of the motivations to clarify the Virasoro block asymptotics in the light cone limit and the Regge limit in \cite{Kusuki2018,Kusuki2018a, Kusuki2018b, Kusuki2018c}.

The study of the Virasoro conformal block is also motivated by the black hole information paradox \cite{Fitzpatrick2016a, Chen2017a, Fitzpatrick2017,Chen2017}. The fusion matrix approach explains the emergence of the forbidden singularities and their resolution, and also it tells us that the late time behavior of the {\it full-order} Virasoro block behaves as $t^{-\fr{3}{2}}$ instead of the exponential decay\cite{Collier2018}. Both of the results are necessary, in order to maintain unitary time evolution of pure state, and was numerically conjectured in \cite{Chen2017}. We would also like to mention that the results (1.8) and (1.10) in \cite{Fitzpatrick2017} can be straightforwardly obtained by using the light cone singularity (\ref{eq:LC}).

There is an important tool to analyze the Virasoro block, called conformal bootstrap. The conformal bootstrap utilizes OPE associativity of operators, which gives constraints on OPE coefficients and operator spectrum. The numerical application of conformal bootstrap has led to significant breakthroughs in investigating low-lying operators \cite{Rattazzi2008, Rychkov2009, Kos2014, Simmons-Duffin2015, Collier2016}, the upper bound on the gap from the vacuum \cite{Hellerman2009, Friedan2013, Collier2016} and the uniqueness of Liouville CFT \cite{CollierKravchukLinYin2017}. Although it is very difficult to analytically solve the bootstrap equation, we can sometimes do so in certain asymptotic regimes \cite{Cardy1986a, Kraus2016, Das2017, Cardy2017, Kusuki2018, Hikida2018, Romero-Bermudez2018, Brehm2018, Kusuki2018a, Mukhametzhanov2019}.  Against this backdrop, the fusion matrix approach provides a new analytic method to access the CFT data by making use of the light-cone singularity of the Virasoro blocks. The light-cone bootstrap was first implemented in \cite{Fitzpatrick2013a, Komargodski2013}, and there have been subsequent advancements \cite{Alday2015,Kaviraj2015,Kaviraj2015a, Alday2017a, Simmons-Duffin2017,Alday2017,Sleight2018} as regards the light-cone bootstrap in higher-dimensional CFTs. Nevertheless, there has been little knowledge of two-dimension CFTs with the application of this method \cite{Fitzpatrick2014}.
This is because there had been currently no explicit forms of the light-cone singularity of Virasoro blocks. But now we have the explicit form of the light cone singularity, and we can solve the light-cone bootstrap equation \cite{Kusuki2018c, Collier2018}. Actually, as explained in Section \ref{sec:LCB}, the result in \cite{Kusuki2018c, Collier2018} can be expressed by the following statement,
\begin{quote}
Fusion rules at large spin $\ca{V}_{\a} \times \ca{V}_{\b}|_{l \to \infty} $ in a pure CFT approach the fusion rules of Liouville CFTs.\footnote{We do not mean that the large spin OPE coefficients in a pure CFT can be described by the DOZZ formula.}
\end{quote}
This is the 2D counterpart of the fact that in higher-dimensional CFTs ($d\geq3$), the twist spectrum in the OPE between two primary states at large spin approaches the spectrum of double-trace states in {\it generalized free theories}. 
Actually, these asymptotic fusion rules can be naturally understood in the bulk.

In this paper, we also generalize the above results for $4$-point conformal blocks to any $n$-point
\footnote{
The $n$-point Virasoro block is also studied in \cite{Banerjee2016} (see also \cite{Alkalaev2016a}), which is the multi-point generalization of Appendix E of \cite{Fitzpatrick2014} . The global limit of the $n$-point conformal block is given in \cite{Rosenhaus2019}.
}
 in Section \ref{sec:nLC}.
There are a lot of applications of this generalization and as one example, we use them to evaluate the $n$-th Renyi entropy as explained in the next subsection. To avoid the complicated calculations, we only present the results in Section \ref{sec:nLC} and present the detailed calculations in Appendix \ref{app:n-point}.

Hopefully, our fusion and monodromy matrix approaches will give answers to the interesting questions in 2D CFTs, besides the applications presented herein. For example, 
we believe that the approaches are also useful to consider the generalization of the sparseness condition \cite{Hartman2014, Mukhametzhanov2019} from a torus partition function to a four point function. There is already such a work based on the global conformal symmetry  \cite{Kraus2018}. We expect that our approach reveals its Virasoro version.

\subsection{Entanglement Growth}

One useful measure of entanglement is entanglement entropy, which is defined as
\begin{equation}
S_A=-\tr \rho_A \log \rho_A,
\end{equation}
where $\rho_A$ is a reduced density matrix for a subsystem $A$, obtained by tracing out its complement. The Renyi entropy is a generalization of the entanglement entropy, which is defined as
\begin{equation}
S_A^{(n)}=\fr{1}{1-n} \log \tr \rho_A^n,
\end{equation}
and the limit $n\to1$ of the Renyi entropy defines the entanglement entropy $S_A$.

\begin{figure}[t]
 \begin{center}
  \includegraphics[width=10.0cm,clip]{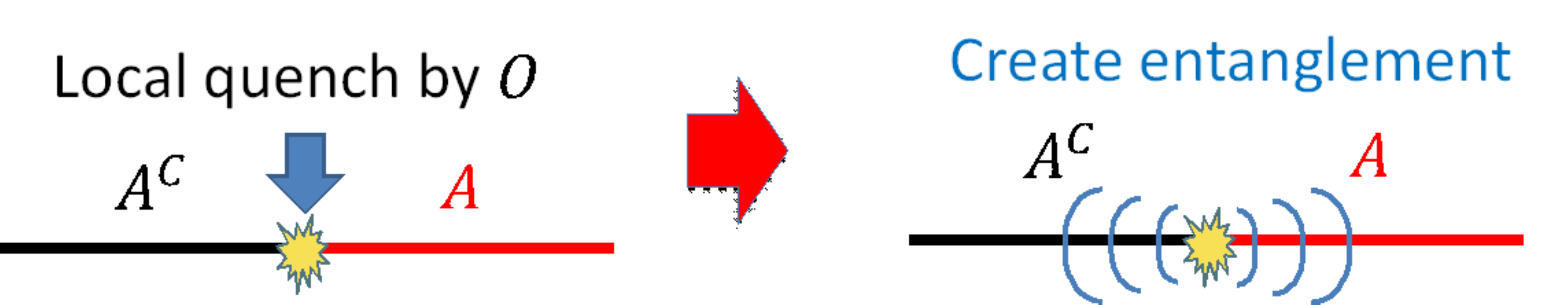}
 \end{center}
 \caption{We probe the growth of entanglement between $A$ and $A^C$ after a local quench in this setup.}
 \label{fig:setup}
\end{figure}

Our interest in this paper is the dynamics of the entanglement entropy. In particular, we consider entanglement entropy for a locally excited state  $\ket{\Psi}$, which is defined by acting with a local operator $ O(-l)$ on the CFT vacuum $\ket{0}$ in the following manner,\footnote{We would like to stress that
$\ep$ in (\ref{lopw}) is the ultraviolet (UV) cut off of the local excitations and should be distinguished from the UV cut off (i.e.,the lattice spacing) of the CFT itself.}
\be\label{lopw}
\ket{\Psi(t)}=\s{\ca{N}}\ex{-\ep H-iHt} O(-l)\ket{0}, 
\ee
where $\ca{N}$ is the normalization factor. The infinitesimally small parameter $\ep>0$ provides UV regularization as the truly localized operator has infinite energy. We choose the subsystem $A$ to be the half-space and induce excitation in its complement, thus creating additional entanglements between them (see Figure \ref{fig:setup}). Then the main quantity of interest is the growth of entanglement entropy compared to the vacuum:
 \be\label{eq:difs}
 \Delta S^{(n)}_A(t)=S^{(n)}_A(\ket{\Psi(t)})-S^{(n)}_A(\ket{0}).
 \ee
This quantity is expected to capture the chaotic natures of CFTs. The entanglement entropy in RCFT is studied in \cite{He2014,Numasawa2016}, which shows that the growth of the entanglement entropy approaches constant in the late time limit. On the other hand, the entanglement entropy in holographic CFTs shows a logarithmic growth \cite{Nozaki2013, Asplund2015}.
This difference is thought to be due to the chaotic nature of holographic CFTs. In other words, we expect that this late time behavior can also be used as a criterion of chaotic nature of a given quantum field theory. (See also  \cite{Caputa2014a, David2016, Caputa2017, He2017, Guo2018, Shimaji2018, Apolo2018}, which revealed the growth of the entanglement entropy after a local quench in other setups.) 

However, since there has been little knowledge of holographic CFTs, we have to rely on some assumptions and approximations in the calculation of the entanglement entropy in holographic CFTs. In particular, most of such calculations are obtained by making use of the HHLL approximation \cite{Fitzpatrick2014,Fitzpatrick2015}, therefore, those results are limited to be perturbative in $\fr{1}{c}$. Moreover, that approximation does not allow us to study conformal blocks with heavy-heavy-heavy-heavy external operators, which means that we cannot make use of it to study $n$-th Renyi entropy with $n \geq 2$. 

\subsubsection{Geometric Part of Entanglement Growth}

\begin{figure}[t]
 \begin{center}
  \includegraphics[width=17.0cm, trim=50 100 50 100,clip]{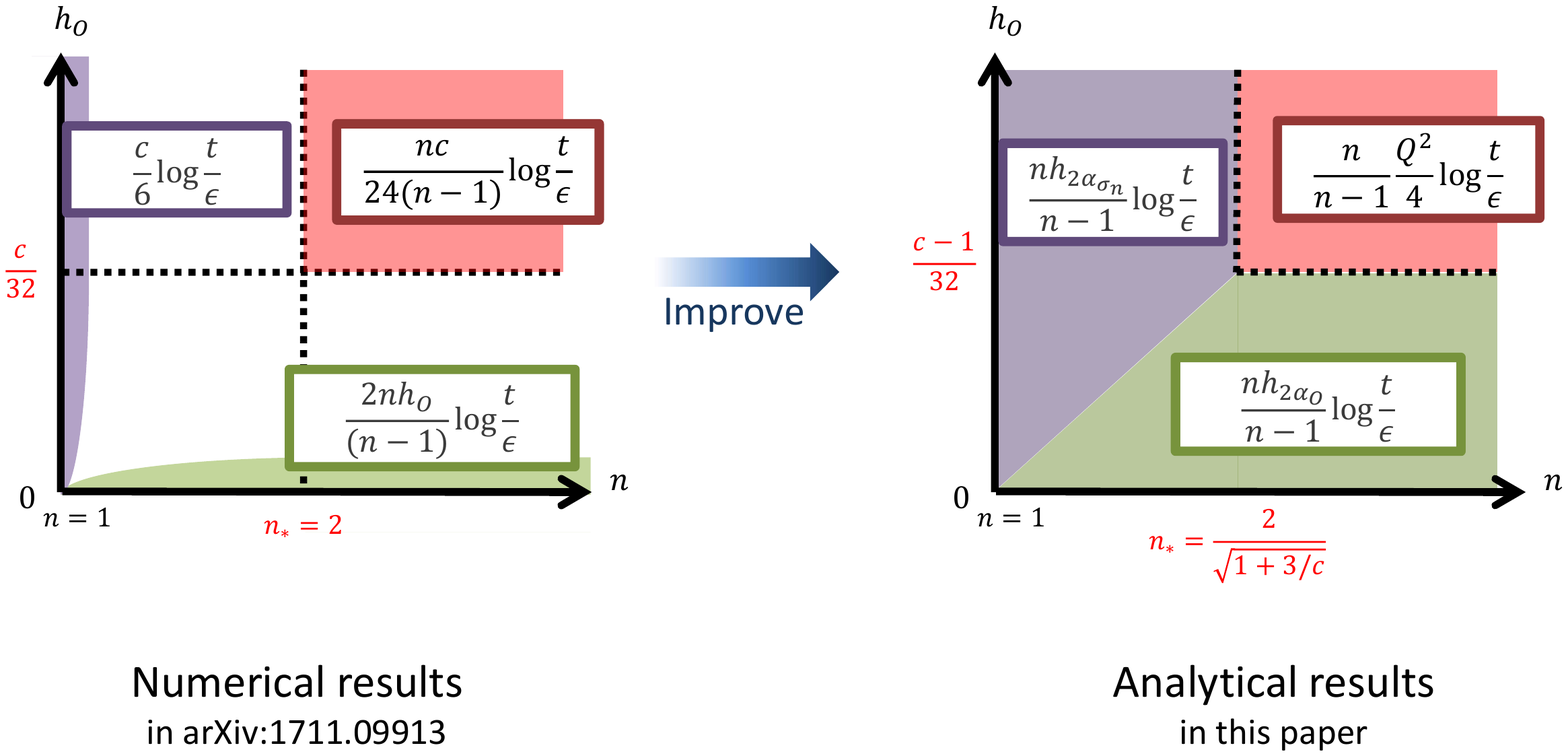}
 \end{center}
 \caption{This figure shows the growth of the Renyi entanglement entropy compared to the vacuum, $\D S^{(n)}_A(t)$. (Left) From the numerical calculations in \cite{Kusuki2018b}, we shed light on the $n$ and $h_O$ dependence of  $\D S^{(n)}_A(t)$ in the purple, green and red regions. However, we do not have any information in the white region. (Right) Our new findings in this paper. Here we introduce the Liouville notations; $h_a=\a(Q-\a)$ and $c=1+6Q^2$.}
 \label{fig:numerical}
\end{figure}

Recently, one of the authors gave numerical results of the leading contribution of the $n$-th Renyi entropy for a locally excited state by using the Zamolodchikov recursion relation \cite{Kusuki2018b}, which suggests that the time dependence of the entanglement entropy after a local quench with conformal dimension $h_O$ can be expressed as in the left figure \ref{fig:numerical}.
The growth of the  $n$-th ($n\geq2$) Renyi entropy compared to the vacuum $\D S^{(n)}_A(t)$ for a light local quench $h_O \ll c$ is given by
\begin{equation}\label{eq:preS1}
\D S^{(n)}_A(t)=\fr{2nh_O}{n-1} \log \fr{t}{\e} + \ca{O}\pa{\pa{t/\e}^0} ,
\end{equation}
whereas the $n$-th ($n\geq2$) Renyi entropy for a heavy local quench $h_0 \geq \fr{c-1}{32}$ shows a {\it universal} growth as
\begin{equation}\label{eq:preS2}
\D S^{(n)}_A(t)=\fr{nc}{24(n-1)} \log \fr{t}{\e} + \ca{O}\pa{\pa{t/\e}^0} .
\end{equation}
On the other hand, the entanglement entropy (i.e.,the limit $n \to 1$ of the Renyi entropy) is just given by the known behavior  \cite{ Asplund2015},
\begin{equation}\label{eq:preS3}
\D S_A(t)=\fr{c}{6} \log \fr{t}{\e} + \ca{O}\pa{\pa{t/\e}^0}  .
\end{equation}
One of our main purposes is to give an analytic proof of these numerical observations and improve the results to perfectly understand the $n$ and $h_O$ dependence of $\D S^{(n)}_A(t)$ in the whole region, including the white region in the left figure \ref{fig:numerical}.
We have to emphasize that our previous result \cite{Kusuki2018b} also relies on the large $c$ limit. In this article, we will relax this assumption, in that, we will study the growth of the Renyi entropy in {\it any} CFTs with $c>1$ and without chiral Virasoro primaries.

The key point to accomplish this work is to understand the {\it light cone singularity} of the Virasoro conformal blocks (i.e.,the limit $z \to 1$ of the blocks) and the {\it Regge singularity} (i.e., the limit $z \to 0$ after picking a monodromy as $(1-z) \to \ex{-2\pi i}(1-z)$).
 This light cone limit is often seen in studying the dynamics of the Lorentzian spacetime, as well as the Lorentzian late time limit of the entanglement entropy after a local quench. Therefore, we have to understand the light cone singularity for our purpose.
The explicit form of the light cone singularity had been unknown until quite recently, but one of the authors investigated this singularity numerically  \cite{Kusuki2018b,Kusuki2018}  and analytically\cite{Kusuki2018a}, and from our findings,
one of the authors succeeded in understanding perfectly the light cone singularity by using the {\it fusion matrix} \cite{Kusuki2018c, Collier2018}.
We will review this fusion matrix approach to derive the light cone singularity in Section \ref{sec:LC}.
With this backdrop, we can proceed in the analytic investigation beyond the previous works resorting to the HHLL approximation \cite{Asplund2015}, and in this article, we utilize this explicit form of the light cone singularity to calculate the Renyi entropy after a local quench in the late time limit. Note that the light cone singularity also appears in other several setups \cite{Asplund2015a, Kusuki2017}, which had already discussed in previous paper \cite{Kusuki2018c}.

Although the light cone singularity is useful to calculate the Renyi entropy, it is not capable of evaluating the $n$-th ($n<2$) Renyi entropy for a technical reason. If we want to go beyond $n\geq2$, we need to know the Regge limit singularity. For this purpose, before calculating the Renyi entropy, we first give the explicit form of the Regge singularity by the use of the {\it monodromy matrix} in Section \ref{sec:Regge}. Consequently, we completely clarify the behavior of the Renyi entropy growth, including the white region in the left figure \ref{fig:numerical}. Our analytic results are summarized in the right figure \ref{fig:numerical}, which are perfectly consistent with our previous numerical results.

The highlight of our results is that the logarithmic growth can be seen not only in large $c$ but also in finite $c$.
It had been expected that the logarithmic growth should be {\it resolved} if taking the non-perturbative effects in $1/c$ into account and the full-order analysis leads to the time dependence of the entanglement entropy that approaches a finite value.
However, our results show the logarithmic growth as in the right figure \ref{fig:numerical}. Another key point is that the Renyi entropy growth at late time has three phases as seen in the right figure \ref{fig:numerical}. In other words, the derivative of the Renyi entropy in $n$ or $h_O$ is not continuous at $n_*=\fr{2}{\s{1+3/c}}$ and $h_O=\fr{c-1}{32}$. This contradicts the assumption that the Renyi entropy would be analytic in $n$. Therefore, we have to consider this exception if we use the replica method to evaluate the entanglement entropy at least in our setup (see \cite{Metlitski2009,Belin2013,Belin2015,Belin2017,Dong2018}, in which we also face with a similar exception).Some other important properties of the Renyi entropy in pure CFTs are summed up in Section \ref{subsec:REEinpureCFT}.

\subsubsection{Topological Part of Entanglement Growth}

We are also interested in the constant part of the Renyi entropy, at late time. Roughly speaking, the late time form of $\D S^{(n)}_A(t)$ is,
\begin{equation}
\D S^{(n)}_A(t)={S^{(n)}}^{\text{geo}}_O \log \fr{t}{\e} +{S^{(n)}}^{\text{top}}_{ O}+ \ca{O}\pa{\pa{t/\e}^{-1}},
\end{equation}
where the coefficient ${S^{(n)}}^{\text{geo}}_O $ is given as in the right figure \ref{fig:numerical} and
the constant part ${S^{(n)}}^{\text{top}}_{ O}$ does not depend on time. 
Actually, as shown in Section \ref{sec:REE}, this constant part is relevant to a fusion matrix (or a monodromy matrix).
This constant part is interesting for the following reasons:

\begin{enumerate}
\item
The Renyi entropy for an excited state in RCFTs had investigated in \cite{He2014}, which states that the Renyi entropy takes the following form;
\begin{equation}\label{eq:RCFTEE}
\D S^{(n)}_A(t)=-\log {\bold F}_{0, 0} [ O] =\log d_{ O} ,
\end{equation}
where $ {\bold F}_{0, 0} [ O]  $ denotes the fusion matrix 
$ {\bold F}_{0, 0} [ O] = {\bold F}_{0, 0} 
   \left[
    \begin{array}{cc}
    \a_{ O}   & \a_{ O}   \\
     \a_{ O}   &   \a_{ O} \\
    \end{array}
  \right]$
and $d_{ O} $ is the quantum dimension. That is, the entropy $\D S^{(n)}_A(t)$ can be interpreted as a measure of  a quantum dimension of $O$. On the other hand, the entropy $\D S^{(n)}_A(t)$ in pure CFTs grows logarithmically and in particular, it diverges in the limit $\frac{t}{\epsilon}\rightarrow \infty$.
We do not have clear physical interpretation of this diverging entropy $\D S^{(n)}_A(t)$ in pure CFTs, unlike in the RCFT case.

However, we find that the fusion matrix also appears in the constant part ${S^{(n)}}^{\text{top}}_{ O}$ ($n\geq2$) in pure CFTs, as shown in Section \ref{subsec:nthREE}. Therefore, we expect that this part plays a role similar to (\ref{eq:RCFTEE}). For general $n$, we show that $\lim_{n \to 1} {S^{(n)}}^{\text{top}}_{ O}$ can be expressed by the {\it monodromy matrix} of the cyclic orbifold $\mathcal{M}^n/\mathbb{Z}_n$, instead of the fusion matrix. We expect that this monodromy matrix can be re-expressed by the fusion matrix, but we leave it in the future.

\item In the limit where $h_{ O} \gg c$, the constant part of the entanglement entropy $\lim_{n \to 1}{S^{(n)}}^{\text{top}}_{ O}$ gives the {\it Cardy entropy},
\begin{equation}
{S^{(1)}}^{\text{top}}_{ O} \ar{h_{ O} \gg c} S_{\text{BH}}(O)=2\pi\s{\fr{c}{6}\pa{h_{ O}-\fr{c}{24}}}.
\end{equation}
It means that ${S^{(1)}}^{\text{top}}_{ O}$ has a physical interpretation as the Bekenstein-Hawking entropy. 
In this article, we show that this B-H entropy comes from the monodromy matrix.
\end{enumerate}

From these viewpoints, it is naturally expected that there is a certain relation between the quantum dimension and the B-H entropy, because both of them come from a special element of the monodromy matrix.
\footnote{The relation between the monodromy matrix and the fusion matrix is shown by (\ref{eq:MtoQD}).}
Possibly, our result gives a natural explanation of the conjecture in \cite{Jackson2015}: For irrational CFTs, ${S^{(n)}}^{\text{top}}_{ O}$ may be defined by
\begin{equation}
{S^{(n)}}^{\text{top}}_{ O} =-\log  \lim_{\beta \to 0} \beta^2 {\bold F}_{\b,\b}[ O].
\end{equation}
The motivation for this definition is because we can obtain the Cardy entropy from this definition,
\begin{equation}\label{eq:Cardy}
{S^{(n)}}^{\text{top}}_{ O} \ar{h_{ O} \gg c} S^{\text{cardy}}_{ O}.
\end{equation}
However, this interesting story which is an analog of that in RCFTs, is not justified straightforwardly. This is because we have to introduce an artificial regularization of the fusion matrix. On the other hand, our result straightforwardly gives the Cardy entropy. In other words, our approach may provide the complete understanding of the relation between the entanglement entropy ${S^{(n)}}^{\text{top}}_{ O}$ and the B-H entropy.

We comment that the fusion matrix has a physical interpretation as {\it three-point function}.
This fact comes from the analytic light cone bootstrap \cite{Kusuki2018c, Collier2018} as explained in Section \ref{sec:LCB}.
Roughly speaking, the fusion matrix (or the residue of that) is related to the OPE coefficient with a large spin primary operator $ O_p$ (with large $h_p$) as
\begin{equation}
 {\bold F}_{0, \bar{h}_p} [ O] \sim C_{ O  O p}.
\end{equation}
This relation may provide a key to understanding why the constant part of the growth $\D S_A(t)$ is given by the fusion matrix.

\subsection{Out-of-Time Ordered Correlator}

One criterion known to characterize quantum chaos is the so-called {\it out-of time ordered correlator} (OTOC). The OTOC is defined by the inner product between two states: $W(t)V\left|0\right>_\beta$ and $VW(t)\left|0\right>_\beta$ where the operators $W$ and $V$ are separated in space by $x$ and in Lorentzian time $t$, and $\beta$ is an inverse temperature\footnote{The definition that we take here is the original that appeared in the context of black holes \cite{Shenker2015}. Clearly, OTOC can be generalized to zero temperature and arbitrary operators.}. 
In this paper, we will deal with the normalized OTOC as
\be\label{eq:introCb}
C_\beta(x,t)\equiv\frac{\braket{V^\dagger W^\dagger(t)V W(t)}_\beta}{\s{\braket{V^\dagger W^\dagger(t)W(t) V }_\beta \braket{ W^\dagger(t)V^\dagger V  W(t) }_\beta }}.
\ee
In the time regime of our interest ($t \gg \b$), the denominator is factorized as a simple form $\braket{V^\dagger V}_\beta \braket{W^\dagger W}_\beta$.
\footnote{In fact, there is an exception of this decomposition rule of the four point function. This is discussed in Section \ref{subsec:LiouvilleOTOC}.}
We usually obtain a Lorentzian correlator by the analytic continuation of a Euclidean correlator so that the product of operators is time-ordered. On the other hand, the OTOC is given by the unusual analytic continuation \cite{Roberts2015}. More precisely,
we consider the analytic continuation of the correlator  $\langle V^\dagger(z_1,\bar{z}_1))  V(z_2,\bar{z}_2)) W^\dagger(z_3,\bar{z}_3)) W(z_4,\bar{z}_4))\rangle_\beta$ so that the order of the operators is given by the form $\langle V^\dagger W^\dagger(t)V W(t)\rangle_\beta$. This is realized by setting the insertion points of the operators on the thermal cylinder as
\begin{equation}
\begin{aligned}
z_1&=\ex{\frac{2\pi}{\beta}(t+ \ii \epsilon_1)},\qquad \bar{z}_1=\ex{-\frac{2\pi}{\beta}(t+ \ii \epsilon_1)},\\
z_2&=\ex{\frac{2\pi}{\beta}(t+\ii \epsilon_2)},\qquad \bar{z}_2=\ex{-\frac{2\pi}{\beta}(t+\ii \epsilon_2)},\\
z_3&=\ex{\frac{2\pi}{\beta}(x+\ii \epsilon_3)},\qquad \bar{z}_3=\ex{\frac{2\pi}{\beta}(x- \ii \epsilon_3)}, \\
z_4&=\ex{\frac{2\pi}{\beta}(x+\ii \epsilon_4)},\qquad \bar{z}_4=\ex{\frac{2\pi}{\beta}(x-\ii \epsilon_4)},
\end{aligned}
\end{equation}
where we analytic continued the correlator to the real time under the assumption $\epsilon_1<\epsilon_3<\epsilon_2<\epsilon_4$.
From these expression, we can find that the time evolution of the cross ratio $z=\frac{z_{12}z_{34}}{z_{13}z_{24}}$ is given by the analytic continuation $(1-z) \to \ex{-2\pi i} (1-z)$. It means that the late time behavior of the OTOC is given by the {\it Regge limit} of the correlator.
This Regge limit simplifies the calculation of the OTOC and consequently, we obtain a lot of information from the OTOC;
in holographic CFTs, we can show that the OTOC exponentially decay in the late time \cite{Roberts2015, Perlmutter2016}, whereas 
this exponential decay cannot be seen in non-chaotic CFTs, where the OTOC approaches non-zero constant \cite{Caputa2016, Gu2016, Fan2018} or decays polynomially \cite{Caputa2017a}. From this viewpoint, we expect that the late time behavior of the OTOC may be a criterion of chaotic nature of a given quantum field theory, in addition to the existing arguments on the Lyapunov exponent \cite{Roberts2015, Maldacena2016, Fitzpatrick2016b} (see also \cite{Turiaci2019}).

Inopportunely, there had been the same problem as the calculation of the entanglement entropy; the explicit form of the Regge singularity had been unknown, therefore, most of the works rely on some assumptions, for example the HHLL approximation. In this paper, we give the explicit form of the Regge singularity by studying the {\it monodromy matrix}, which is related to the fusion matrix. Consequently, we obtain the analytic result about the late time behavior of the OTOC.

In fact,  our monodromy matrix approach can also shed light on the late time behavior of the Liouville OTOC.
The Liouville CFT attracts much attention as a tractable example of irrational CFTs because it has known many developments \cite{Teschner2001a}, compared to other irrational CFTs. Moreover, it has the connection to the 3D gravity \cite{McGough2013,Jackson2015} in some sense. We do not have any idea to interpret this Liouville CFT/gravity connection in the context of the AdS/CFT correspondence, nevertheless, we believe that the Liouville CFT plays an important role in understanding the quantum gravity and the AdS/CFT correspondence (for relevant references, see \cite{Caputa2017b}). In this background, it is expected that the study of the Liouville OTOC gives a key to understanding the Liouville CFT more clearly, therefore, we also studies the Liouville OTOC, apart from the OTOC in the pure CFT.

Although we focus only on the late time behavior of the OTOC, our technique might be also useful to study the OTOC in the scrambling regime. As pointed out in \cite{Liu2018, Hampapura2018}, we have to take the non-vacuum contributions into account to obtain the OTOC in the scrambling time. In \cite{Chang2018}, it is investigated how the non-vacuum contributions appear in the OTOC by using the Zamolodchikov recursion relation. The technical procedure in \cite{Chang2018} is very similar to the study of the entanglement growth \cite{Kusuki2018b} because both of them consider the Regge asymtotics of the correlator.
The analytic proof of conjectures in \cite{Kusuki2018b} from the Zamolodchikov recursion relation is given by the fusion matrix approach (or more precisely, the monodromy matrix approach in Section \ref{subsec:Regge}), which is developed in this paper. Therefore, it is natural to expect that our fusion matrix approach should be also useful to clarify the non-vacuum contributions of the OTOC. We will discuss this issue in more details in Section \ref{subsec:Scrambling}.

\section{Light Cone Bootstrap}

\subsection{Light Cone Singularity from Fusion Matrix}\label{sec:LC}
\newsavebox{\boxpc}
\sbox{\boxpc}{\includegraphics[width=120pt]{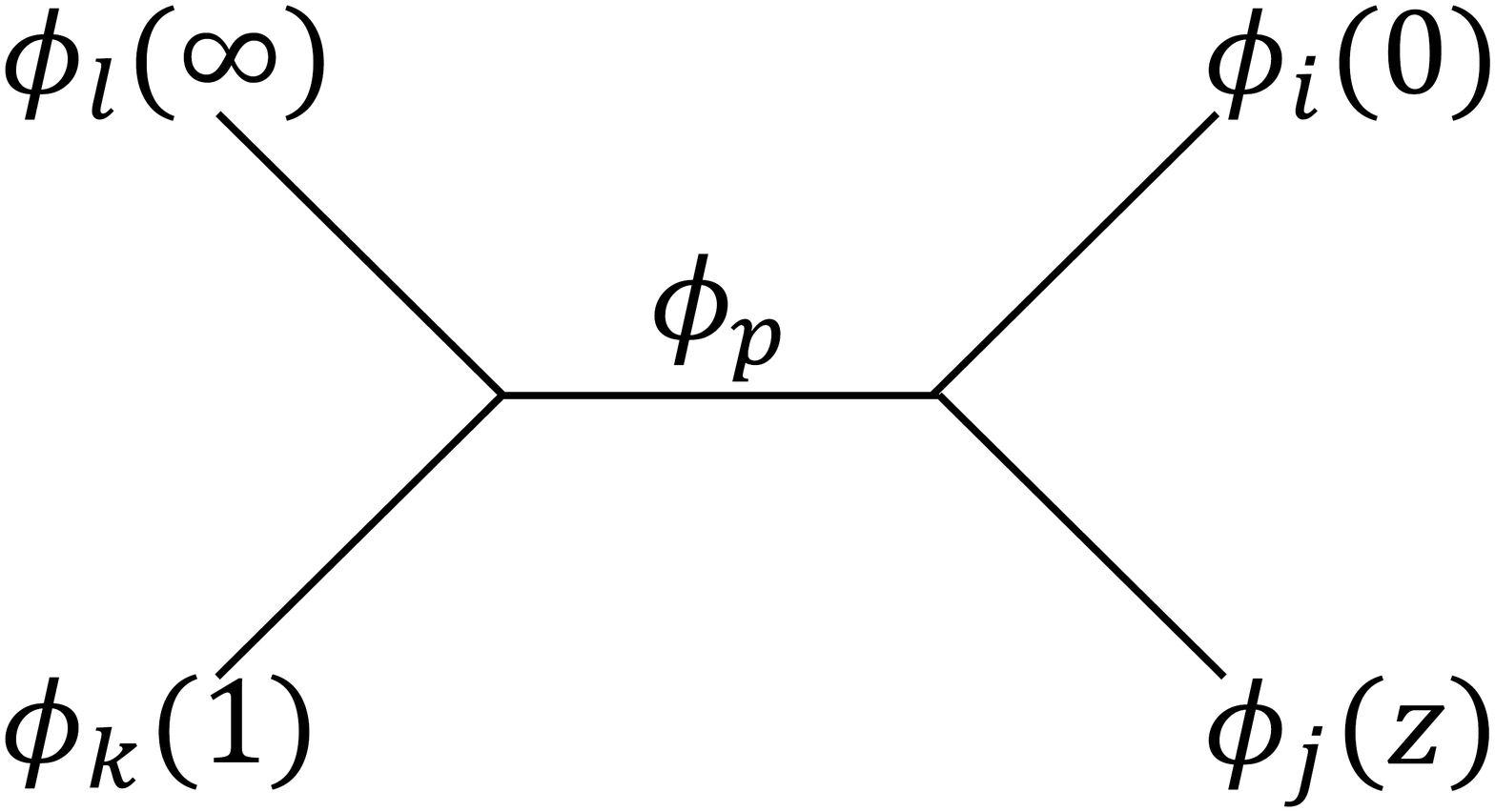}}
\newlength{\pcw}
\settowidth{\pcw}{\usebox{\boxpc}} 

We will give the light cone singularity of the Virasoro conformal blocks.
\footnote{ In this paper, the word {\it light cone limit} has two meanings as follows: For a chiral sector, it means the limit $z \to 1$, whereas for the full correlator, it means the limit $z \to 1$ and $\bar{z} \to 0$ (or  $z \to 0$ and $\bar{z} \to 1$).}
In this paper, we introduce the notation of the Virasoro block as follows,

\begin{equation}\label{eq:blockdef}
 \ca{F}^{ji}_{kl}(h_p|z) \equiv \parbox{\pcw}{\usebox{\boxpc}}.
\end{equation}
We first focus on a special case where $h_i=h_j=h_A$ and $h_k=h_l=h_B$ (but the other case is also given later by (\ref{eq:bootstrapBAAB})).
The key point is that there are invertible fusion transformations between s and t-channel conformal blocks.
In more details, the fusion transformation of the vacuum conformal block is given by the following form,
\footnote{More precisely, if $ c\leq25 $, only the $(n,m)=(0,0)$ term contributes to the sum. However, this difference is not important in this paper.}
\begin{equation}\label{eq:bootstrapAABB2}
\begin{aligned}
&\ca{F}^{AA}_{BB}(0|z)\\
&=
 \sum_{\substack{\a_{n,m}<\fr{Q}{2} \\ n,m \in \bb{Z}_{\geq0}}}\ \text{Res}\pa{   -2\pi i 
  {\bold F}_{0, \a_t} 
   \left[
    \begin{array}{cc}
    \a_A   & \a_A  \\
     \a_B  &   \a_B\\
    \end{array}
  \right]
  \ca{F}^{AB}_{AB}(h_{\a_t}|1-z);\a_t=\a_{n,m}}\\
&+
\int_{\fr{Q}{2}+0}^{\fr{Q}{2}+i \infty} \dd \a_t {\bold F}_{0, \a_t} 
   \left[
    \begin{array}{cc}
    \a_A   & \a_A  \\
     \a_B  &   \a_B\\
    \end{array}
  \right]
  \ca{F}^{AB}_{AB}(h_{\a_t}|1-z),
\end{aligned}
\end{equation}
where $\a_{n,m}\equiv\a_A+\a_B+mb+nb^{-1}$ and kernel $ {\bold F}_{\a_s, \a_t} $ denotes the {\it fusion matrix} (or {\it crossing matrix}), and we introduce the notations usually found in Liouville CFTs,
\begin{equation}
c=1+6Q^2, \ \ \ \ \ Q=b+\fr{1}{b}, \ \ \ \ \ h_i=\a_i(Q-\a_i), \ \ \ \ \ \bar{h}_i=\bar{\a}_i(Q-\bar{\a}_i).
\end{equation}
We would like to emphasize that the discrete terms come from the poles of the fusion matrix (see Appendix A of \cite{Kusuki2018c}).
The explicit form of the fusion matrix is presented in Appendix \ref{app:FM}. Since the asymptotic behavior of the Virasoro blocks is given by
\begin{equation}
\ca{F}^{ji}_{kl}(h_p|z) \ar{z \to 0} z^{h_p-h_i-h_j},
\end{equation}
the light cone singularity is given by
\begin{equation}\label{eq:LC}
\begin{aligned}
&\ca{F}^{AA}_{BB}(0|z)\\
&\ar{z \to 1}
 \sum_{\substack{\a_{n,m}<\fr{Q}{2} \\ n,m \in \bb{Z}_{\geq0}}}\ \text{Res}\pa{   -2\pi i 
  {\bold F}_{0, \a_t} 
   \left[
    \begin{array}{cc}
    \a_A   & \a_A  \\
     \a_B  &   \a_B\\
    \end{array}
  \right]; \a_t=\a_{n,m}}
(1-z)^{h_{\a_{n,m}}-h_A-h_B}  \\
&+
\int_{\fr{Q}{2}+0}^{\fr{Q}{2}+i \infty} \dd \a_t {\bold F}_{0, \a_t} 
   \left[
    \begin{array}{cc}
    \a_A   & \a_A  \\
     \a_B  &   \a_B\\
    \end{array}
  \right]
  (1-z)^{h_{\a_t}-h_A-h_B} .
\end{aligned}
\end{equation}
This light cone singularity is first conjectured in \cite{Kusuki2018b,Kusuki2018} by using the Zamolodchikov recursion relation (see Section \ref{sec:recursion}), shown in \cite{Kusuki2018a} by using the monodromy method and shown in \cite{Kusuki2018c, Collier2018} by using the fusion matrix.

We will note that the leading behavior of the integral part in the light cone limit is given by the root of the integral $\a_t=\fr{Q}{2}$, which reduces the integral form to 
\begin{equation}\label{eq:intsingularity}
\begin{aligned}
\int_{\fr{Q}{2}+0}^{\fr{Q}{2}+i \infty} \dd \a_t {\bold F}_{0, \a_t} 
  & \left[
    \begin{array}{cc}
    \a_A   & \a_A  \\
     \a_B  &   \a_B\\
    \end{array}
  \right]
  (1-z)^{h_{\a_t}-h_A-h_B}\\
&\ar{z \to 1}
   \del_{\a_t}^2 
	\pa{
	{\bold F}_{0, \a_t}	
	\left.
   \left[
    \begin{array}{cc}
    \a_A   & \a_A  \\
     \a_B  &   \a_B\\
    \end{array}
  \right]}
\right|_{\a=\fr{Q}{2}}
 \int_{\fr{Q}{2}+0}^{\fr{Q}{2}+i \infty} \dd \a_t 
	\pa{\a_t-\fr{Q}{2}}^2
  (1-z)^{h_{\a_t}-h_A-h_B} \\
&=
-i\del_{\a_t}^2 
	\pa{
	{\bold F}_{0, \a_t}	
	\left.
   \left[
    \begin{array}{cc}
    \a_A   & \a_A  \\
     \a_B  &   \a_B\\
    \end{array}
  \right]}
\right|_{\a=\fr{Q}{2}}
\fr{\s{\pi}}{8}(1-z)^{\fr{Q^2}{4}-h_A-h_B}\pa{-\log(1-z)}^{-\fr{3}{2}}.
\end{aligned}
\end{equation}
Interestingly, this behavior can be universally seen for the conformal block with $\a_A+\a_B>\fr{Q}{2}$ (where the discrete terms in (\ref{eq:bootstrapAABB2}) disappear) and in fact, we had already shown this universality from the Zamolodchiklov recursion relation  \cite{Kusuki2018b,Kusuki2018}.

\subsection{Light-Cone Conformal Bootstrap}\label{sec:LCB}
The light cone structure (\ref{eq:LC}) made significant progress in the analytic bootstrap in 2D CFTs \cite{Kusuki2018c, Collier2018}. 
From this viewpoint, the fusion matrix has a physical interpretation as {\it three-point function}. We will review \cite{Kusuki2018c, Collier2018} in this section.

Here we  assume that our unitary CFT  has a central charge $c>1$ and there is no chiral Virasoro primaries  (hence, no extra currents apart from the Virasoro current). We call such a CFT a {\it pure CFT}. By definition, pure CFTs are classified as irrational CFTs. It is believed that pure CFTs are ubiquitous and have holographic duals of generic quantum theories of gravity in AdS${}_3$ (see \cite{Collier2016,CollierKravchukLinYin2017}). 

The conformal bootstrap equation is the relation between different OPE coefficients $C_{ijk}$ as
the following form,
\begin{equation}
\begin{aligned}\label{eq:bootstrap}
&\sum_p C_{12p}C_{34p} \ca{F}^{21}_{34}(h_p|z)\overline{\ca{F}^{21}_{34}}(\bar{h}_p|\bar{z})\\
&=\sum_p C_{12p}C_{23p} \ca{F}^{23}_{14}(h_p|1-z)\overline{\ca{F}^{23}_{14}}(\bar{h}_p|1-\bar{z}).
\end{aligned}
\end{equation}
To solve this bootstrap equation at large spin, we focus on the case wherein $h_1=h_2=h_A$ and $h_3=h_4=h_B$ in the limit $z \to 0$.
In the pure CFT, the left-hand side of this equation can be approximated by the vacuum block as
\begin{equation}
\ca{F}^{AA}_{BB}(0|z)\overline{\ca{F}^{AA}_{BB}}(0|\bar{z})=\int \dd h \dd \bar{h}  \ \ \rho_{AB} (h,\bar{h})   \ca{F}^{AB}_{AB}(h|1-z)\overline{\ca{F}^{AB}_{AB}}(\bar{h}|1-\bar{z}).
\end{equation}
where we define $\rho_{AB} (h,\bar{h}) \equiv \sum_p \pa{C_{ABp}}^2\delta(h-h_p)\delta(\bar{h}-\bar{h}_p)  $. We freeze the holomorphic part by the limit $z \to 0$ and factor out the anti-holomorphic part as
\begin{equation}\label{eq:bootstrapAABB}
\overline{\ca{F}^{AA}_{BB}}(0|\bar{z})=\int  \dd \bar{h}  \ \ \rho_{AB} (\infty,\bar{h}) \overline{\ca{F}^{AB}_{AB}}(\bar{h}|1-\bar{z}).
\end{equation}
Here, we define the {\it large spin spectral density},
\begin{equation}
\rho_{AB} (\infty,\bar{h}) \equiv \lim_{z \to 0} \int \dd h  \ \ \rho_{AB} (h,\bar{h}) \fr{ \ca{F}^{AB}_{AB}(h|1-z)}{\ca{F}^{AA}_{BB}(0|z)}.
\end{equation}
Upon comparing (\ref{eq:bootstrapAABB}) and (\ref{eq:bootstrapAABB2}), we find that the twist spectrum in the OPE between $O_A$ and $O_B$ at large spin can be expressed as
\begin{equation}\label{eq:LCresult}
\begin{aligned}
 \bar{h}&=\left\{
    \begin{array}{ll}
   \{ \bar{\a}_{n,m}(Q-\bar{\a}_{n,m}) \ \ \ \text{for $n,m \in \bb{Z}_{\geq 0}$ s.t. $\bar{\a}_{n,m}<\fr{Q}{2}$} \} \cup \kagi{\fr{c-1}{24},\infty}, \\
	\hspace{5cm} (\text{if }  \bar{\a}_A+\bar{\a}_B<\fr{Q}{2}) ,\\ \\
    \kagi{  \fr{c-1}{24},\infty}, \hspace{3.7cm} ( \text{if } \bar{\a}_A+\bar{\a}_B\geq\fr{Q}{2})  ,\\
    \end{array}
  \right.\\
\end{aligned}
\end{equation}
where $\bar{\a}_{n,m}\equiv\bar{\a}_A+\bar{\a}_B+mb+nb^{-1}$. 
This approach is the so-called {\it light-cone bootstrap}.

This spectrum has a simple interpretation.
If we consider the analytic continuation of the Liouville OPE to $\a\notin [0,Q]$, we can obtain the following fusion rules \cite{Ribault2014},
\begin{equation}\label{eq:fusion rule}
\begin{aligned}
\ca{V}_{\a_A} \times \ca{V}_{\a_B} =
 \sum_{\substack{\a_{n,m}<\fr{Q}{2} \\ n,m \in \bb{Z}_{\geq0}}} \ca{V}_{\a_{n,m}}
+ \int^{\fr{Q}{2}+i \infty}_{\fr{Q}{2}+0} \dd \a \ \ca{V}_{\a},
\end{aligned}
\end{equation}
where $\ca{V}_\a$ denotes a primary operator characterized  by a conformal dimension $h=\a(Q-\a)$.
Upon inspecting this fusion rule, we can straightforwardly observe that the twist spectrum (\ref{eq:LCresult}) exactly matches the spectrum originating from the fusion between primary operators in the (extended) Liouville CFT (\ref{eq:fusion rule}).  It means that 
the twist spectrum at large spin approaches that of the {\it (extended) Liouville CFT}, in the same way that the twist spectrum at large spin in higher-dimensional CFTs ($d\geq3$) approaches the spectrum of double-trace states in a {\it generalized free theory}.

Moreover, the large spin spectral density with $\bar{h}<\fr{c-1}{24}$ is given by
\begin{equation}
\rho_{AB} (\infty,\bar{h})=
\sum_{\substack{\bar{\a}_{n,m}<\fr{Q}{2} \\ n,m \in \bb{Z}_{\geq0}}} \text{Res}\pa{   -2\pi i 
  {\bold F}_{0, \ba_t} 
   \left[
    \begin{array}{cc}
    \ba_A   & \ba_A  \\
     \ba_B  &   \ba_B\\
    \end{array}
  \right]; \ba_t=\ba_{n,m}}\delta\pa{\bar{h}-\bar{h}_{\ba_{n,m}}},
\end{equation}
whilst the spectral density with  $\bar{h}\geq\fr{c-1}{24}$,
\begin{equation}
\rho_{AB} (\infty,\bar{h}_{\ba_t})=\fr{1}{2\sqrt{\bar{h}_{\ba_t}-\fr{Q^2}{4}}}
  {\bold F}_{0, \ba_t} 
   \left[
    \begin{array}{cc}
    \ba_A   & \ba_A  \\
     \ba_B  &   \ba_B\\
    \end{array}
  \right].
\end{equation}

Note that in the same way, we can relate the density of primary states to the modular S-matrix.
We define the degeneracy of primary states $\rho (h,\bar{h})$ as $\rho (h,\bar{h})=\sum_{i,j} D_{i,j} \delta(h-h_i) \delta(\bar{h}-\bar{h}_j)$, where the function $D_{i,j}$ denotes the degeneracy of primary operators of weight $(h_i,\bar{h}_j)$. Then, the partition function can be expressed by
\begin{equation}
Z(\tau,\bar{\tau})=\sum_{i,j} D_{i,j} \chi_{h_i}(\tau)\bar{\chi}_{\bar{h}_j}(\bar{\tau})
=\int \dd h \dd \bar{h} \ \ \rho(h,\bar{h}) \chi_h(\tau)\bar{\chi}_{\bar{h}}(\bar{\tau}).
\end{equation}
If we define the {\it large spin spectrum}  as
\begin{equation}
\rho (\infty, \bar{h}) \equiv \lim_{\tau \to i\infty}  \int \dd h \ \ \rho(h,\bar{h}) \fr{\chi_h \pa{-\fr{1}{\tau}}}{\chi_0 (\tau)},
\end{equation}
then by using the modular bootstrap, we find that the large spin spectrum is given by
\begin{equation}\label{eq:large spin spectrum}
\rho(\infty,\bar{h})=S(\bar{h},0).
\end{equation}
where $S(h,0)$ is the modular S-matrix,
\begin{equation}
\begin{aligned}
S(\bar{h},0)=2\s{2}\pa{\bar{h}-\fr{c-1}{24}}^{-\fr{1}{2}} \sinh \pa{2 \pi b \s{\bar{h}-\fr{c-1}{24}}} \\
\times \sinh \pa{2 \pi b^{-1} \s{\bar{h}-\fr{c-1}{24}}}.
\end{aligned}
\end{equation}
In particular, the asymptotic behavior of this spectrum reproduces the Cardy formula \cite{Cardy1986a},
\begin{equation}
\rho(\infty,\bar{h}) \ar{\bar{h} \to \infty} \ex{2\pi \s{\fr{c-1}{6} \pa{\bar{h}-\fr{c-1}{24}}}}.
\end{equation}
More details can be found in Appendix of \cite{Kusuki2018c}.
We have to mention that the relation (\ref{eq:large spin spectrum}) holds only if we consider a CFT with $c>1$ and without chiral Virasoro primaries. If we want to consider minimal models, instead of the large spin spectrum  (\ref{eq:large spin spectrum}), we should use the following definition,
\begin{equation}
\sigma (\infty, \bar{h}) \equiv \lim_{\tau \to i\infty}  \int \dd h \ \ \sigma(h,\bar{h}) \fr{\ex{ \fr{2\pi i \bar{h}}{\tau}}}{\chi_0 (\tau)},
\end{equation}
where $\sigma(b,\bar{h})$ is the degeneracy of primary and {\it descendant} operators of weight $(h,\bar{h})$. We can obtain the same relation as
\begin{equation}\label{eq:large spin spectrum2}
\sigma(\infty,\bar{h})=S(\bar{h},0).
\end{equation}

\subsection{Twist Spectrum as Two Particle State in AdS${}_3$} 

The light-cone conformal bootstrap affords us the large spin spectrum (\ref{eq:LCresult}), which has discrete and continuous sectors. The bulk dual of the continuous sector is also just the spinning BTZ black hole. The discrete spectrum can be interpreted as spinning two-particle states in $AdS_3$. This interpretation can be naturally understood by the helical multi-particle solutions in $AdS_3$ \cite{Hulik2017, Hulik2018}. 
We focus on chiral solutions to pure 3D gravity,
\begin{equation}
\dd s^2=l^2 \pa{-(\dd t-A)^2+\ex{-2 \phi} \dd z \dd \bar{z}},
\end{equation}
where function $\phi$ and the one-form $A$ are defined on the base space parameterized by $z, \bar{z}$ (see \cite{Hulik2017} for more details).
The curves of constant $z$ in this background are geodesics, and in particular, these curves describe a spinning particle trajectory in global AdS${}_3$. 
This means that two-particle states can be described by the Liouville theory in some sense.

From the Brown--Henneaux boundary of these two-(scalar) particle solutions with mass $2 h_A$ and $2 h_B$ in AdS${}_3$, we can obtain the relation $\bar{h}=\bar{\a}_{0,0}(Q-\bar{\a}_{0,0})$, which is the lowest-energy state in the twist spectrum (\ref{eq:LCresult}). It is naturally expected that at the quantum level, the spectrum of this two-particle state can be described by the fusion rules of the extended Liouville CFT (\ref{eq:fusion rule}), and therefore, we can conclude that the twist spectrum of this two-particle state is given as $\bar{h}=\bar{\a}_{n,m}(Q-\bar{\a}_{n,m}), \ \ \ n,m\in \bb{Z}_{\geq 0} \ \ \text{s.t.} \ \ \bar{h}<\fr{c-1}{24}$.

We comment on the difference between three- and higher-dimensional AdS.
It is naturally expected that at large spin, these two particles are well separated, as the
interactions between two objects become negligible at large angular momentum.
In fact, this expectation holds true in higher dimensions, and the twist spectrum is merely given as $\tau=\tau_A+\tau_B+2n, \ \ \ n>0$. This result can be translated into the fact that the twist spectrum at large spin approaches that of double-trace states in a generalized free theory, which is the dual of free quantum field theories in AdS${}_{d\geq4}$. However, gravitational interactions in AdS${}_3$ create a deficit angle, and their effect can be detected even at infinite separation. Therefore, we can observe the universal anomalous dimension $\delta\tau=\tau-\pa{\tau_A+\tau_B+2n}$ in the dual CFT, which is always negative for non zero $\tau_A$ and $\tau_B$ because of the gravitational interaction. Our result suggests that this universal anomalous dimension is governed by the Liouville theory instead of free quantum field theories. This might be closely related to \cite{Jackson2015}.

\subsection{Singularity of $n$-point Block}\label{sec:nLC}

In Section \ref{sec:LCB}, we give the light cone singularity of the 4-point conformal blocks. In this section, we will generalize this result to $n$-point conformal blocks. Any $n$-point function can be decomposed in terms of the $n$-point {\it conformal partial waves} \footnote{
Here we make distinction between {\it conformal block} and {\it conformal partial wave}. Up until now, we have called a conformal partial wave as a conformal block only when it depends on the cross ratios.
}
 in comb channel as
\begin{equation}
\begin{aligned}
&\braket{ O_n (z_{n},\bar{z}_{n})  O_{n-1} (z_{n-1},\bar{z}_{n-1})  \dots  O_2 (z_2 ,\bar{z}_2) O_1 (z_1 ,\bar{z}_1)}\\
&=\sum_{p_1, p_2, \dots}  C_{12p_1} C_{p_1 3 p_2} C_{p_2 4 p_3} \cdots C_{p_{n-4}, n-2, p_{n-3} } C_{p_{n-3}, n-1, n } \abs{\ca{F}^{h_1,h_2,\dots, h_n}_{h_{p_1}, h_{p_2},\dots, h_{p_{n-3}}}(z_1,z_2, \dots, z_{n})}^2,
\end{aligned}
\end{equation}
where the comb channel is defined by
\newsavebox{\boxpg}
\sbox{\boxpg}{\includegraphics[width=250pt]{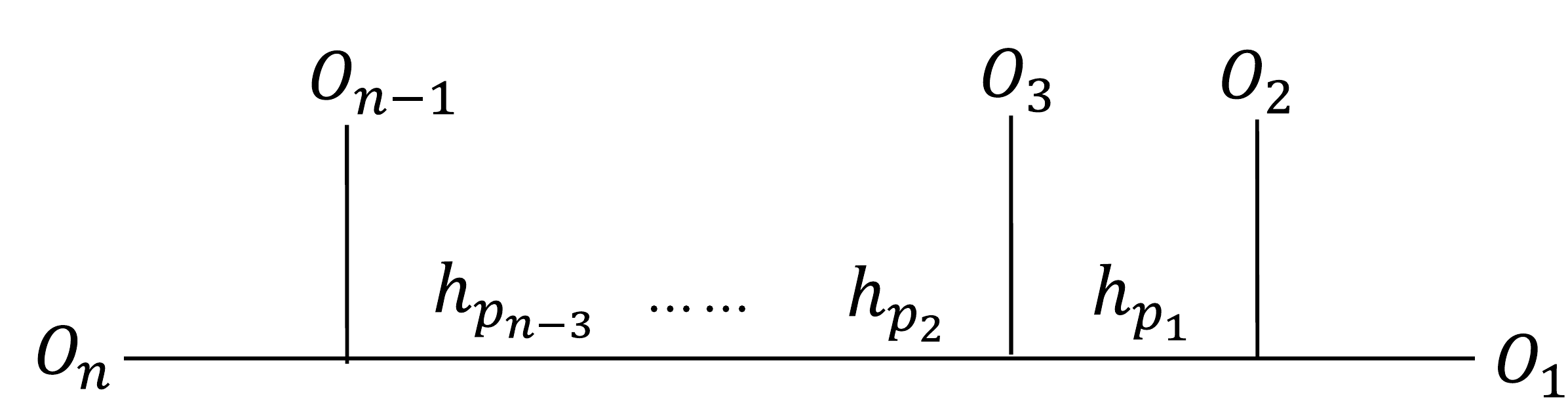}}
\newlength{\pgw}
\settowidth{\pgw}{\usebox{\boxpg}} 

\begin{equation}\label{eq:wavedef}
\ca{F}^{h_1,h_2,\dots, h_n}_{h_{p_1}, h_{p_2},\dots, h_{p_{n-3}}} (z_1,z_2, \dots, z_{n}) \equiv \parbox{\pgw}{\usebox{\boxpg}}.
\end{equation}
In Appendix \ref{app:n-point}, we provide the more explicit definition of $n$-point partial waves(see (\ref{eq:pants})).

The singularity of the $n$-point partial wave can be also given by using the fusion matrix. The result is as follows,

\begin{equation}\label{eq:n-point singularity2}
\begin{aligned}
&\ca{F}^{h_1,h_2,\dots,h_{k},h_{k+1},h_{k+2},h_{k+3}\dots h_n}_{h_{p_1}, h_{p_2},\dots,h_{p_{k-1}}, h_{p_{k}}, h_{p_{k+1}}, \dots   , h_{p_{n-3}}}(z_1,z_2, \dots,z_k, z_{k+1}, z_{k+2}, z_{k+3}, \dots , z_{n})\\
&\ar{z_{k+2, k+1}\to 0}\left\{
    \begin{array}{ll}
\text{Res}\pa{   -2\pi i  \ 
	{\bold F}_{\a_{p_{k}}, \a_{q_{k}}} \left[
	    \begin{array}{cc}
	     \a_{k+1}  &   \a_{p_{k-1}}\\
	    \a_{k+2}   & \a_{p_{k+1}}  \\
	    \end{array}
	  \right] 
	  ; \a_{q_k}=\a_{p_{k-1},p_{k+1}}} 
	(z_{k+2}-z_{k+1})^{h_{p_{k-1},p_{k+1}}-h_{k+1}-h_{k+2}}  \\
	 \times  
	\ca{F}^{h_1,h_2,\dots,h_{k}, h_{p_{k-1},p_{k+1}}  ,h_{k+3}\dots h_n}_{h_{p_1}, h_{p_2},\dots,h_{p_{k-1}}, h_{p_{k+1}}, \dots   , h_{p_{n-3}}}(z_1,z_2, \dots,z_k, z_{k+1}, z_{k+3}, \dots , z_{n}), 
	\hspace{1cm} \text{if } \a_{p_{k-1},p_{k+1}}<\fr{Q}{2},\a_{k+1, k+2}   ,\\ \\ 
\text{Res}\pa{   -2\pi i  \ 
	{\bold F}_{\a_{p_{k}}, \a_{q_{k}}} \left[
	    \begin{array}{cc}
	     \a_{k+1}  &   \a_{p_{k-1}}\\
	    \a_{k+2}   & \a_{p_{k+1}}  \\
	    \end{array}
	  \right] 
	  ; \a_{q_k}=\a_{k+1,k+2}} 
	(z_{k+2}-z_{k+1})^{h_{k+1,k+2}-h_{k+1}-h_{k+2}}  \\
	 \times  
	\ca{F}^{h_1,h_2,\dots,h_{k}, h_{k+1,k+2}  ,h_{k+3}\dots h_n}_{h_{p_1}, h_{p_2},\dots,h_{p_{k-1}}, h_{p_{k+1}}, \dots   , h_{p_{n-3}}}(z_1,z_2, \dots,z_k, z_{k+1}, z_{k+3}, \dots , z_{n}), 
	\hspace{1cm} \text{if } \a_{k+1, k+2} <\fr{Q}{2}, \a_{p_{k-1},p_{k+1}}  ,\\ \\ 
-i\del_{\a_{q_k}}^2 \left. {\bold F}_{\a_{p_{k}}, \a_{q_{k}}} \left[
	    \begin{array}{cc}
	     \a_{k+1}  &   \a_{p_{k-1}}\\
	    \a_{k+2}   & \a_{p_{k+1}}  \\
	    \end{array}
	  \right]  \right|_{\a_{q_k}=\fr{Q}{2}}   \fr{\s{\pi}}{8} (z_{k+2}-z_{k+1})^{\fr{Q^2}{4}-h_{k+1}-h_{k+2}}\pa{-\log (z_{k+2}-z_{k+1})}^{-\fr{3}{2}} \\	 		\times  
	\ca{F}^{h_1,h_2,\dots,h_{k}, \fr{Q^2}{4}  ,h_{k+3}\dots h_n}_{h_{p_1}, h_{p_2},\dots,h_{p_{k-1}}, h_{p_{k+1}}, \dots   , h_{p_{n-3}}}(z_1,z_2, \dots,z_k, z_{k+1}, z_{k+3}, \dots , z_{n}), 
	\hspace{1cm} \text{if } \a_{k+1, k+2}, \a_{p_{k-1},p_{k+1}}  >\fr{Q}{2} ,\\ \\ 
    \end{array}
  \right.\\
&(k=2,3,\cdots, n-4)
\end{aligned}
\end{equation}
where $\a_{i,j} \equiv \a_i+\a_j$ and $h_{i,j} \equiv \a_{i,j}(Q-\a_{i,j})$. We postpone the detailed calculations to Appendix \ref{app:n-point}. In the above, we restrict ourselves to the special limits $z_{k+2,k+1} \to 0$. Actually, we can evaluate more general limits by using the {\it braiding matrix} ${\bold B}_{\a_p, \a_q}^{(\e)}$, which is defined by
\footnote{In fact, the braiding matrix can be re-expressed in terms of the fusion matrix (see (\ref{eq:FtoB})).}
\newsavebox{\boxpe}
\sbox{\boxpe}{\includegraphics[width=120pt]{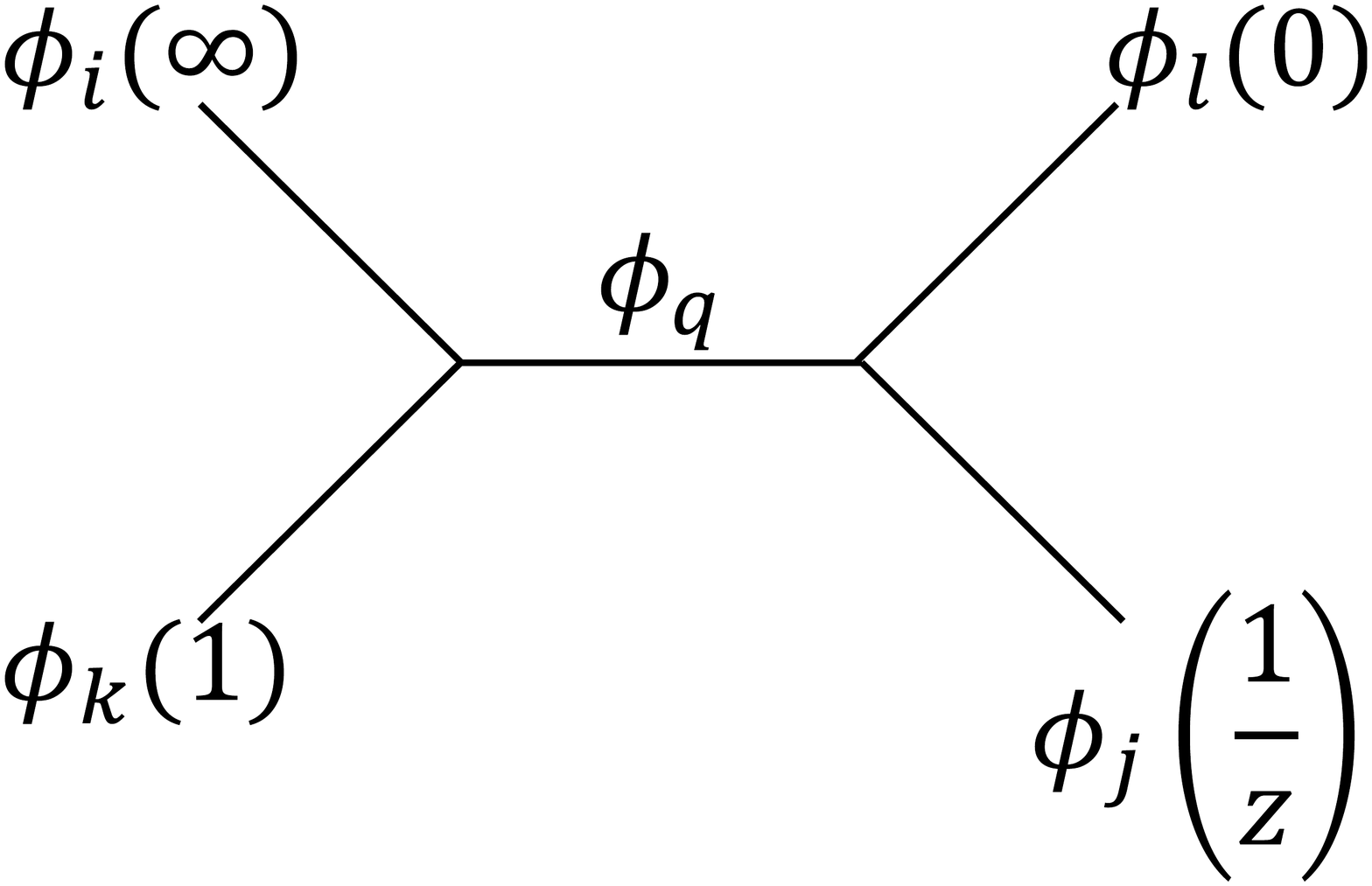}}
\newlength{\pew}
\settowidth{\pew}{\usebox{\boxpe}} 
\begin{equation}\label{eq:Bdef}
\parbox{\pcw}{\usebox{\boxpc}}=
\sum_{\a_q}
 {\bold B}_{\a_p, \a_q}^{(\e)}
   \left[
    \begin{array}{cc}
    \a_j   & \a_i  \\
     \a_k  &   \a_l\\
    \end{array}
  \right]
\parbox{\pew}{\usebox{\boxpe}}.
\end{equation}
Below we only present an important special result from the braiding matrix and present the more general results and the detailed calculation in Appendix \ref{app:n-point}, as the calculation is complicated.

In Section \ref{subsec:nthREE}, we need the following asymptotics,
\begin{equation}
\begin{aligned}
\lim_{\e \to 0}\ca{F}^{h_{\a},h_{\b}, \cdots , h_{\b},h_{\a}}_{h_{\a},\cdots,h_{\a}}\pa{z_1, z_2, z_3, \cdots, z_{n-1}, z_1+\e},
\hspace{1cm}
\b=\left\{
    \begin{array}{ll}
   2\a    ,& \text{if }  \a<\fr{Q}{4}   ,\\
      \fr{Q}{2}   ,& \text{if }   \a>\fr{Q}{4}   ,\\
    \end{array}
  \right.\\
\end{aligned}
\end{equation}
and this asymptotic behavior can be expressed by
\begin{equation}\label{eq:edge limit2}
\begin{aligned}
& \ca{F}^{h_{\a},h_{\b}, \cdots , h_{\b},h_{\a}}_{h_{\a},\cdots,h_{\a}}\pa{z_1, z_2, z_3, \cdots, z_{n-1}, z_1+\e}\\
&\ar{\e  \to 0}\left\{
    \begin{array}{ll}
	\pa{\prod_{k=1}^{n-4}\int_{\bb{S}}\dd \a_{q_{k}} \ {\bold B}_{\a, \a_{q_{k}}}
	   \left[
   	 \begin{array}{cc}
     	      \a   &     \a_{q_{k-1}}\\
     	      2\a  &  \a \\
   	 \end{array}
  	\right]}
\text{Res}\pa{   -2\pi i  \ 
{\bold B}_{\a, \a_{q_{n-3}}}
	   \left[
   	 \begin{array}{cc}
     	      \a   &     \a_{q_{n-4}}\\
     	      2\a  &  \a \\
   	 \end{array}
  	\right]
	  ; \a_{q_{n-3}}=2\a}  \\
 \hspace{1cm} \times  
	\ca{F}^{h_{2\a},h_{2\a} \cdots , h_{2\a},h_{2\a}}_{h_{q_1},h_{q_2}, \cdots,h_{q_{n-4}}  }   \pa{z_2, z_3, \cdots, z_{n-1}, z_1}
	\e^{h_{2\a}-2h_{\a}}, 
	\hspace{4cm} \text{if } \a<\fr{Q}{4}  ,\\ \\ 
	\pa{\prod_{k=1}^{n-4}\int_{\bb{S}}\dd \a_{q_{k}} \ {\bold B}_{\a, \a_{q_{k}}}
	   \left[
   	 \begin{array}{cc}
     	      \a   &     \a_{q_{k-1}}\\
     	      \fr{Q}{2}  &  \a \\
   	 \end{array}
  	\right]}
\pa{-i  \fr{\s{\pi}}{8} \del_{\a_{q_{n-3}}}^2 \left. 
	{\bold B}_{\a, \a_{q_{n-3}}}
	   \left[
   	 \begin{array}{cc}
     	      \a   &     \a_{q_{n-4}}\\
     	      \fr{Q}{2}  &  \a \\
   	 \end{array}
  	\right]  \right|_{\a_{q_{n-3}}=\fr{Q}{2}}   }\\
 \hspace{1cm} \times  
	\ca{F}^{\fr{Q^2}{4} ,\fr{Q^2}{4} \cdots , \fr{Q^2}{4}, \fr{Q^2}{4} }_{h_{q_1},h_{q_2}, \cdots,h_{q_{n-4}}  }   \pa{z_2, z_3, \cdots, z_{n-1}, z_1} 
	\e^{\fr{Q^2}{4}-2h_a}\pa{-\log \e}^{-\fr{3}{2}} , 
	\hspace{2.2cm} \text{if } \a>\fr{Q}{4},
    \end{array}
  \right.\\
\end{aligned}
\end{equation}
where $\a_{q_0}=0$.
To avoid cumbersome expressions, we introduce the following shorthand notation,
\begin{equation}
\begin{aligned}
& \bb{f}^{h_{\a},h_{\b}, \cdots , h_{\b},h_{\a}}_{h_{\a},\cdots,h_{\a}}\pa{z_1, z_2, z_3, \cdots, z_{n-1}, z_1}
&\equiv \left\{
    \begin{array}{ll}
\lim_{\e \to 0}
	\fr{\ca{F}^{h_{\a},h_{\b}, \cdots , h_{\b},h_{\a}}_{h_{\a},\cdots,h_{\a}}\pa{z_1, z_2, z_3, \cdots, z_{n-1}, z_1-\e}}{	\e^{h_{2\a}-2h_{\a}} },
	\hspace{1cm} \text{if } \a<\fr{Q}{4}  ,\\ \\ 
\lim_{\e \to 0}
	\fr{\ca{F}^{h_{\a},\fr{q^2}{4}, \cdots , \fr{Q^2}{4},h_{\a}}_{h_{\a},\cdots,h_{\a}}\pa{z_1, z_2, z_3, \cdots, z_{n-1}, z_1-\e}}{\e^{\fr{Q^2}{4}-2h_a}\pa{-\log \e}^{-\fr{3}{2}} },
	\hspace{1cm} \text{if } \a>\fr{Q}{4},
    \end{array}
  \right.\\
\end{aligned}
\end{equation}
where we flip the sign of $\e$ in $\ca{F}^{h_{\a},h_{\b}, \cdots , h_{\b},h_{\a}}_{h_{\a},\cdots,h_{\a}}$ for later convenience. We will call it the {\it regularized partial wave}.
Note that in the test mass limit ($c \to \infty$ with $h_\a$ fixed),  the regularized partial wave with a special set of $z_i$'s reduces to a simple form as
\begin{equation}
 \bb{f}^{h_{\a},h_{\b}, \cdots , h_{\b},h_{\a}}_{h_{\a},\cdots,h_{\a}}
\pa{\ex{2 \pi i \fr{1}{n} }, \ex{2 \pi i \fr{2}{n} } , \ex{2 \pi i \fr{3}{n} }  ,\ex{2 \pi i \fr{4}{n} } ,\cdots, \ex{2 \pi i \fr{n-1}{n} }  , \ex{2 \pi i \fr{n}{n} }  , \ex{2 \pi i \fr{1}{n} }}
\ar{\fr{h_\a}{c} \to 0}
(-1)^{nh_\a}\pa{\fr{ 1}{2 \sin\pa{\fr{\pi}{n}}}}^{2n h_\a}.
\end{equation}
For later use, we introduce the following notation,
\begin{equation}\label{eq:bbf}
\begin{aligned}
\bb{f}(h_\a) 
\equiv 
\fr{\bb{f}^{h_{\a},h_{\b}, \cdots , h_{\b},h_{\a}}_{h_{\a},\cdots,h_{\a}}
{\pa{\ex{2 \pi i \fr{1}{n} }, \ex{2 \pi i \fr{2}{n} } , \ex{2 \pi i \fr{3}{n} }  ,\ex{2 \pi i \fr{4}{n} } ,\cdots, \ex{2 \pi i \fr{n-1}{n} }  , \ex{2 \pi i \fr{n}{n} }  , \ex{2 \pi i \fr{1}{n} }}  }}{
(-1)^{nh_\b}\pa{\fr{ 1}{2 \sin\pa{\fr{\pi}{n}}}}^{2n h_\b}}
\end{aligned}
\end{equation}
In particular, this quantity approaches 1 in the test mass limit, that is, $ \lim_{\fr{h_\a}{c} \to 0} \bb{f}(h_\a) = 1$. Moreover we expect that the identity $ \bb{f}(h_\a) = 1$ is satisfied for any $\a$ but there is no proof for now. 

These fusion transformations in multi-point conformal partial waves have a number of potential applications. In Section \ref{sec:REE}, we will apply them to evaluating the evaluation of the Renyi entropy via the replica method.

\section{Regge Singularity from Monodromy Matrix Approach}\label{sec:Regge}

\subsection{Regge Singularity from Monodromy Matrix}\label{subsec:Regge}

In 2D CFTs, the {\it Regge limit} is defined by the limit $z, \bar{z}\to 0$ after picking up a monodromy around $z=1$ as $(1-z) \to \ex{-2\pi i}(1-z)$  \cite{Cornalba2007,Cornalba2007a} (see also \cite{Cornalba2007b, Costa2012}).
\footnote{Like the light cone limit, the word {\it Regge limit} also has two meanings as follows: For a chiral sector, it means the limit $z \to 0$ after picking a monodromy around $z =1$, whereas for the full correlator, it means the limit $z \to 0$ with the monodromy around $z=1$ and $\bar{z} \to 0$ without  crossing any branch cuts of the conformal block.}
 In the same way as the light cone singularity, we can also derive this Regge singularity from the monodromy matrix.

\begin{figure}[t]
 \begin{center}
  \includegraphics[width=12.0cm,clip]{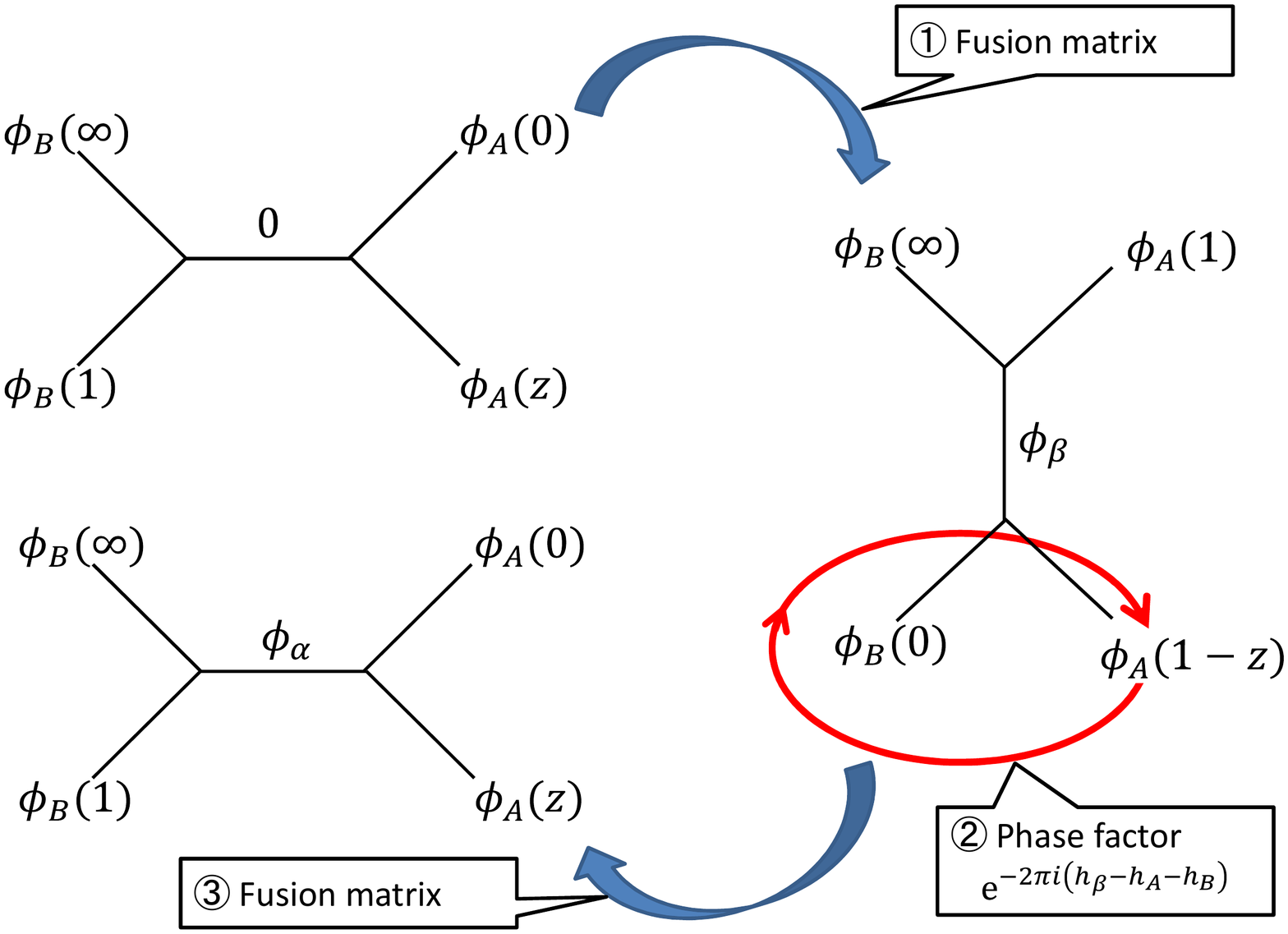}
 \end{center}
 \caption{The monodromy matrix is given by three steps; (1) fusion transformation, (2) picking up a trivial phase factor, and (3) fusion transformation.}
 \label{fig:monomatrix}
\end{figure}

Let us denote the contribution of the monodromy around $z=1$ to the conformal block by the monodromy matrix as
\begin{equation}\label{eq:monodromy matrix}
\begin{aligned}
\ca{F}^{AA}_{BB}(h_p|z)
\ar{(1-z) \to \ex{-2\pi i}(1-z)}
\int_{\mathbb{S}'} \dd \a \ {\bold M}_{\a_p, \a}
   \left[
    \begin{array}{cc}
    \a_A   & \a_A  \\
     \a_B  &   \a_B\\
    \end{array}
  \right]
  \ca{F}^{AA}_{BB}(h_{\a}|z),
\end{aligned}
\end{equation}
where the contour $\bb{S}'$ runs from $\fr{Q}{2}$ to $\fr{Q}{2}+ i\infty$, and also runs anti-clockwise around $\a=2\a_A+mb+nb^{-1}<\fr{Q}{2}$ and $\a=2\a_B+mb+nb^{-1}<\fr{Q}{2}$. 
In fact, this monodromy matrix can be expressed in terms of the fusion matrix as
\footnote{This definition is equivalent to that introduced in \cite{Moore1989} (see (\ref{eq:M=BB})).}
\begin{equation}\label{eq:defmono}
 {\bold M}_{\a_p, \a}
   \left[
    \begin{array}{cc}
    \a_A   & \a_A  \\
     \a_B  &   \a_B\\
    \end{array}
  \right]
=
\int_{\mathbb{S}} \dd \b \ 
\ex{-2\pi i (h_\b-h_A-h_B)}
 \ {\bold F}_{\a_p, \b}
   \left[
    \begin{array}{cc}
    \a_A   & \a_A  \\
     \a_B  &   \a_B\\
    \end{array}
  \right]
 \ {\bold F}_{\b, \a}
   \left[
    \begin{array}{cc}
    \a_A   & \a_B  \\
     \a_A  &   \a_B\\
    \end{array}
  \right],
\end{equation}
where the contour $\bb{S}$ runs from $\fr{Q}{2}$ to $\fr{Q}{2}+ i\infty$, and also runs anti-clockwise around $\b=\a_A+\a_B+mb+nb^{-1}<\fr{Q}{2}$ for $m, n\in \mathbb{Z}_{\geq 0}$.
From the $t$-channel representation, the contribution from the monodromy around $z=1$ is simply given by the phase factor, therefore, we can re-express the monodromy matrix by the simple form in terms of the fusion matrix (see Figure \ref{fig:monomatrix}). We will verify this relation between the monodromy matrix and the fusion matrix directly in the Ising model in Appendix \ref{app:Ising}.

At this stage, we find that in the same way as the fusion transformation (\ref{eq:bootstrapAABB2}), there are contributions from the pole of the monodromy matrix associated to those of the fusion matrix in the integrand. To proceed further, we need the fusion transformation law of the conformal block $\ca{F}^{AB}_{AB}(h_{\a_s}|z)$, instead of  the block $\ca{F}^{AA}_{BB}(0|z)$.
\begin{equation}\label{eq:bootstrapBAAB}
\begin{aligned}
&\ca{F}^{AB}_{AB}(h_{\a_s}|z)\\
&=
 \sum_{\substack{\a_{n,m}^{(I)}<\fr{Q}{2} \\ n,m \in \bb{Z}_{\geq0}\\ I=A,B}}\ \text{Res}\pa{   -2\pi i 
  {\bold F}_{\a_s, \a_t} 
   \left[
    \begin{array}{cc}
    \a_A   & \a_B \\
     \a_A  &   \a_B\\
    \end{array}
  \right]
  \ca{F}^{AA}_{BB}(h_{\a_t}|1-z);\a_t=\a_{n,m}^{(I)}}\\
&+
\int_{\fr{Q}{2}+0}^{\fr{Q}{2}+i \infty} \dd \a_t {\bold F}_{\a_s, \a_t} 
   \left[
    \begin{array}{cc}
    \a_A   & \a_B  \\
     \a_A  &   \a_B\\
    \end{array}
  \right]
  \ca{F}^{AA}_{BB}(h_{\a_t}|1-z),
\end{aligned}
\end{equation}
where $\a_{n,m}^{(I)}\equiv 2\a_I+mb+nb^{-1}$. 
In particular, we find that the light cone singularity of this type of the block is given by
\begin{equation}\label{eq:BAABlightcone}
\begin{aligned}
\ca{F}^{AB}_{AB}(h_{\a_s}|z)
&\ar{z \to 1} \left\{
    \begin{array}{ll}
   \text{Res}\pa{    -2\pi i  
  {\bold F}_{\a_s, \a_t}^T [\a_A,\a_B]
  ; \a_t=2\a_A} (1-z)^{h_{2\a_A}-2h_A}   ,& \text{if } \a_A<\fr{Q}{4},\a_B   ,\\
     \text{Res}\pa{    -2\pi i
  {\bold F}_{\a_s, \a_t}^T [\a_A,\a_B]
  ; \a_t=2\a_B} (1-z)^{h_{2\a_B}-2h_A}   ,& \text{if } \a_B<\fr{Q}{4},\a_A   ,\\
    -i\del_{\a_t}^2 \left. {\bold F}_{\a_s,\a_t}^T [\a_A,\a_B] \right|_{\a_t=\fr{Q}{2}}   \fr{\s{\pi}}{8} (1-z)^{\fr{Q^2}{4}-2h_A}\pa{-\log (1-z)}^{-\fr{3}{2}}   ,& \text{if } \a_A,\a_B>\fr{Q}{4} ,\\
    \end{array}
  \right.\\
\end{aligned}
\end{equation}
where we abbreviate the fusion matrix 
$  {\bold F}_{\a_s, \a_t} 
   \left[
    \begin{array}{cc}
    \a_A   & \a_B  \\
     \a_A  &   \a_B\\
    \end{array}
  \right]  $
as $  {\bold F}_{\a_s, \a_t}^T [\a_A,\a_B] $, which is consistent with our previous result \cite{Kusuki2018} (see Section \ref{sec:recursion}).

From this pole structure of the fusion matrix  $  {\bold F}_{\a_s, \a_t}^T [\a_A,\a_B] $, we can find that the pole structure of the monodromy matrix is given by
\begin{equation}\label{eq:monodromy matrix2}
\begin{aligned}
\ca{F}^{AA}_{BB}(h_p|z)
&\ar{(1-z) \to \ex{-2\pi i}(1-z)}
 \sum_{\substack{\a_{n,m}^{(I)}<\fr{Q}{2} \\ n,m \in \bb{Z}_{\geq0}\\ I=A,B}}\ \text{Res}\pa{   -2\pi i 
  {\bold M}_{\a_p, \a} 
   \left[
    \begin{array}{cc}
    \a_A   & \a_A  \\
     \a_B  &   \a_B\\
    \end{array}
  \right]
  \ca{F}^{AA}_{BB}(h_{\a}|z);\a=\a_{n,m}^{(I)}}\\
&+\int_{\fr{Q}{2}+0}^{\fr{Q}{2}+i \infty} \dd \a \ {\bold M}_{\a_p, \a}
   \left[
    \begin{array}{cc}
    \a_A   & \a_A  \\
     \a_B  &   \a_B\\
    \end{array}
  \right]
  \ca{F}^{AA}_{BB}(h_{\a}|z).
\end{aligned}
\end{equation}
In particular, this result leads to the explicit form of the Regge limit as
\begin{equation}\label{eq:Reggesing}
\begin{aligned}
\ca{F}^{AA}_{BB}(h_p|z)
&\ar{\text{Regge limit}} \left\{
    \begin{array}{ll}
    \text{Res}\pa{   -2\pi i 
  {\bold M}_{\a_p, \a} [\a_A,\a_B]
  ; \a=2\a_A} z^{h_{2\a_A}-2h_A}   ,& \text{if } \a_A<\fr{Q}{4},\a_B   ,\\
    \text{Res}\pa{   -2\pi i 
  {\bold M}_{\a_p, \a} [\a_A,\a_B]
  ; \a=2\a_B} z^{h_{2\a_B}-2h_A}   ,& \text{if } \a_B<\fr{Q}{4},\a_A   ,\\
    -i\del_{\a}^2 \left. {\bold M}_{\a_p,\a}[\a_A,\a_B] \right|_{\a=\fr{Q}{2}}   \fr{\s{\pi}}{8} z^{\fr{Q^2}{4}-2h_A}\pa{-\log z}^{-\fr{3}{2}}   ,& \text{if } \a_A,\a_B>\fr{Q}{4} ,\\
    \end{array}
  \right.\\
\end{aligned}
\end{equation}
where we used the saddle point analysis as in (\ref{eq:intsingularity}).
For simplicity, we abbreviate the monodromy matrix
 $ {\bold M}_{\a_p, \a} 
   \left[
    \begin{array}{cc}
    \a_A   & \a_A  \\
     \a_B  &   \a_B\\
    \end{array}
  \right]$
 as $  {\bold M}_{\a_p, \a} [\a_A,\a_B] $.

This result exactly reproduces the numerical results of the Regge singularity \cite{Kusuki2018b}.
In particular, when we set $h_A=\fr{c}{24}\pa{1-\fr{1}{n^2}}$ and $h_B=n h_O$, we obtain the Renyi entanglement entropy after a local quench, which are perfectly consistent with the numerical observations  (\ref{eq:preS1})$\sim$(\ref{eq:preS3}).
We will explain it in more details later.
Note that the Regge singularity for $\a_A,\a_B>\fr{Q}{4} $ had been derived by using the light cone singularity, which is the so-called {\it Regge limit universality}  \cite{Kusuki2018c}.

\subsection{HHLL Virasoro Block and Regge Singularity}
As a consistency check of (\ref{eq:Reggesing}), we consider the Regge limit of  the HHLL conformal block at large $c$, focusing on the residue of the monodromy matrix $ 
  {\bold M}_{\a_p, \a} [\a_L,\a_H]$ at the leading pole $\a= 2\a_L$, where we have $\a_L  \approx  b h_L$ and $\a_H=O(b^{-1})$.
\footnote{In this paper, “$\approx$” means an approximation by extracting a leading contribution.}
 The residue can be expressed as
  
\begin{equation}
\begin{aligned}
   &\text{Res}\pa{   
  {\bold M}_{\a_p, \a} [\a_L,\a_H]
  ; \a=2\a_L}\\
&=\int_{\mathbb{S}} \dd\b \    
  {\bold F}_{\a_p, \b} \left[
    \begin{array}{cc}
    \a_L   & \a_L  \\
     \a_H  &   \a_H\\
    \end{array}
  \right]\ex{-2\pi i(h_{\b}-h_L-h_H)}\text{Res}\pa{{\bold F}_{\b, \a} \left[
    \begin{array}{cc}
    \a_L   & \a_H  \\
     \a_L  &   \a_H\\
    \end{array}
  \right]  ; \a=2\a_L}.  
\end{aligned}
\end{equation}
 
 Let us first consider the vacuum block with two heavy operators beyond the threshold and two light operators, namely we assume $\a_p=0$, $\a_L=bh_L+O(b^2)$ and $\a_H=\fr{Q}{2}+ib^{-1}p$. 
 In the HHLL limit,
 $\text{Res}\pa{   
  {\bold F}_{\b, \a} \left[
    \begin{array}{cc}
    \a_L   & \a_H  \\
     \a_L  &   \a_H\\
    \end{array}
  \right]  ; \a=2\a_L}$ does not have poles in terms of $\b$, and the poles of ${\bold F}_{\a_p, \b} \left[
    \begin{array}{cc}
    \a_L   & \a_L  \\
     \a_H  &   \a_H\\
    \end{array}
  \right]$ are all simple and located at $\b=\a_H+\a_L+mb$ with $m\in \mathbb{Z}_{\geq 0}$. Therefore, by summing up the residues at large $c$ and using (B.10) and (6.6) of \cite{Collier2018}, we have
 
    \begin{equation}\label{eq:HHLLMonodromy}
     \text{Res}\pa{   -2 \pi i
  {\bold M}_{0, \a} [\a_L,\a_H]
  ; 2\a_L}
\ar{\text{HHLL limit}}
 \pa{\fr{ip}{\sinh\pa{2\pi p}}}^{2h_L}
=\pa{\fr{i}{2} \fr{\g_H}{\sin \pa{\pi \g_H}}}^{2h_L},
\end{equation}
with $\gamma_H=\sqrt{1-\fr{24h_{H}}{c}}$. 
Then the Regge limit of the HHLL block is given by
\begin{equation}
\ca{F}^{LL}_{HH}(0|z)
\ar{\text{Regge limit}}  
\pa{\fr{i}{2} \fr{\g_H}{\sin \pa{\pi \g_H}}}^{2h_L},
\end{equation}
which coincides exactly with the Regge limit of the HHLL block obtained in \cite{Fitzpatrick2014}. 

Let us turn our eyes on nonzero, order one $h_p$ case, with $\a_L\approx bh_L$ and $\a_H\approx \eta b^{-1}~(0\leq \eta\leq \fr{1}{2})$. In this case, $\text{Res}\pa{   
  {\bold F}_{\b, \a} \left[
    \begin{array}{cc}
    \a_L   & \a_H  \\
     \a_L  &   \a_H\\
    \end{array}
  \right]  ; \a=2\a_L}$ does not have poles in terms of $\b$ again. On the other hand, ${\bold F}_{\a_p, \b} \left[
    \begin{array}{cc}
    \a_L   & \a_L  \\
     \a_H  &   \a_H\\
    \end{array}
  \right]$ has double and simple poles at $\b=\a_H+\a_L+mb$. By summation of these residues and double residues in the HHLL limit, using (B.10) and (6.12) of \cite{Collier2018}, one can show
    
  \begin{equation}\label{eq:HHLLMonodromyGeneral}
     \text{Res}\pa{   - 2 \pi i
  {\bold M}_{\a_p, \a} [\a_L,\a_H]
  ; \a=2\a_L}\ar{\fr{h_L}{c}\to0}  \ex{2\pi ih_L(1-\gamma_H)}\pa{\fr{1-\ex{-2\pi i\gamma_{H}}}{\gamma_H}}^{-2h_L+h_p}{}_2F_1[h_p,h_p,2h_p; 1-\ex{-2\pi i\gamma_H}].
\end{equation}
Then the Regge limit of the HHLL conformal block is given by

\begin{equation}
\ca{F}^{LL}_{HH}(h_p|z)
\ar{\text{Regge limit}} \ex{2\pi ih_L(1-\gamma_H)}\pa{\fr{1-\ex{-2\pi i\gamma_{H}}}{\gamma_H}}^{-2h_L+h_p}{}_2F_1[h_p,h_p,2h_p; 1-\ex{-2\pi i\gamma_H}],\end{equation}
       which coincides exactly with the Regge limit of the HHLL conformal block with a generic intermediate state obtained in \cite{Fitzpatrick2015}. 
       
\subsection{General Solutions to Zamolodchikov Recursion Relation} \label{sec:recursion}

Besides the light cone singularity, the Regge singularity has also analyzed by the Zamonodchikov recursion relation \cite{Kusuki2018b}.
\footnote{This recursion relation is first derived in  \cite{Zamolodchikov1987,Zamolodchikov1984}, developed by  \cite{Cho2017a} and recently used in the context of the bootstrap \cite{EsterlisFitzpatrickRamirez2016, BaeLeeLee2016, Chen2017, LinShaoSimmons-DuffinWangYin2017,CollierKravchukLinYin2017} and the calculation of the entanglement entropy or the OTOC \cite{Kusuki2018b, Ruggiero2018, Chang2018}.} 
 We will review our previous results and check the consistency with our new result for the Regge singularity (\ref{eq:Reggesing}).

In general, the conformal block can be decomposed into a universal part $\Lambda^{21}_{34}(h_p|q)$  and a non-trivial part $H^{21}_{34}(h_p|q)$ as
\begin{equation}
\ca{F}^{21}_{34}(h_p|z)=\Lambda^{21}_{34}(h_p|q)H^{21}_{34}(h_p|q),\ \ \ \ \ \ q(z)=\ex{-\pi \fr{K(1-z)}{K(z)}},
\end{equation}
where $K(x)$ is the elliptic integral of the first kind. The explicit form of $\Lambda^{21}_{34}(h_p|q)$ is
\begin{equation}\label{eq:pre}
 \Lambda^{21}_{34}(h_p|q)=(16q)^{h_p-\frac{c-1}{24}}z^{\frac{c-1}{24}-h_1-h_2}(1-z)^{\frac{c-1}{24}-h_2-h_3}
(\theta_3(q))^{\frac{c-1}{2}-4(h_1+h_2+h_3+h_4)}.
\end{equation}
The non-trivial part $H^{21}_{34}(h_p|q)$ is recursively defined as follows. For simplicity, we expand $H^{21}_{34}(h_p|q)$ as
\be\label{eq:seriesH}
H^{21}_{34}(h_p|q)=1+\sum_{k=1}^\infty c_k(h_p) q^{k},
\ee
then the coefficients $ c_k(h_p)$ can be calculated by the following recursion relation,
\footnote{The relation between this recursion relation for $ c_k(h_p)$ and the original recursion relation \cite{Zamolodchikov1987,Zamolodchikov1984} can be found in \cite{Chen2017}, which gives an explanation how to solve the recursion relation efficiently in a very nice way.}
\begin{equation}\label{eq:ck}
	c_k(h_p) = \sum_{i=1}^k \sum_{\substack{m=1, n=1\\mn=i}} \frac{R_{m,n}}{h_p - h_{m,n}} c_{k-i}(h_{m,n}+mn),
\end{equation}
where $R_{m,n}$ is a constant in $q$, which is defined by
\begin{equation}\label{eq:Rmn}
R_{m,n}=2\fr{
\substack{m-1\\ \displaystyle{\prod} \\p=-m+1\\p+m=1 (\text{mod } 2) \  } \ 
\substack{n-1\\ \displaystyle{\prod} \\q=-n+1\\q+n=1 (\text{mod } 2) }
\pa{\la_2+\la_1-\la_{p,q}}\pa{\la_2-\la_1-\la_{p,q}}\pa{\la_3+\la_4-\la_{p,q}}\pa{\la_3-\la_4-\la_{p,q}}}
{\substack{
\substack{m \\ \displaystyle{\prod} \\k=-m+1 } \ \ 
\substack{n \\ \displaystyle{\prod} \\l=-n+1 }\\
(k,l)\neq(0,0), (m,n)
}
 \la_{k,l}}.
\end{equation}
In the above expressions, we used the notations,
\begin{equation}
\begin{aligned}
&c=1+6\pa{b+\fr{1}{b}}^2,  \hspace{16ex}   h_i=\fr{c-1}{24}-\l_i^2,\\
&h_{m,n}=\fr{1}{4}\pa{b+\fr{1}{b}}^2-\lambda_{m,n}^2,  \hspace{10ex}  \lambda_{m,n}=\fr{1}{2} \pa{\fr{m}{b}+nb}.
\end{aligned}
\end{equation}

Our previous numerical computations suggest that the solution $c_n$ for large $n$ takes the simple {\it Cardy-like} form of
\begin{equation}\label{eq:cn}
c_n \sim\xi_n  n^{\a} \ex{A\s{n}} \ \ \ \ \ \ \ \ \text{for large $n\gg c$},
\end{equation}
where 
\begin{equation}\label{eq:xin}
\begin{aligned}
\xi_n&=\left\{
    \begin{array}{ll}
      \d_{n,\text{even}} \times \sgn\BR{\pa{h_A-\fr{c-1}{32}}\pa{h_B-\fr{c-1}{32}}}^{\fr{n}{2}} ,& \text{for }  \ca{F}^{AA}_{BB}(h_p|z)   ,\\
      1 ,& \text{for } \ca{F}^{BA}_{BA}(h_p|z)   .\\
    \end{array}
  \right.\\
\end{aligned}
\end{equation}
The parameters $A$ and $\a$ are non-trivial, depending on the external conformal dimensions and the central charge as follows,
\begin{quote}
{\large $\ca{F}^{AA}_{BB}(h_p|z)$ case:}
(For simplicity, we assume $h_B\geq h_A$, but it does not matter.)\\
\begin{equation}\label{eq:AABBA}
\begin{aligned}
A&=\left\{
    \begin{array}{ll}
    2\pi\s{\fr{Q^2}{4}-h_A-h_B+2\a_A\a_B}   ,& \text{if } h_A, h_B>\fr{c-1}{32}  ,\\
    \pi\s{\fr{Q^2}{4}-2h_A}   ,& \text{if } h_B>\fr{c-1}{32}>h_A  ,\\
   0  ,& \text{if } \fr{c-1}{32}>h_A, h_B  ,\\
    \end{array}
  \right.\\
\end{aligned}
\end{equation}

\begin{equation}\label{eq:AABBal}
\begin{aligned}
\a&=\left\{
    \begin{array}{ll}
    2(h_A+h_B)-\fr{c+5}{8}   ,& \text{if } h_A, h_B>\fr{c-1}{32}  ,\\
    4(h_A+h_B)-\fr{c+9}{4}  ,& \text{otherwise }   .\\
    \end{array}
  \right.\\
\end{aligned}
\ \ \ \ \ \ \ \ \ \ \ \ \ \ \ \ \ \ \ \ \ \ 
\end{equation}

{\large $\ca{F}^{BA}_{BA}(h_p|z)$ case:} 
(For simplicity, we assume $h_B>h_A$, but it does not matter.)\\
\begin{equation}\label{eq:ABBAA}
\begin{aligned}
A&=\left\{
    \begin{array}{ll}
    2\pi\s{\fr{Q^2}{4}-h_{2\a_A}}   ,& \text{if } h_A<\fr{c-1}{32}  ,\\
   0   ,& \text{if } h_A>\fr{c-1}{32}  ,\\
    \end{array}
  \right.\\
\end{aligned}
\end{equation}

\begin{equation}\label{eq:ABBAal}
\begin{aligned}
\a&=\left\{
    \begin{array}{ll}
    2(h_A+h_B)-\fr{c+5}{8}   ,& \text{if } h_A<\fr{c-1}{32} ,\\
    4(h_A+h_B)-\fr{c+9}{4} ,& \text{if } h_A>\fr{c-1}{32}   .\\
    \end{array}
  \right.\\
\end{aligned}
\ \ \ \ \ \ \ \ \ \ 
\end{equation}
\end{quote}

We find that the summands in $H^{BA}_{BA}(h_p|q)$ are always positive, therefore, we can use the approximation,
\begin{equation}\label{eq:sumcn}
\begin{aligned}
\sum_{n\in\bb{Z}_{\geq0}} \xi_n n^{\a}\ex{A\s{n}} q^n 
&\ar{q \to 1}\left\{
    \begin{array}{ll}
     \ex{-\fr{A^2}{4\log q}}\pa{\log q}^{-\fr{3}{2}-2\a}  ,& \text{if } A>0  ,\\
     \pa{\log q}^{-1-\a}  ,& \text{if } A=0  .\\
    \end{array}
  \right.\\
\end{aligned}
\end{equation}
Inserting the expression for $A$ (\ref{eq:ABBAA}) into this approximation form provides the light cone singularity,
\begin{equation}
\begin{aligned}
\ca{F}^{AB}_{AB}(h_p|z)
&\ar{z \to 1}
\left\{
    \begin{array}{ll}
     (1-z)^{h_{2\a_A}-2h_A}  ,& \text{if } h_A<\fr{c-1}{32}   ,\\
     (1-z)^{\fr{Q^2}{4}-2h_A} \pa{\log(1-z)}^{-\fr{3}{2}}  ,& \text{if }  h_A>\fr{c-1}{32}, \\
    \end{array}
  \right.\\
\end{aligned}
\end{equation}
where we used the asymptotics of the elliptic nome,
\begin{equation}
\log q(z) \ar{z\to 1} \fr{\pi^2}{\log \fr{1-z}{16}}
\end{equation}
These behaviors perfectly match (\ref{eq:BAABlightcone}). 
Alternatively, we can also say that our analytic derivation of the asymptotic properties supports our conjecture for the general solution to the Zamolodchikov recursion relation.

The situation for the block $\ca{F}^{AA}_{BB}(h_p|z)$  is more complicated because of the intricate sign pattern (\ref{eq:xin}).
For $h_A, h_B>\fr{c-1}{32}$ or $h_A,h_B<\fr{c-1}{32}$, we find $\xi_n q^n$ with $q \sim 1$ to be always positive, therefore, we can approximate $H^{AA}_{BB}(h_p|q)$ in the same way as $H^{AB}_{AB}(h_p|q)$,
\begin{equation}\label{eq:sumcn2}
\begin{aligned}
\sum_{n=0,2,4,\cdots } \xi_n n^{\a}\ex{A\s{n}} q^n 
&\ar{q \to 1}\left\{
    \begin{array}{ll}
     \ex{-\fr{A^2}{4\log q}}\pa{\log q}^{-\fr{3}{2}-2\a}  ,& \text{if } A>0  ,\\
     \pa{\log q}^{-1-\a}  ,& \text{if } A=0  .\\
    \end{array}
  \right.\\
\end{aligned}
\end{equation}
This provides the light cone singularity as
\begin{equation}
\begin{aligned}
\ca{F}^{AA}_{BB}(h_p|z)
&\ar{z \to 1}
\left\{
    \begin{array}{ll}
     (1-z)^{-2\a_A\a_B}  ,& \text{if } h_A,h_B<\fr{c-1}{32}   ,\\
     (1-z)^{\fr{Q^2}{4}-h_A-h_B} \pa{\log(1-z)}^{-\fr{3}{2}}  ,& \text{if }  h_A,h_B>\fr{c-1}{32}, \\
    \end{array}
  \right.\\
\end{aligned}
\end{equation}
which is consistent with (\ref{eq:LC}) and (\ref{eq:intsingularity}).
On the other hand, the sign pattern $\xi_n$ for $h_B>\fr{c-1}{32}>h_A$ spoils the approximation (\ref{eq:sumcn2}) in the limit $q \to 1$. Nevertheless, we can derive interesting results from the general solution to the recursion relation. In fact, the {\it Regge asymptotics} of the elliptic nome is given by
\begin{equation}
q(z) \ar{\text{Regge limit}} i \ex{\fr{\pi^2}{4\log\fr{z}{16}}} \sim i,
\end{equation}
therefore, $\xi_{2n} q^{2n}$ is always positive in the Regge limit. As a result, we obtain the Regge limit of the block as
\begin{equation}
\begin{aligned}
\ca{F}^{AA}_{BB}(h_p|z)
&\ar{\text{Reege limit}}
    \ca{O}(1)  ,& \text{if } h_B>\fr{c-1}{32}>h_A   ,\\
\end{aligned}
\end{equation}
which is consistent with the Regge limit of the HHLL Virasoro block. But now we know the more accurate asymptotics in the Regge limit from the monodromy matrix approach (\ref{eq:Reggesing}).  That is, we can go beyond the HHLL approximation in our way.
From the Regge asymptotics (\ref{eq:Reggesing}), we can deduce that the improved version of (\ref{eq:AABBA}) is
\begin{equation}
\begin{aligned}
A&=\left\{
    \begin{array}{ll}
    2\pi\s{\fr{Q^2}{4}-h_A-h_B+2\a_A\a_B}   ,& \text{if } h_A, h_B>\fr{c-1}{32}  ,\\
    \pi\s{\fr{Q^2}{4}-h_{2\a_A}}   ,& \text{if } h_B>\fr{c-1}{32}>h_A  ,\\
   0  ,& \text{if } \fr{c-1}{32}>h_A, h_B.\\
    \end{array}
  \right.\\
\end{aligned}
\end{equation}

\begin{figure}[h]
 \begin{center}
  \includegraphics[width=8.0cm,clip]{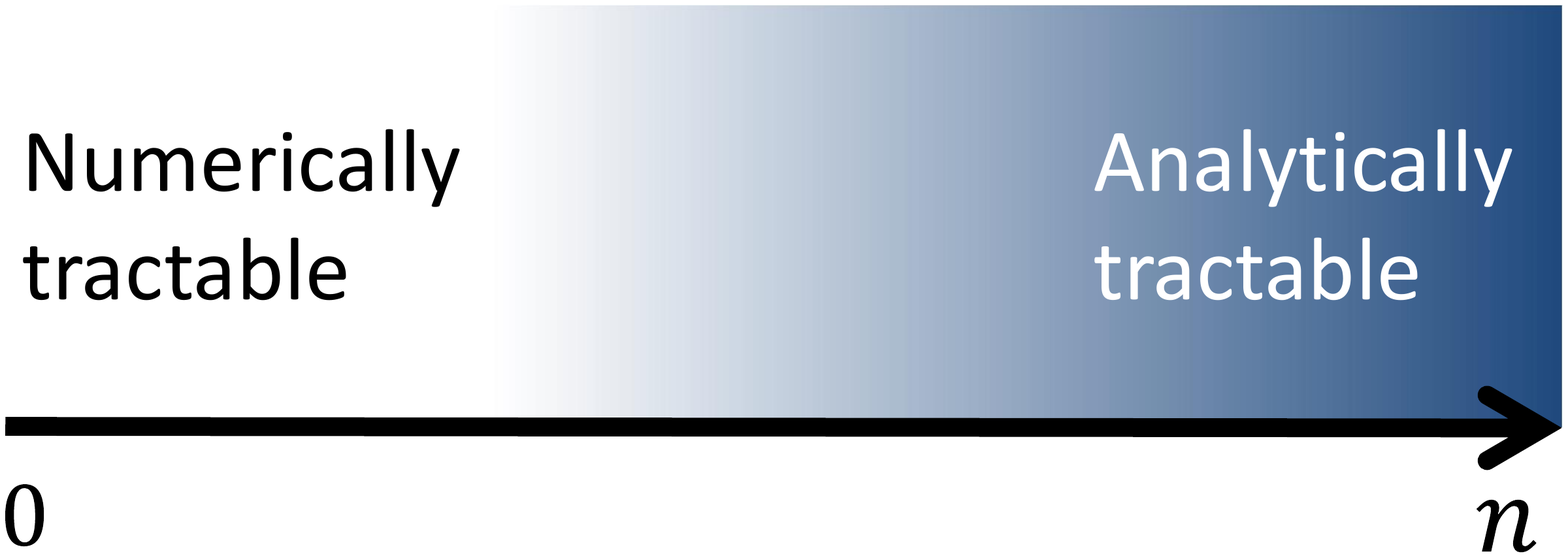}
 \end{center}
 \caption{The coefficients $c_n$ for small $n$ can be calculated numerically. On the other hand, the coefficients for large $n$ can be analytically evaluated by using the fusion or the monodromy matrix approaches. }
 \label{fig:tractable}
\end{figure}

We would like to mention that there results for the solution to the recursion relation are useful to investigate unknown Virasoro blocks by the recursion relation. The computational time complexity to calculate $c_n$ up to $n=N$ by the method \cite{Chen2017} is roughly $\ca{O}(N^3 (\log N)^2)$. Therefore, we do not want to calculate $c_n$ for very large $n$. Now that we have the asymptotics of $c_n$ for large $n$,  we can use it instead of calculating them. On the other hand, the coefficients $c_n$ for lower $n$ can be calculated easily by the numerical method. That is, combining numerical calculation and analytical asymptotics, we can investigate unkown Virasoro blocks very efficiently (see Figure \ref{fig:tractable}).

\section{Bulk Interpretation of Light Cone and Regge limit Singularities}

\begin{figure}[h]
 \begin{center}
  \includegraphics[width=7.0cm,clip]{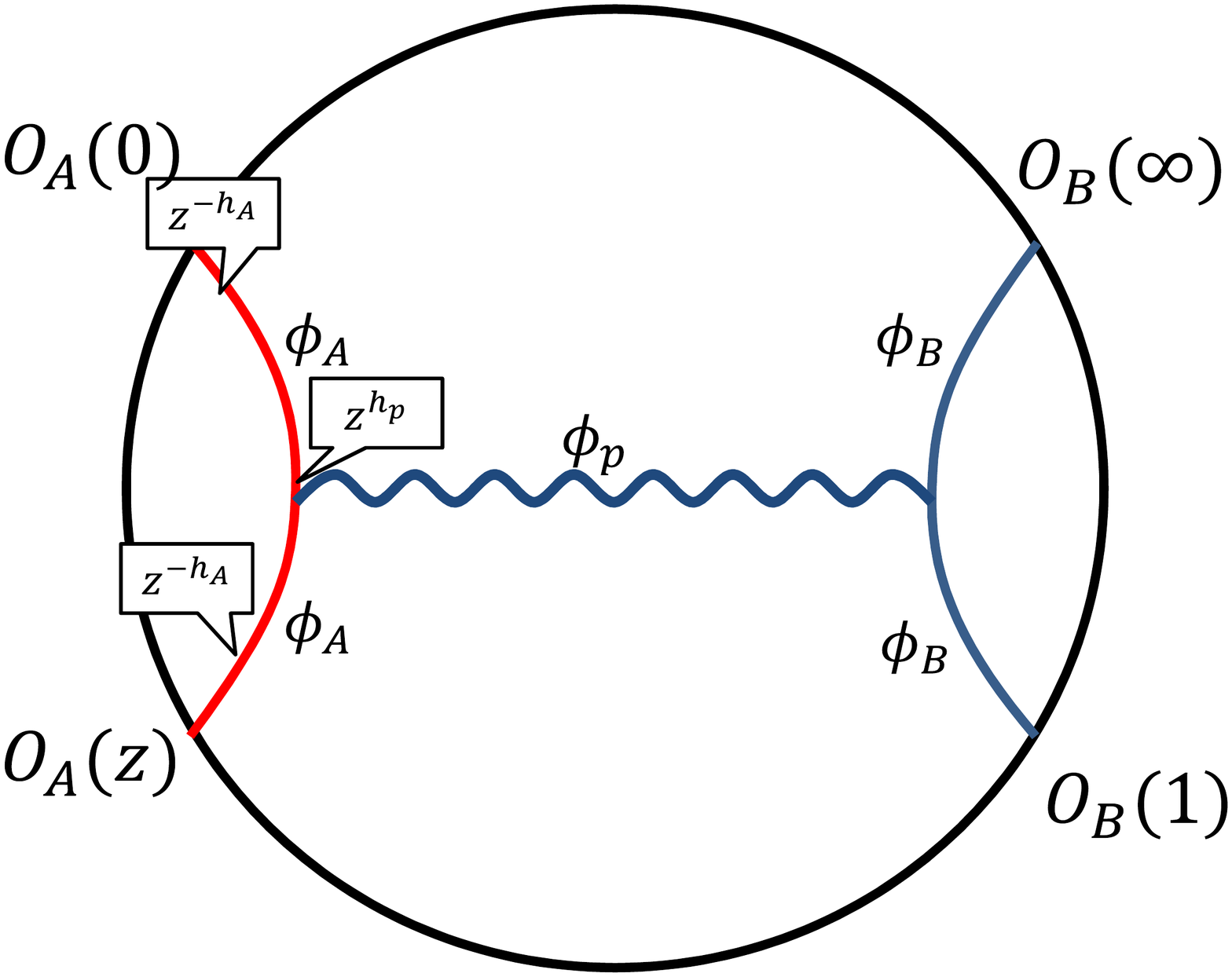}
  \includegraphics[width=7.0cm,clip]{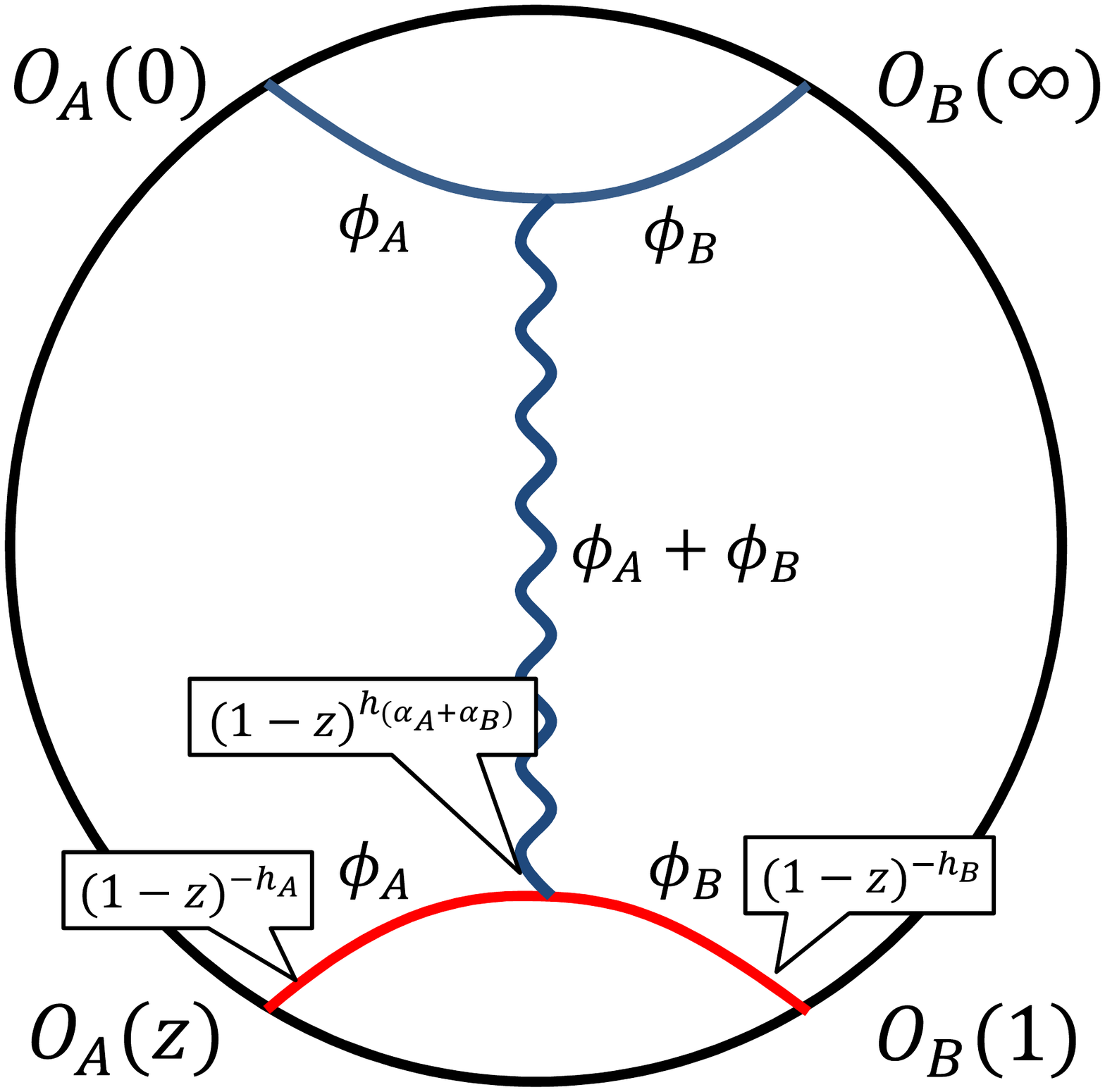}
  \includegraphics[width=7.0cm,clip]{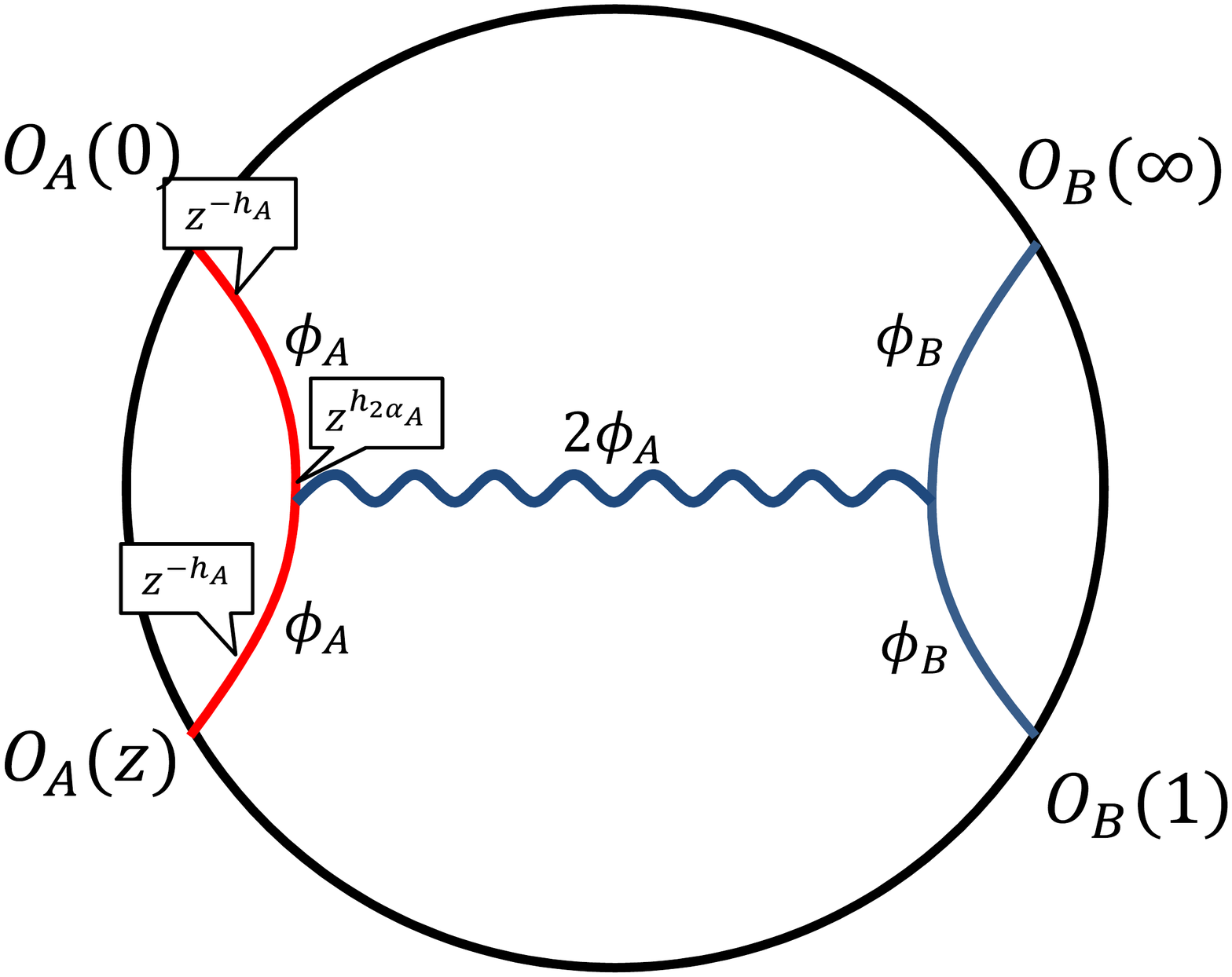}
 \end{center}
 \caption{The simple interpretation of the light cone limit and Regge limit singularities. The left upper figure shows the trivial OPE singularity $z^{h_p-2h_A}$. The upper right figure shows the light cone singularity  (\ref{eq:LC}), which predicts that the interaction between $O_A$ and $O_B$ is characterized by the linear combination of their deficit angles $\{\phi_A, \phi_B\}$, or equivalently, by the fusion rule of the Liouville CFT (\ref{eq:fusion rule}). The lower figure shows the Regge limit singularity (\ref{eq:Reggesing}).}
 \label{fig:Poincare}
\end{figure}

The results for the light cone and the Regge limit singularities are given by  (\ref{eq:LC}), (\ref{eq:BAABlightcone}) and (\ref{eq:Reggesing}). We can simply interpret them by using diagrams as in Figure \ref{fig:Poincare}. 
To make clearer, we exhibit the explicit relation between the Liouville momentum and the deficit angle; $\phi=\fr{4\pi}{Q}\a$.
We expect that the OPE limit of the Virasoro block is obtained by a product of the propagations in AdS${}_3$ without taking account of the backreaction.
For example, the trivial OPE limit ($z \to 0$) can be calculated by a product of three parts $z^{-h_A}$, $z^{-h_A}$ and $z^{h_p}$, which reproduces the well-known OPE singularity $z^{h_p-2h_A}$ (see the upper left of Figure \ref{fig:Poincare}). On the other hand, it is naturally expected from the expression of the light cone singularity that if we take $z$ close to $1$, then two operators $O_A$ and $O_B$ fuse into a particular operator, which is described by the sum of deficit angles $\phi_A+\phi_B$ and the conformal block is given by  a product of three parts $(1-z)^{-h_A}$, $(1-z)^{-h_A}$ and $(1-z)^{h_{\a_A+\a_B}}$ as in the upper right of Figure \ref{fig:Poincare}. In the same way, the Regge limit is described by a diagram with the intermediate deficit angle, $\min\{2\phi_A, 2\phi_B\}$, as in the lower of Figure \ref{fig:Poincare}. These are just natural and simple diagram expressions and we do not have any evidence for this interpretation. However, we believe that this interpretation (in term of the deficit angle fusion) would capture the essence of the relation between the Virasoro block and the geodesics. We plan to return this challenge of understanding the connection in the future.

\section{Renyi Entropy after Local Quench}\label{sec:REE}

\begin{figure}[t]
 \begin{center}
  \includegraphics[width=9.0cm]{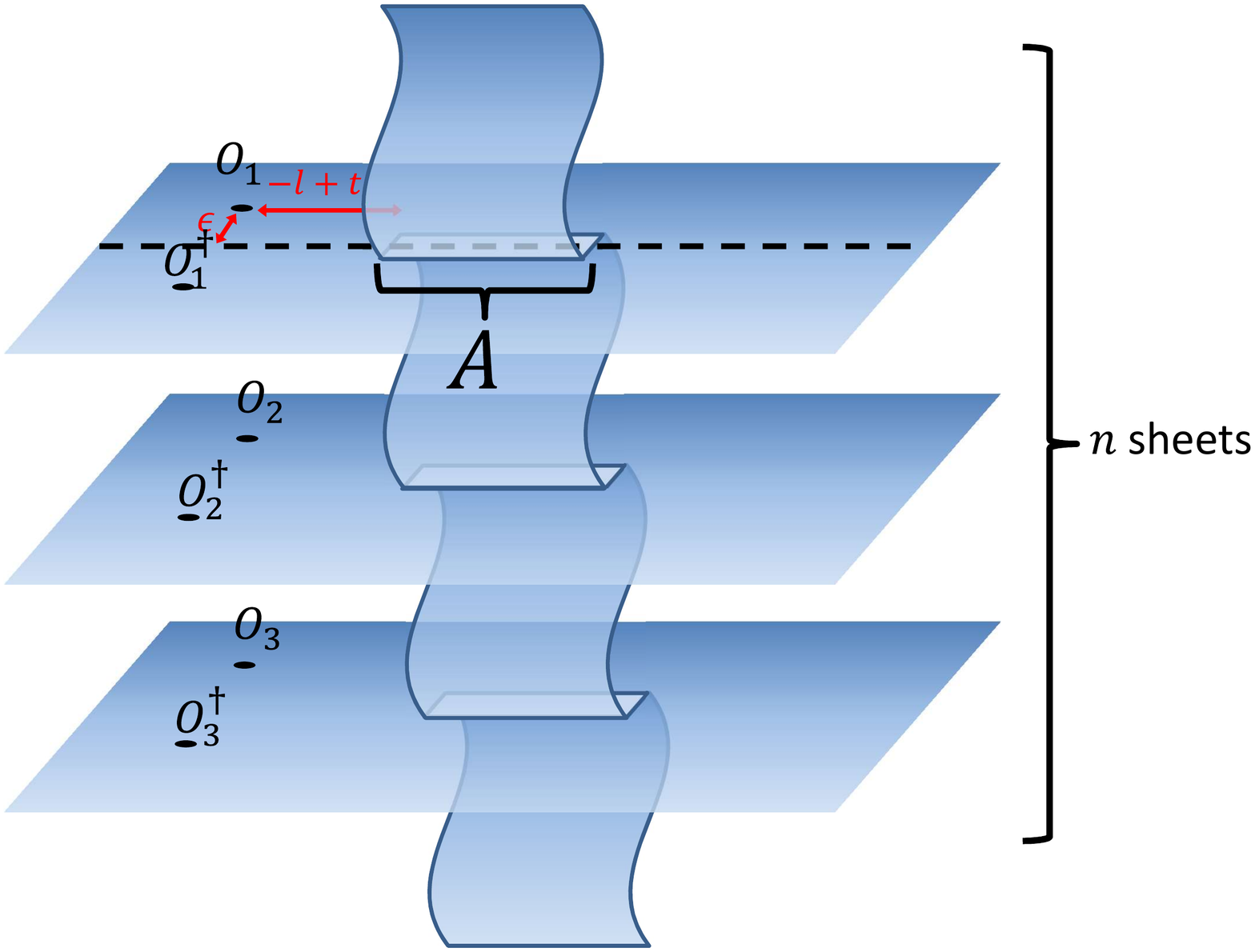}
 \end{center}
 \caption{The $n$-sheeted manifold $\Sigma_n$. The holomorphic insertion points of operators are given by (\ref{eq:coor}) on each sheet. We label the operator on $i$-th sheet as $O_i$.}
 \label{fig:sheet}
\end{figure}

\begin{figure}[t]
 \begin{center}
  \includegraphics[width=7.0cm]{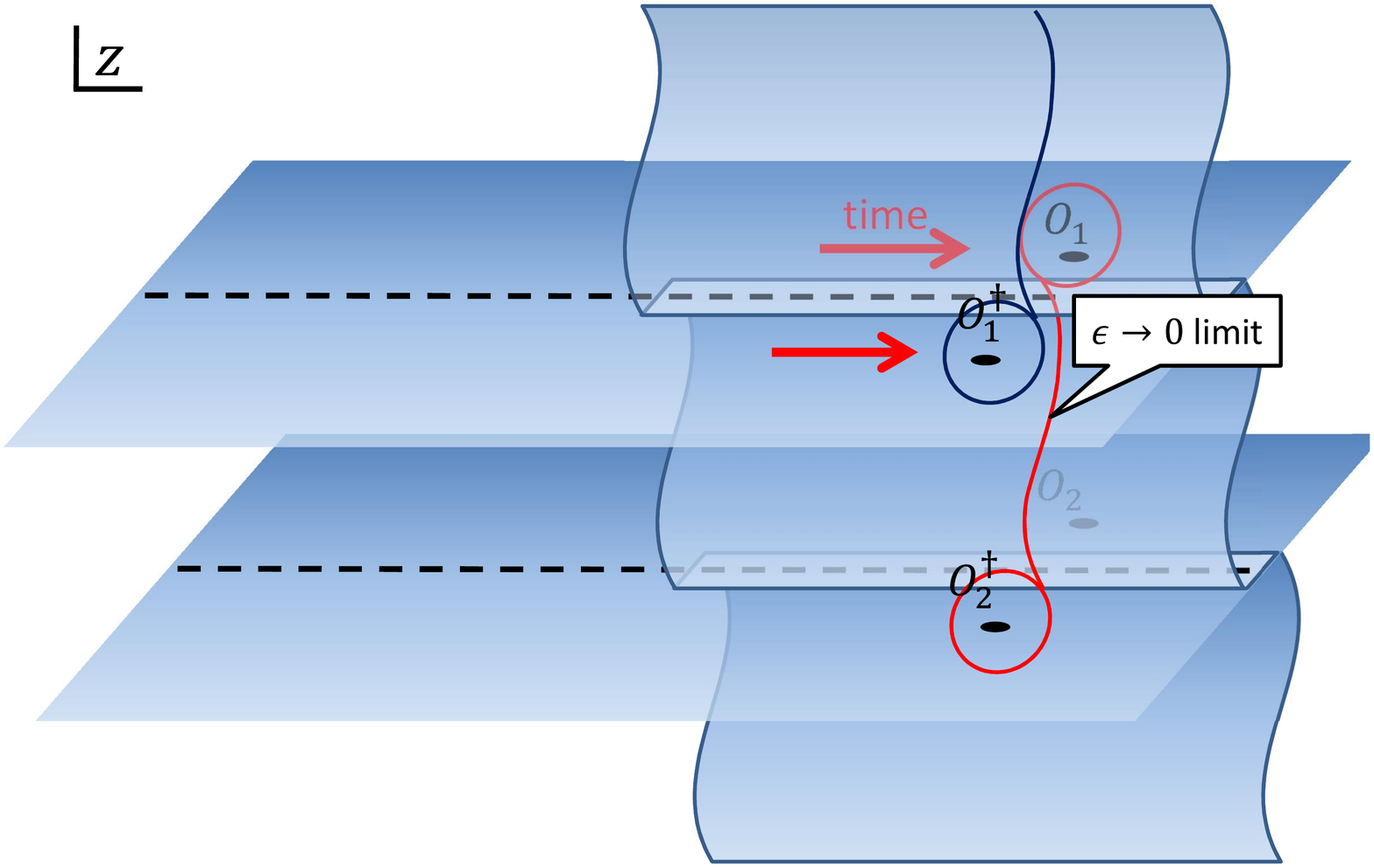}
  \includegraphics[width=7.0cm]{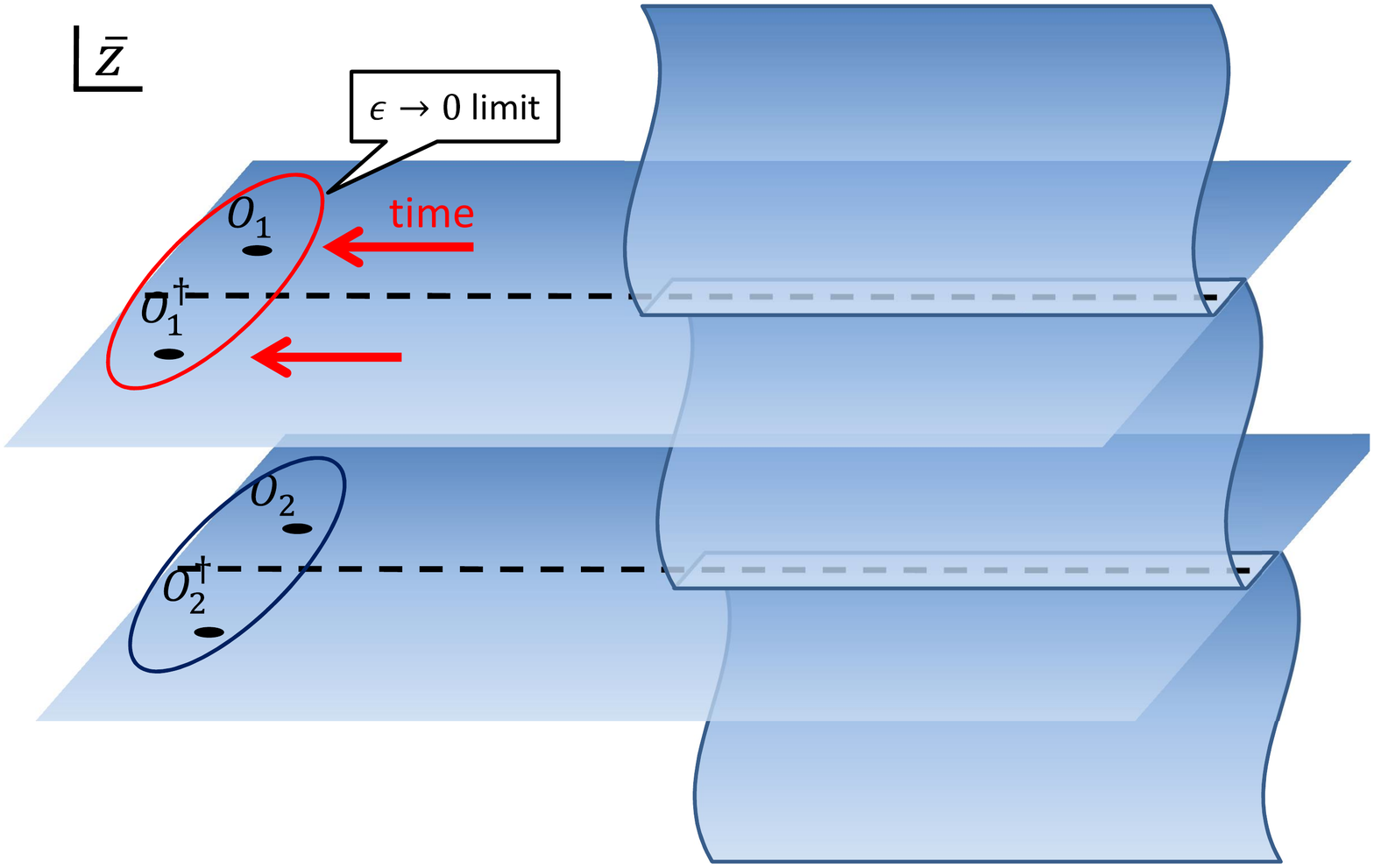}
 \end{center}
 \caption{The holomorphic and anti-holomorphic insertion points of local operators in the late time ($t>l$).}
 \label{fig:sheet2}
\end{figure}

Now, we are ready to calculate the Renyi entanglement entropy.
As mentioned in the introduction, we will consider a locally excited state,
\begin{equation}
\rho(t)=\ca{N}\ex{-\ep H-iHt} O(-l)\ket{0}\bra{0} \dg{O}(-l) \ex{-\ep H+iHt}=\ca{N} O(w_1,\bar{w}_1)\ket{0}\bra{0}\dg{O}(w_2,\bar{w}_2),
\end{equation}
where $\ca{N}$ is a normalization factor $\tr \rho(t)=1$ and we define
\begin{equation}\label{eq:coor}
\begin{aligned}
w_1=-i(\e+it)-l, \ \ \bar{w}_1=i(\e+it)-l,\\
w_2=i(\e-it)-l, \ \ \bar{w}_2=-i(\e-it)-l.
\end{aligned}
\end{equation}
In the path integral formalism, the Renyi entanglement entropy can be described as  \cite{Nozaki2014}
\begin{equation}\label{eq:tr1}
\tr \rho_A^n=\fr{\int_{\Sigma_n}\ca{D}\phi\  O_1 {\dg O}_1  O_2 {\dg O}_2 \cdots\ex{-S}}{\pa{\int_{\Sigma_1}\ca{D}\phi\ O {\dg O}\ex{-S}}^n}
\end{equation}
where $\Sigma_n$ is the $n$-sheeted manifold and we label the operator on $i$-th sheet as $O_i$ (see Figure \ref{fig:sheet}).
From this path integral description of $\tr \rho_A^n$, we get the growth of the entanglement entropy in terms of the operator formalism as
\begin{equation}\label{eq:DS}
\begin{aligned}
\D S_A^{(n)}
&=\fr{1}{1-n}\pa{\log \fr{\int_{\Sigma_n}\ca{D}\phi\  O_1 {\dg O}_1  O_2 {\dg O}_2 \cdots\ex{-S}}{\pa{\int_{\Sigma_1}\ca{D}\phi\ O {\dg O}\ex{-S}}^n}-\log \fr{\int_{\Sigma_n}\ca{D}\phi\  \ex{-S}}{\pa{\int_{\Sigma_1}\ca{D}\phi\ \ex{-S}}^n}}\\
&=\fr{1}{1-n}\pa{\log \fr{\int_{\Sigma_n}\ca{D}\phi\  O_1 {\dg O}_1  O_2 {\dg O}_2 \cdots\ex{-S}}{\int_{\Sigma_n}\ca{D}\phi\   \ex{-S}}-\log \fr{\pa{\int_{\Sigma_1}\ca{D}\phi\ O {\dg O}\ex{-S}}^n}{\pa{\int_{\Sigma_1}\ca{D}\phi\ \ex{-S}}^n}}\\
&=\fr{1}{1-n}\log \fr{\ave{ O(w_1^{(1)},\bar{w}_1^{(1)})  {\dg O}(w_2^{(1)},\bar{w}_2^{(1)})   O(w_1^{(2)},\bar{w}_1^{(2)})  {\dg O}(w_2^{(2)},\bar{w}_2^{(2)})  \cdots}_{\Sigma_n}}{\ave{O(w_1,\bar{w}_1) {\dg O}(w_2,\bar{w}_2)}_{\Sigma_1}^n},
\end{aligned}
\end{equation}
where we label the insertion points of the operators on $i$-th sheet as $w^{(i)}$. Therefore, all we need to do is to calculate the $2n$-point function on the $n$-sheeted manifold. Here, we focus on the late time ($t>l$) behavior of the Renyi entanglement entropy.
Looking at the time dependence of the insertion points (\ref{eq:coor}) shows that the $\e \to 0$ limit is NOT the usual OPE limit but the light cone OPE limit. Figure \ref{fig:sheet2} shows the insertion points of the local operators in the late time ($t>l$). The $\e \to 0$ limits of the anti-holomorphic parts are formally given by
\begin{equation}
\braket{  O_1 {\dg O}_1  O_2 {\dg O}_2 \cdots} \ar{\e \to 0} \braket{  O_1 {\dg O}_1} \braket{ O_2 {\dg O}_2} \cdots,
\end{equation}
whereas the $\e \to 0$ limits of the holomorphic parts are 
\begin{equation}
\braket{  O_1 {\dg O}_1  O_2 {\dg O}_2 \cdots} \ar{\e \to 0} \braket{  O_1 {\dg O}_2} \braket{ O_2 {\dg O}_3} \cdots.
\end{equation}
This type of limit is called the light cone limit. For example, the light cone limit of a four point function $\braket{O_4(\infty) O_3(1) O_2(z,\bar{z}) O_1(0)}$ is defined in terms of the cross ratio as $(z,\bar{z}) \to (1,0)$ (see Section \ref{sec:LCB}), which means that the holomorphic part of $O_2$ approaches $O_1$, whereas the anti-holomorphic part of $O_2$ approaches $O_3$, instead of $O_1$.
This light cone limit leads to a nontrivial singularity, so-called the light cone singularity.

In the limit where $\bar{w}_1^{(m)} \ar{\e \to 0} \bar{w}_2^{(m)} $, the $2n$-point function can be approximate by the vacuum Virasoro block as 
\begin{equation}\label{eq:nvac}
\fr{\ave{ O(w_1^{(1)},\bar{w}_1^{(1)})  {\dg O}(w_2^{(1)},\bar{w}_2^{(1)})   O(w_1^{(2)},\bar{w}_1^{(2)})  {\dg O}(w_2^{(2)},\bar{w}_2^{(2)})  \cdots}_{\Sigma_n}}{\ave{O(w_1,\bar{w}_1) {\dg O}(w_2,\bar{w}_2)}_{\Sigma_1}^n}
\ar{ \e \to 0 } 
\ca{C}_n \ca{F}(0|w_i^{(m)})\overline{\ca{F}}(0|\bar{w}_i^{(m)}),
\end{equation}
where  $\ca{C}_n$ is the universal part of the correlation function (which is explained later) and the channel of the  block is chosen as the following form,

\newsavebox{\boxpd}
\sbox{\boxpd}{\includegraphics[width=200pt]{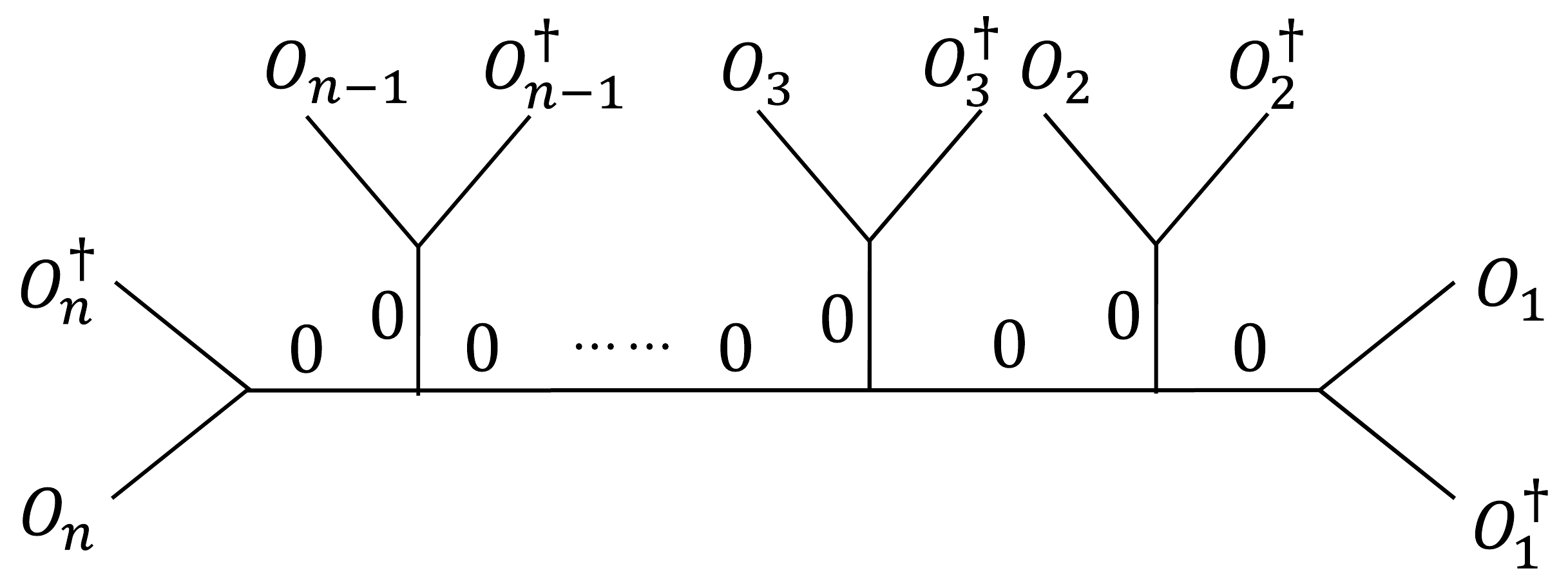}}
\newlength{\pdw}
\settowidth{\pdw}{\usebox{\boxpd}} 

\newsavebox{\boxpa}
\sbox{\boxpa}{\includegraphics[width=280pt]{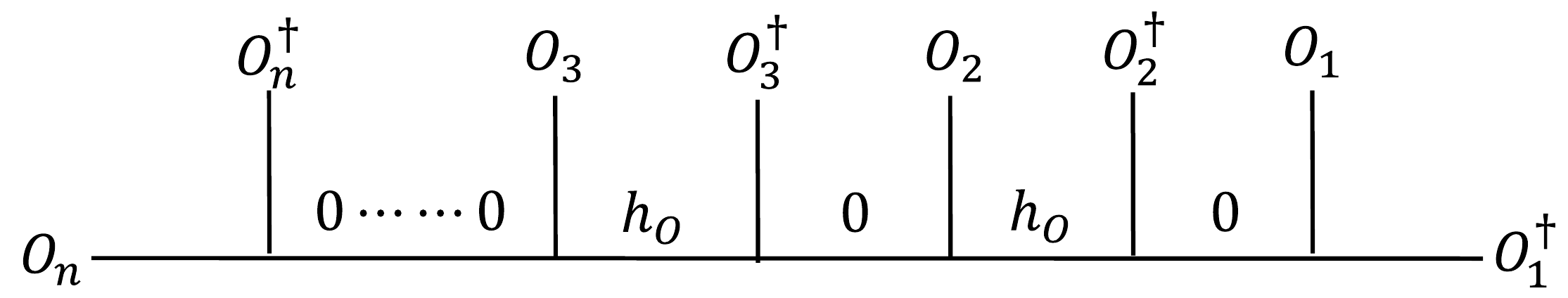}}
\newlength{\paw}
\settowidth{\paw}{\usebox{\boxpa}} 

\begin{equation}
\begin{aligned}
\ca{F}(0|w_i^{(m)}) &\equiv \parbox{\pdw}{\usebox{\boxpd}}\\
&=\parbox{\paw}{\usebox{\boxpa}}.
\end{aligned}
\end{equation}
In the second line, we re-expressed the block by the trivial fusion transformation.
Here we have to mention that the approximation (\ref{eq:nvac}) is justified only in a pure CFT (which is defined in Section \ref{sec:LCB}).

To proceed further, we need an explicit asymptotic form of the Virasoro conformal blocks in the light cone limit,
which was unknown until the recent works of numerical estimations \cite{Kusuki2018,Kusuki2018b} and analytic proofs \cite{Kusuki2018a,Kusuki2018c, Collier2018}. In this article, we make use of the results \cite{Kusuki2018a,Kusuki2018c, Collier2018} for our purpose.

\subsection{2nd Renyi Entropy after Local Quench}\label{subsec:2ndREE}

We first give the concrete calculation of the $2$nd Renyi entropy in the following.
After that, we will generalize the results to any $n$. For simplicity, we assume that the interval $A$ is semi infinite.
In this case, we can map the $n$-sheeted manifold $\Sigma_n$ to a sphere $\Sigma_1$ by the conformal transformation,
\begin{equation}\label{eq:replica map}
w=z^n,
\end{equation}
which maps the insertion points $\{ w_i^{(m)}\}$ on the 2-sheet manifold to
\begin{equation}\label{eq:w}
\begin{aligned}
z_1^{(1)} = -z_1^{(2)} &= \sqrt{ -i \e+t-l  },  \\
z_2^{(1)} = -z_2^{(2)} &= \sqrt{ i \e+t-l  } , \\
\bar{z}_1^{(1)} = -\bar{z}_1^{(2)} &= \sqrt{ i \e-t-l  },  \\
\bar{z}_2^{(1)} = -\bar{z}_2^{(2)} &= \sqrt{ -i \e-t-l  } .
\end{aligned}
\end{equation}
When considering the time dependence, we have to take the branch cut of $z=\s{w}$ into account.
It means that  if it is after $t=l$, the coordinate $z_1^{(1)} ( = -z_1^{(2)} ) = \sqrt{ -i \e+t-l  }$ is replaced by $\ex{\pi i}\sqrt{ -i \e+t-l  }$, whereas the other coordinates are left unchanged. In the following, we particularly focus on the regime $t>l$.

If we define the cross ratio as
\begin{equation}
\begin{aligned}
z&=\fr{(z_1^{(1)}-z_2^{(1)})(z_1^{(2)}-z_2^{(2)})}{(z_1^{(1)}-z_1^{(2)})(z_2^{(1)}-z_2^{(2)})}=\fr{1}{2}\pa{1-\fr{l-t}{\s{(l-t)^2+\e^2}}},\\
\bar{z}&=\fr{(\bz_1^{(1)}-\bz_2^{(1)})(\bz_1^{(2)}-\bz_2^{(2)})}{(\bz_1^{(1)}-\bz_1^{(2)})(\bz_2^{(1)}-\bz_2^{(2)})}=\fr{1}{2}\pa{1-\fr{l+t}{\s{(l+t)^2+\e^2}}},
\end{aligned}
\end{equation}
then we can show that the four point function in (\ref{eq:DS}) can be expressed by
\begin{equation}
\fr{\braket{ O(w_1^{(1)},\bar{w}_1^{(1)})  {\dg O}(w_2^{(1)},\bar{w}_2^{(1)})   O(w_1^{(2)},\bar{w}_1^{(2)})  {\dg O}(w_2^{(2)},\bar{w}_2^{(2)}) }_{\Sigma_2}}{\ave{O(w_1,\bar{w}_1) {\dg O}(w_2,\bar{w}_2)}_{\Sigma_1}^2}
=\pa{z(1-z)}^{2h_O} \pa{\bar{z}(1-\bar{z})}^{2\bar{h}_O} G(z,\bar{z}) ,
\end{equation}
where $G(z,\bar{z})$ is the four point function
\be
G(z,\bar{z})=\braket{ O(\infty)\dg{O}(1)O(z, \bar{z}),\dg{O}(0)}.
\ee

From the expression of the cross ratio, we can find that in the late time ($t>l$),
\begin{equation}
\begin{aligned}
z&=1-\fr{1}{4}\pa{\fr{\e}{l-t}}^2+\ca{O}(\e^4),\\
\bar{z}&=\fr{1}{4}\pa{\fr{\e}{l+t}}^2+\ca{O}(\e^4),
\end{aligned}
\end{equation}
which mean that the $\e \to 0$ limit is just the light cone limit $(z, \bar{z}) \to (1,0)$.
Therefore, we can approximate the function $G(z,\bar{z})$ by the vacuum block,
\begin{equation}
G(z,\bar{z}) \ar{\e \to 0} \ca{F}^{{\dg O}O}_{O{\dg O}}(0|z)  \overline{\ca{F}^{{\dg O}O}_{O{\dg O}}}(0|\bar{z}).
\end{equation}

We has already shown that the light cone singularity of the Virasoro block is given by the form (\ref{eq:LC}).
In particular, the leading term of the Virasoro block in the light cone limit is given by
\begin{equation}
\begin{aligned}
\ca{F}^{{\dg O}O}_{O{\dg O}}(0|z) 
&\ar{z \to 1} \left\{
    \begin{array}{ll}
    \text{Res}\pa{   -2\pi i 
  {\bold F}_{0, \a} [O]
  ; \a=2\a_O} z^{h_{2\a_O}-2h_O}   ,& \text{if } h_O<\fr{c-1}{32}   ,\\
    \fr{\s{\pi}}{8} \del_{\a}^2 \left. {\bold F}_{0,\a}[O] \right|_{\a=\fr{Q}{2}} z^{\fr{Q^2}{4}-2h_O}\pa{\log z}^{-\fr{3}{2}}   ,& \text{otherwise } ,\\
    \end{array}
  \right.\\
\end{aligned}
\end{equation}
where  $ {\bold F}_{0, \a} [ O]  $ denotes the fusion matrix 
$ {\bold F}_{0, \a} [ O] = {\bold F}_{0, \a} 
   \left[
    \begin{array}{cc}
    \a_{ O}   & \a_{ O}   \\
     \a_{ O}   &   \a_{ O} \\
    \end{array}
  \right]$.

From the asymtptoic form of the Virasoro block,
\begin{equation}
\ca{F}^{ji}_{kl}(h_p|z) \ar{z \to 0} z^{h_p-h_i-h_j},
\end{equation}
we can find
\begin{equation}
\begin{aligned}
\pa{z(1-z)}^{2h_O} \pa{\bar{z}(1-\bar{z})}^{2\bar{h}_O} G(z,\bar{z}) 
&\ar{\e \to 0}  \left\{
    \begin{array}{ll}
    \text{Res}\pa{   -2\pi i 
  {\bold F}_{0, \a} [O]
  ; \a=2\a_O} \pa{\fr{\e}{2t}}^{2h_{2\a_O}}   ,& \text{if } h_O<\fr{c-1}{32}   ,\\
    \fr{\s{\pi}}{8} \del_{\a}^2 \left. {\bold F}_{0,\a}[O] \right|_{\a=\fr{Q}{2}} \pa{\fr{\e}{2t}}^{\fr{Q^2}{2}}\pa{\log \pa{\fr{\e}{2t}}^2}^{-\fr{3}{2}}   ,& \text{otherwise } .\\
    \end{array}
  \right.\\
\end{aligned}
\end{equation}
where we set $l=0$ for simplicity.
As a result, the growth of the $2$nd Renyi entropy after a light local quench ($h_O<\fr{c-1}{32}$)  is given by
\begin{equation}
\D S_A^{(2)}(t)\ar{\fr{t}{\e} \to \infty}4\a_O(Q-2\a_O)\log\fr{2t}{\e}- \log  \BR{-2\pi i\  \text{Res}\pa{{\bold F}_{0,\a_t}[O];  \a_t= 2\a_O}}.
\end{equation}
In particular, if expanding this at small $\fr{h_O}{c}$, the result reduces to
\begin{equation}
\D S_A^{(2)}\ar{\fr{h_O}{c}\ll1}4h_O\log\fr{t}{\e}.
\end{equation}
This result in the light limit is consistent with the result in \cite{Caputa2014a}.
The growth for a heavy local quench ($h_O>\fr{c-1}{32}$) is more interesting, that is, it has the following {\it universal} form:
\begin{equation}\label{eq:univREE}
\D S_A^{(2)}(t)\ar{\fr{t}{\e} \to \infty}\fr{Q^2}{2} \log\fr{2t}{\e}+\fr{3}{2}\log\log\pa{\fr{\e}{2t}}^2- \log \pa{ \fr{\s{\pi}}{8} \del_{\a}^2 \left. {\bold F}_{0,\a}[O] \right|_{\a=\fr{Q}{2}}} .
\end{equation}
These results are consistent with our numerical results in \cite{Kusuki2018b}.

We can, therefore, conclude that the $2$nd Renyi entropy after a local quench undergoes a phase transition as the conformal dimension of the local quench is varied, if we restrict ourselves to the pure CFT. That is, in one of the phases, the entropy is monotonically increasing in $h_O$, and in the other phase, it is saturated by the universal form (\ref{eq:univREE}).

\subsection{$n$-th Renyi Entropy after Local Quench}\label{subsec:nthREE}

We will generalize the above results to $n$-th Renyi entropy.
In a similar way as $2$nd, we consider the conformal transformation,
\footnote{This map is slightly different from (\ref{eq:replica map}). In this section, we do not use the cross ratios but evaluate the $n$-point partial wave directly via the $z_i$ coordinates. In such a case, it is more convenient to use the map (\ref{eq:replica map2}) than (\ref{eq:replica map}). }
\begin{equation}\label{eq:replica map2}
z=\pa{\fr{w}{ i \e+t-l }}^{\fr{1}{n}},
\end{equation}
which maps the insertion points $\{ w_i^{(m)}\}$ on the $n$-sheet manifold to
\begin{equation}\label{eq:wn}
\begin{aligned}
z_1^{(k)}  &=\ex{2\pi i \fr{k}{n}} \pa{\fr{ -i \e+t-l }{ i \e+t-l } }^{\fr{1}{n}},  \\
z_2^{(k)} &=\ex{2\pi i \fr{k}{n}} \pa{ \fr{ i \e+t-l }{ i \e+t-l } }^{\fr{1}{n}} , \\
\bar{z}_1^{(k)}  &=\ex{-2\pi i \fr{k}{n}} \pa{ \fr{ i \e-t-l }{ i \e+t-l } }^{\fr{1}{n}},  \\
\bar{z}_2^{(k)} &=\ex{-2\pi i \fr{k}{n}} \pa{  \fr{ -i \e-t-l }{ i \e+t-l } }^{\fr{1}{n}} ,
\end{aligned}
\end{equation}
with a branch cut on the real axis from $0$ to $+\infty$.
As explained in Figure \ref{fig:sheet2}, the $\e \to 0$ limit leads to the light cone limit,
\begin{equation}
\begin{aligned}
z_1^{(k)} - z_2^{(k+1)} &\ar{\e \to 0} -\fr{2 i \e}{n(t-l)} z_1^{(k)}   ,\\
\bar{z}_1^{(k)} - \bar{z}_2^{(k)} &\ar{\e \to 0} -\fr{2 i \e}{n(t+l)} \bar{z}_1^{(k)}    .\\
\end{aligned}
\end{equation}

In this coordinates, the light cone limit of the $2n$-point function can be approximated by the vacuum partial wave as 
\begin{equation}\label{eq:Replican}
\fr{\ave{ O(w_1^{(1)},\bar{w}_1^{(1)})  {\dg O}(w_2^{(1)},\bar{w}_2^{(1)})   O(w_1^{(2)},\bar{w}_1^{(2)})  {\dg O}(w_2^{(2)},\bar{w}_2^{(2)})  \cdots}_{\Sigma_n}}{\ave{O(w_1,\bar{w}_1) {\dg O}(w_2,\bar{w}_2)}_{\Sigma_1}^n}
\ar{ \e \to 0 } 
\ca{C}_n \ca{F}(0|z_i^{(m)})\overline{\ca{F}}(0|\bar{z}_i^{(m)}),
\end{equation}
where we used the following shorthand notation for (\ref{eq:wavedef}),
\begin{equation}
\ca{F}(0|z_i^{(m)}) \equiv \ca{F}^{h_O,h_O,\dots, h_O}_{0 , h_O , 0, h_O, \dots, h_O, 0}(z_2^{(1)} ,z_1^{(1)}, \dots,  z_{2}^{(n)}, z_{1}^{(n)}),
\end{equation}
and we defined
\begin{equation}\label{eq:Cn}
\begin{aligned}
\ca{C}_n&= \pa{\prod_{k=1}^n  \fr{-4 \e^2}{n^2\pa{(t-l)^2+\e^2}}{z_1^{(k)}z_2^{(k)}}}^{h_O} 
 \pa{ \prod_{k=1}^n  \fr{-4 \e^2}{n^2\pa{(t+l)^2+\e^2}}{\bar{z}_1^{(k)}\bar{z}_2^{(k)}}}^{\bar{h}_O}\\
&\ar{\e \to 0}  \pa{\fr{2i \e}{n (t-l)}}^{2nh_O}  \pa{\fr{2i \e}{n (t+l)}  \pa{- \fr{t+l}{t-l}}^{\fr{1}{n}}  }^{2n \bar{h}_O} .
\end{aligned}
\end{equation}
In the right hand side of (\ref{eq:Replican}), the anti-holomorphic part is just given by
\begin{equation}
 \overline{\ca{F}}(0|\bar{z}_i^{(m)}) \ar{\e \to 0}  
 \pa{\fr{2i \e}{n (t+l)}  \pa{- \fr{t+l}{t-l}}^{\fr{1}{n}}  }^{-2n \bar{h}_O},
\end{equation}
whereas the holomorphic part is given by the light cone singularity of the partial wave $\ca{F}(0|z_i^{(m)}) $.

Before considering the most general case, let us first consider the global limit, meaning $O$'s are light operators and $c \to \infty$. Since we are taking the global limit, we have \begin{equation}
\ca{F}(0|z_i^{(m)})\overline{\ca{F}}(0|\bar{z}_i^{(m)})=\prod_{k=1}^n \ave{O(z_1^{(k)},\bar{z}_1^{(k)}) {\dg O}(z_2^{(k)},\bar{z}_2^{(k)})}_{\Sigma_1}.\end{equation} 
This formula immediately tells us that the late time behavior of the Renyi entropy shows
\begin{equation}
\begin{aligned}\D S_A^{(n)}(t)&\ar{\fr{t-l}{\ep}\ar{}\infty}
\fr{2nh_O}{n-1}\text{log}\pa{\fr{n(t-l)\sin\pa{\fr{\pi}{n}}}{\e}}.
\end{aligned}
\end{equation}

Let us consider the most general case. The key tool to study it is the singularity of the conformal partial wave, which have been already given by (\ref{eq:n-point singularity2}). 
Applying this result  (\ref{eq:n-point singularity2}) to the current case , we obtain
      \begin{equation}
\begin{aligned}
&\ca{F}(0|z_i^{(m)})\ar{\e \to 0} \left\{
    \begin{array}{ll}
    \BR{\text{Res}\pa{   -2\pi i  {\bold F}_{0,\a}[O]; \a=2\a_{O}}}^{n-1} 
	\BR{\pa{\fr{2i\ep}{n(t-l)}}^{n-1}\ex{-\fr{2\pi i}{n}}}^{h_{2\a_O}-2h_{O}}\\
  ~~~\times\ca{F}^{h_O,h_{2\a_O},\cdots , h_{2\a_O},h_O}_{h_O,\cdots,h_O}\pa{ z_2^{(1)} , \ex{2 \pi i \fr{2}{n} } , \ex{2 \pi i \fr{3}{n} }  ,\ex{2 \pi i \fr{4}{n} } ,\cdots, \ex{2 \pi i \fr{n-1}{n} }  , \ex{2 \pi i \fr{n}{n} }  ,  z_1^{(n)}}
    ,~~~~~ \text{if } \a_{O}<\fr{Q}{4}  ,\\
    \pa{-i\fr{\s{\pi}}{8} \del_{\a_t}^2 \left. {\bold F}_{0,\a_t}[O] \right|_{\a_t=\fr{Q}{2}}}^{n-1}   \pa{\log \pa{\fr{t-l}{\ep}}}^{-\fr{3}{2}(n-1)}   
	\BR{\pa{\fr{2i\ep}{n(t-l)}}^{n-1}\ex{-\fr{2\pi i}{n}}}^{\fr{c-1}{24}-2h_{O}}\\
  ~~~\times    \ca{F}^{h_O,\fr{Q^2}{4} ,\cdots , \fr{Q^2}{4},h_O}_{h_O,\cdots,h_O}\pa{ z_2^{(1)} , \ex{2 \pi i \fr{2}{n} } , \ex{2 \pi i \fr{3}{n} }  ,\ex{2 \pi i \fr{4}{n} } ,\cdots, \ex{2 \pi i \fr{n-1}{n} }  , \ex{2 \pi i \fr{n}{n} }  ,  z_1^{(n)}},
	   ~~~~~~\text{if } \a_{O}>\fr{Q}{4}.\\
    \end{array}
  \right.\\
  \end{aligned}   
  \end{equation} 
By using the regularized partial wave (\ref{eq:bbf}),  we can perfectly extract the time dependence of the partial wave as
      \begin{equation}
\begin{aligned}
&\ca{F}(0|z_i^{(m)})\ar{\e \to 0} \left\{
    \begin{array}{ll}
    \BR{\text{Res}\pa{   -2\pi i  {\bold F}_{0,\a}[O]; \a=2\a_{O}}}^{n-1} 
	\pa{\fr{\ep}{n(t-l) \sin\fr{\pi}{n}  }}^{nh_{2\a_O}} 
\pa{\fr{2i\ep}{n(t-l)}}^{-2n h_O}
\bb{f}(h_O) ,
    & \text{if } \a_{O}<\fr{Q}{4}  ,\\
    \pa{-i\fr{\s{\pi}}{8} \del_{\a_t}^2 \left. {\bold F}_{0,\a_t}[O] \right|_{\a_t=\fr{Q}{2}}}^{n-1}   \pa{\log \pa{\fr{t-l}{\ep}}}^{-\fr{3}{2}n}   
	\pa{\fr{\ep}{n(t-l)\sin\fr{\pi}{n} }}^{n\fr{Q^2}{4}} 
\pa{\fr{2i\ep}{n(t-l)}}^{-2n h_O}
\bb{f}(h_O), 
	&\text{if } \a_{O}>\fr{Q}{4}.\\
    \end{array}
  \right.\\
  \end{aligned}   
  \end{equation} 	
As a result, we find
\begin{equation}\label{eq:$n$-th}
\begin{aligned}
\D S_A^{(n)}(t)
&\ar{\fr{t-l}{\e} \to \infty}
\left\{
    \begin{array}{ll}
    \fr{n}{n-1}h_{2\a_O}\log\fr{n (t-l) \sin\fr{\pi}{n}}{\e}- \log  \BR{-2\pi i\  \text{Res}\pa{{\bold F}_{0,\a_t}[O]; \a_t=2\a_O}}   
	-\fr{1}{n-1} \log \bb{f} (h_O) , \\
 \hspace{12cm} \text{if }  h_O<\fr{c-1}{32 } ,\\ \\
     \fr{n}{n-1} \fr{Q^2}{4}\log\fr{n (t-l) \sin\fr{\pi}{n}}{\e}
	+\fr{3}{2} \fr{n}{n-1} \log\log\pa{\fr{t-l}{\e}}
	 - \log \pa{ -i \fr{\s{\pi}}{8} \del_{\a}^2 \left. {\bold F}_{0,\a}[O] \right|_{\a=\fr{Q}{2}} }
	 -\fr{1}{n-1} \log \bb{f} (h_O)  ,  \\ 
 \hspace{12cm}  \text{if } h_O>\fr{c-1}{32}.\\
    \end{array}
  \right.\\
\end{aligned}
\end{equation}
In fact, we can also calculate the Renyi entropy in a second independent way (see Section \ref{subsec:twistframe}) and the result is consistent with the above calculation as explained in Section \ref{subsec:largecRenyi}). It justifies our expression of the light cone singularity of $n$-point conformal block.

From the holographic calculation, the late time behavior of the entanglement entropy after a local quench is given by
\begin{equation}
\D S_A(t) \ar{\fr{t}{\e} \to \infty} \fr{c}{6} \log \fr{t}{\e},
\end{equation}
where we set $l=0$ for simplicity.
However, we straightforwardly find that the analytic continuation $n \to 1$ of the $n$-th Renyi entropy  (\ref{eq:$n$-th}) cannot reproduce this holographic result. Moreover, this $n$-th Renyi entropy even diverges as $n \to 1$. 

In many cases, to calculate the entanglement entropy (for example, the replica method), we implicitly assume that the Renyi entropy is analytic in $n$. However, we now find an exception of this assumption in the light cone limit. Therefore, we have to consider this exception if we use the replica method to evaluate the entanglement entropy. We emphasize that this assumption does not contradict with the derivation of the Ryu--Takayanagi formula in \cite{Lewkowycz2013a}, as our result for the Renyi entropy is analytic in the vicinity of $n=1$.

\subsection{Renyi Entropy from Twist Frame}\label{subsec:twistframe}

\begin{figure}[t]
 \begin{center}
  \includegraphics[width=5.0cm]{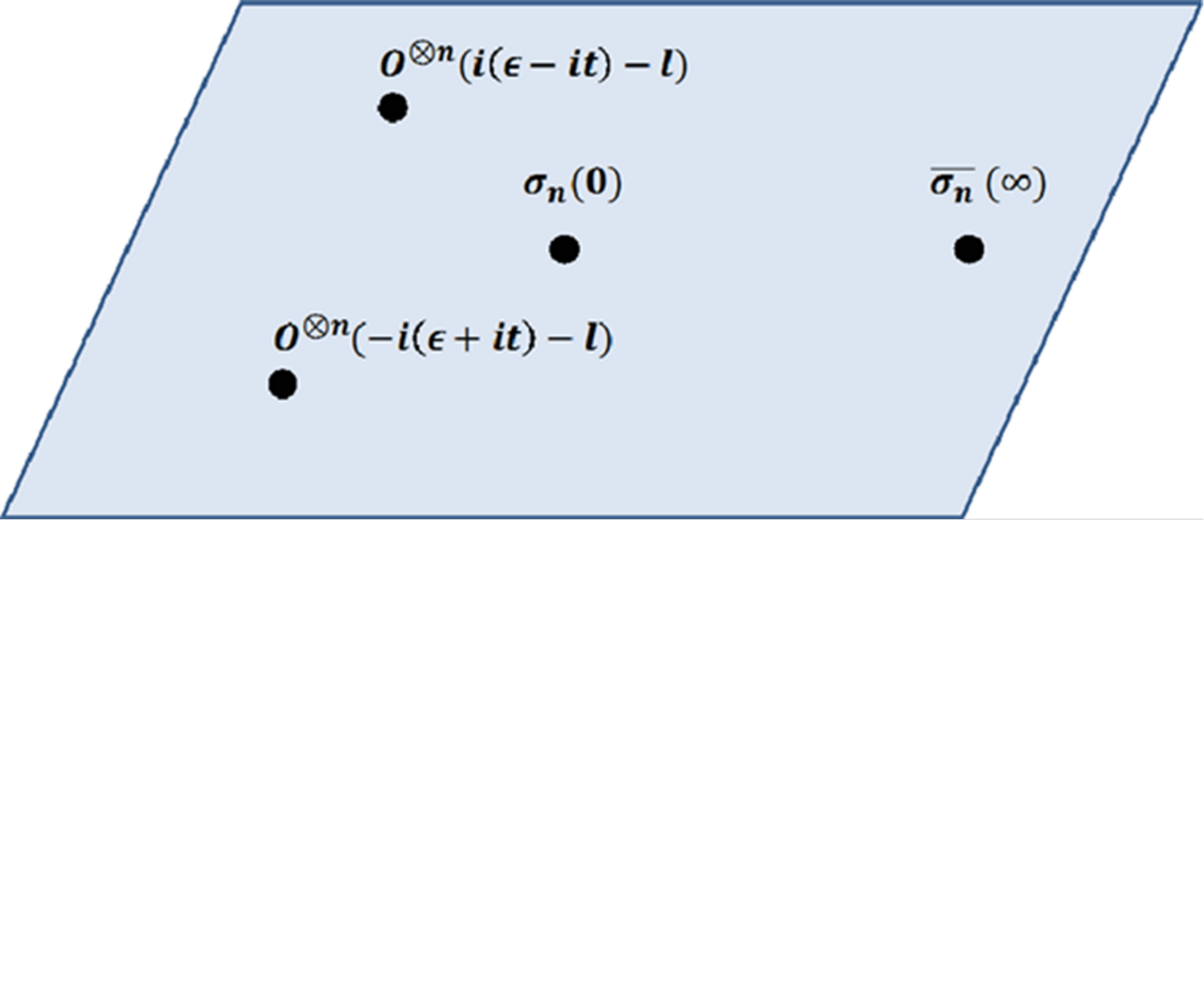}
  \includegraphics[width=9.0cm]{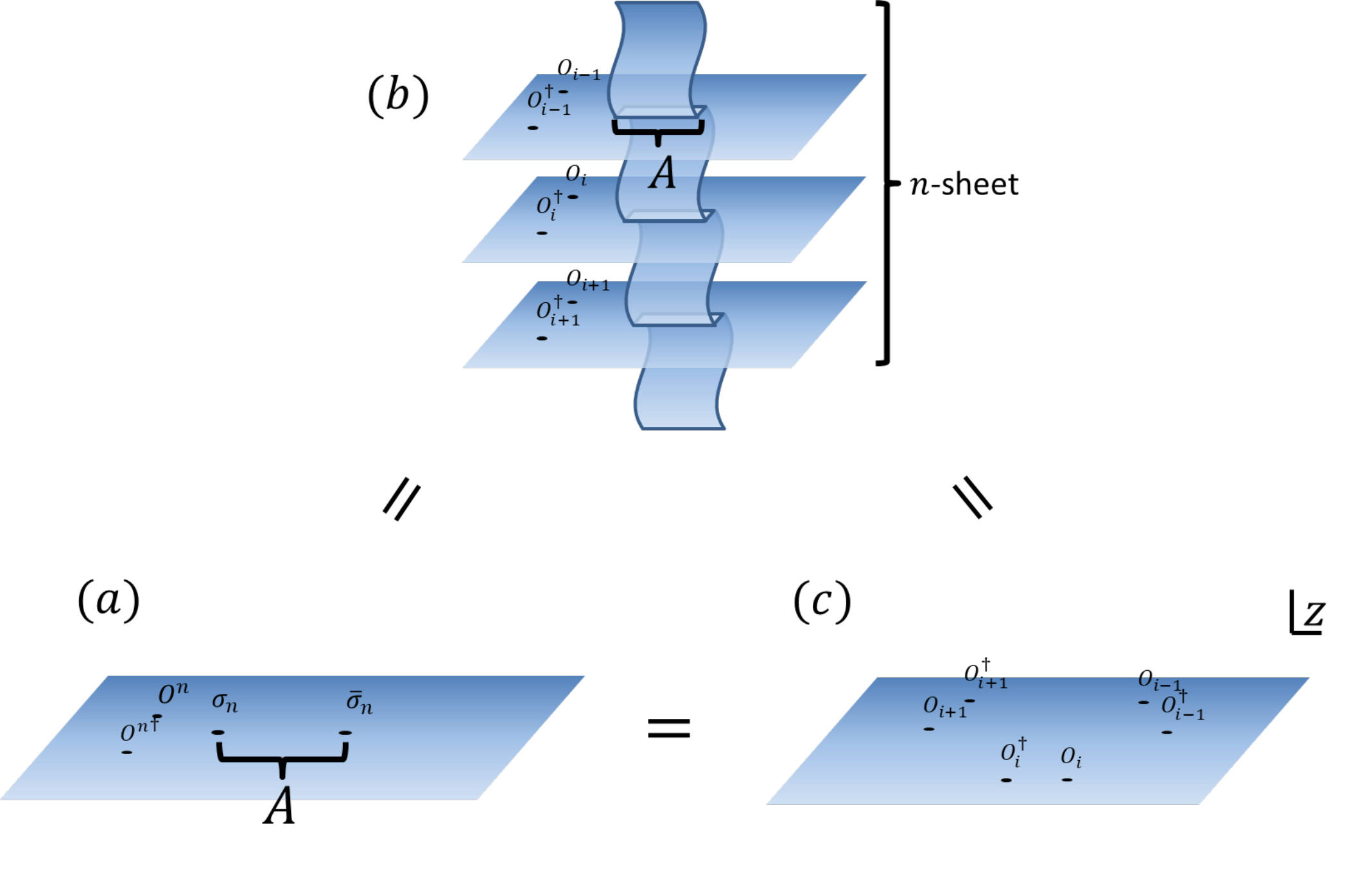}
 \end{center}
 \caption{(Left) The positions of operators in the replica computation (\ref{eq:defREE}). (Right) The equivalence between $(a)$ and $(b)$ explains the relation between a correlator with twist operators in an orbifold theory and a replica manifold. The equivalence between $(b)$ and $(c)$ can be obtained using a conformal map $w=z^n$.}
 \label{fig:pos}
\end{figure}

In fact, this quantity can also be calculated analytically using twist operators as
\begin{equation}\label{eq:defREE}
\Delta S^{(n)}_A=\frac{1}{1-n}\log \frac{\ave{O^{\otimes n}O^{\otimes n}\sigma_n \bar{\sigma_n} }}{\ave{O^{\otimes n}O^{\otimes n}}\ave{\sigma_n \bar{\sigma_n}}},
\end{equation}
where the operator $O^{\otimes n}$ is defined on the cyclic orbifold CFT $\ca{M}^n/\bb{Z}_n$, using the operators in the seed CFT $\ca{M}$ as
\begin{equation}
O^{\otimes n} = O \otimes O \otimes \cdots \otimes O (\equiv \bb{O}).
\end{equation}
The local excitation $O$ is separated by a distance $l$ from the boundary of $A$, as shown in the left of Figure \ref{fig:pos}.
We will introduce some notations on the cyclic orbifold CFT $\ca{M}^n/\bb{Z}_n$ as follows:
\begin{table}[h]
  \begin{tabular}{|l|c|} 
\hline
    $\ca{F}^{(n)}$ & The conformal block associated to the chiral algebra $(\text{Vir})^n/\bb{Z}_n$, instead of the Virasoro. \\
				& (Note that this conformal block is defined on the CFT with the central charge $nc$.)		\\  \hline
    ${\bold F}^{(n)}$ & The fusion matrix associated to the chiral algebra $(\text{Vir})^n/\bb{Z}_n$, instead of the Virasoro. \\ \hline
    ${\bold M}^{(n)}$ & The monodromy matrix associated to the chiral algebra $(\text{Vir})^n/\bb{Z}_n$. \\ \hline
    ${S}^{(n)}$ & The modular S matrix associated to the chiral algebra $(\text{Vir})^n/\bb{Z}_n$. \\ \hline
    $h_{\sigma_n}$ & The conformal dim. of the twist op.; $h_{\sigma_n}=\frac{c}{24}\pa{1-\fr{1}{n^2}}$. \\ \hline
  \end{tabular}
\end{table}
\\
Here, we have to emphasize that the central charge $c$ is defined not on the orbifold CFT $\ca{M}^n/\bb{Z}_n$ but on the original CFT $\ca{M}$.

\begin{figure}[t]
 \begin{center}
  \includegraphics[width=7.0cm]{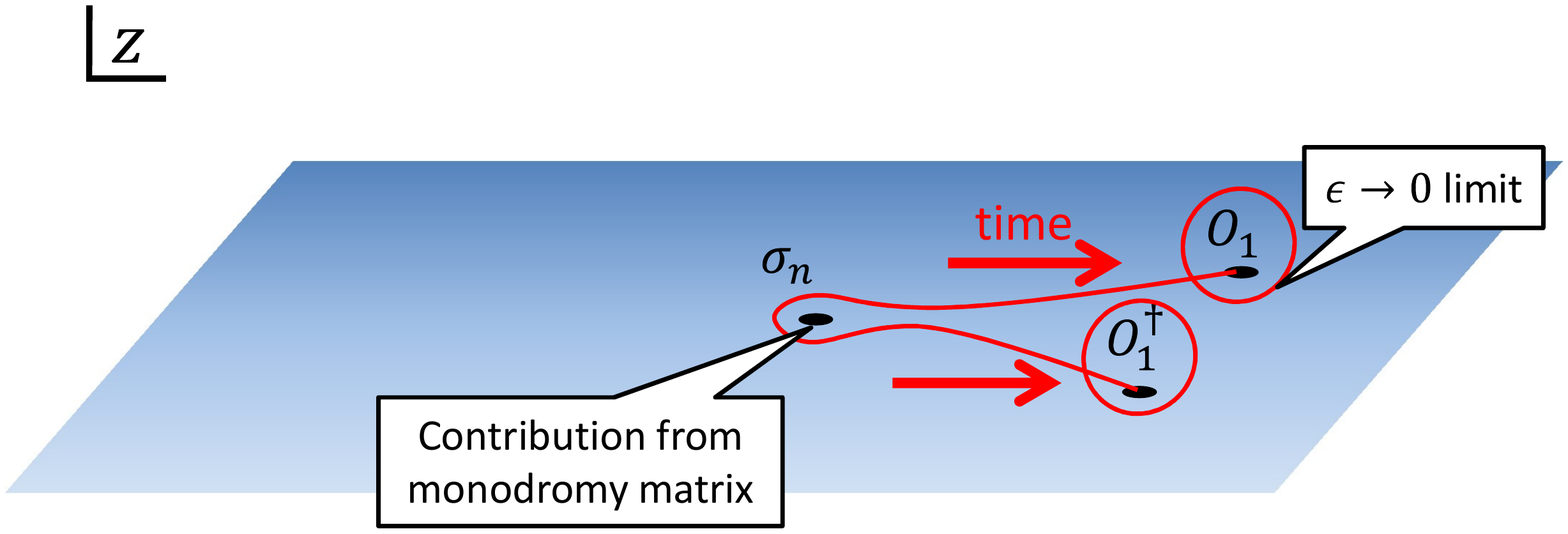}
  \includegraphics[width=7cm]{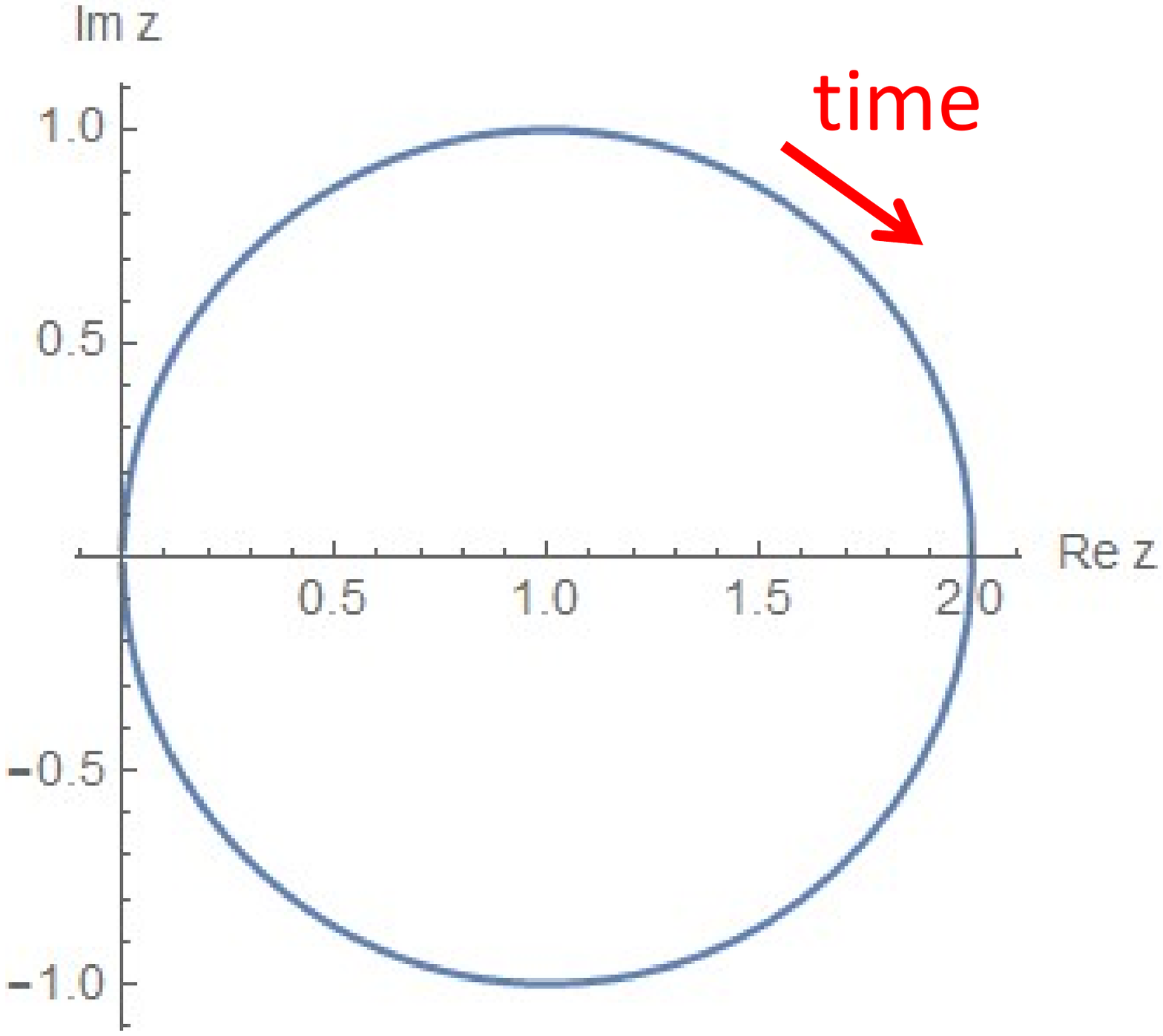}
 \end{center}
 \caption{(Left) The holomorphic insertion points of local operators in the late time ($t>l$). (Right) The time evolution of the cross ratio $z$.}
 \label{fig:sheet3}
\end{figure}

In this {\it twist frame}, the light cone limit of the $n$ point function is translated into the Regge limit of the four point function $\ave{O^{\otimes n}O^{\otimes n}\sigma_n \bar{\sigma_n} }$ (see the left of Figure \ref{fig:sheet3} ). Here we mean the {\it Regge limit} by the OPE limit after picking up a monodromy around another singular point. We will explain it in more familiar expression.
By using the cross ratio $z=\frac{z_{12}z_{34}}{z_{13}z_{24}}$, we can rewrite (\ref{eq:defREE}) as
\begin{equation}\label{eq:GRenyi}
\frac{\ave{O^{\otimes n}O^{\otimes n}\sigma_n \bar{\sigma_n} }}{\ave{O^{\otimes n}O^{\otimes n}}\ave{\sigma_n \bar{\sigma_n}}}=\abs{z^{2h_{\sigma_n}}}^2G(z,\bar{z}),
\end{equation}
where $G(z,\bar{z})$ is the four point function
\be
G(z,\bar{z})=\braket{O^{\otimes n}(\infty) O^{\otimes n}(1)\bar{\sigma}_n(z,\bar{z})\sigma_n(0)},
\ee
and in our setup, the cross ratio $(z,\bar{z})$ is given by
\begin{equation}
z=\frac{2i \e}{l-t+i \e}, ~~~~~\bar{z}=-\frac{2i\e}{l+t-i\e}.
\end{equation}
From these expressions, one finds that the sign of the imaginary part of the cross ratio $z$ changes at $t=l$.  As a result, the holomorphic cross ratio picks up the factor $\ex{-2\pi i}$ at $t=l$ as $1-z \to \ex{-2\pi i}(1-z)$ (see the right of Figure \ref{fig:sheet3}). This does not happen for the anti-chiral coordinate $\bar{z}$.

From now on, we will focus on the late time region ($t>l$).
In the $\e \to 0$ limit, the four point function can be approximated by the vacuum block as
\begin{equation}
\braket{ \sigma_n(0)\bar{\sigma}_n(z)O^{\otimes n}(1)O^{\otimes n}(\infty)}
\ar{\e \to 0}
{\ca{F}^{(n)}}^{\bar{\sigma}_n \sigma_n }_{\bb{O}\dg{\bb{O}}}(0|z)  \overline{{\ca{F}^{(n)}}^{\bar{\sigma}_n \sigma_n }_{\bb{O}\dg{\bb{O}}}} (0|\bar{z}),
\end{equation}
where we used the abbreviation $\bb{O}\equiv O^{\otimes n}$. However, we have to take care of the contribution from picking up  the monodromy around $z=1$. In that, it is more appropriate to express the approximation as
\begin{equation}\label{eq:twistRenyi}
\begin{aligned}
\braket{ \sigma_n(0)\bar{\sigma}_n(z)O^{\otimes n}(1)O^{\otimes n}(\infty)}
&\ar{\e \to 0}
\int \dd \a \ {{\bold M}^{(n)}}_{0, \a} [\sigma_n,\bb{O}]
 {\ca{F}^{(n)}}^{\bar{\sigma}_n \sigma_n }_{\bb{O}\dg{\bb{O}}}(n h_{\a}|z)  \overline{{\ca{F}^{(n)}}^{\bar{\sigma}_n \sigma_n }_{\bb{O}\dg{\bb{O}}}} (0|\bar{z}),
\end{aligned}
\end{equation}
where the kernel ${\bold M}^{(n)}$ is the monodromy matrix associated to the current algebra $(\text{Vir})^n/\bb{Z}_n$, which we will call the {\it replica monodromy matrix}.
The reason why the argument of ${\ca{F}^{(n)}}^{\bar{\sigma}_n \sigma_n }_{\bb{O}\dg{\bb{O}}}(n h_{\a}|z)$ is expressed by $n h_\a$ instead of $h_\a$ will be explained in the following.
Since we have the relation,
\begin{equation}\label{eq:replicatwist}
\fr{\ave{ O(w_1^{(1)},\bar{w}_1^{(1)})  {\dg O}(w_2^{(1)},\bar{w}_2^{(1)})   O(w_1^{(2)},\bar{w}_1^{(2)})  {\dg O}(w_2^{(2)},\bar{w}_2^{(2)})  \cdots}_{\Sigma_n}}{\ave{O(w_1,\bar{w}_1) {\dg O}(w_2,\bar{w}_2)}_{\Sigma_1}^n}=\frac{\ave{O^{\otimes n}O^{\otimes n}\sigma_n \bar{\sigma_n} }}{\ave{O^{\otimes n}O^{\otimes n}}\ave{\sigma_n \bar{\sigma_n}}}.
\end{equation}
We can deduce the following asymptotics of the Regge limit of $ {\ca{F}^{(n)}}^{\bar{\sigma}_n \sigma_n }_{\bb{O}\dg{\bb{O}}}(0|z)$,
\begin{equation}
\begin{aligned}
 {\ca{F}^{(n)}}^{\bar{\sigma}_n \sigma_n }_{\bb{O}\dg{\bb{O}}}(0|z)
& \ar{\text{Regge limit}}  \left\{
    \begin{array}{ll}
   \pa{\text{Res}\pa{  -2\pi i   {\bold F}_{0, \a} [O] ; \a=2\a_O} }^{n-1}
 z^{n(h_{2\a_O}-2h_O)} ,& \text{if } h_O<\fr{c-1}{32}   ,\\
  \pa{  \fr{\s{\pi}}{8} \del_{\a}^2 \left. {\bold F}_{0,\a}[O] \right|_{\a=\fr{Q}{2}} 
    \pa{\log z}^{-\fr{3}{2}}  }^{n-1}
 z^{n\pa{\fr{Q^2}{4}-2h_O}} ,& \text{if } h_O>\fr{c-1}{32} .\\
    \end{array}
  \right.\\
\end{aligned}
\end{equation}
This Regge asymptotics suggests that the pole structure of the replica monodromy matrix is expressed by the following form,
\begin{equation}\label{eq:ZnRegge}
\begin{aligned}
 {\ca{F}^{(n)}}^{\bar{\sigma}_n \sigma_n }_{\bb{O}\dg{\bb{O}}}(0|z)
&\ar{(1-z) \to \ex{-2\pi i}(1-z)}
 \sum_{\substack{\a_{l,m}<\fr{Q}{2} \\ l,m \in \bb{Z}_{\geq0}}}\ \text{Res}\pa{    -2\pi i 
	{{\bold M}^{(n)}}_{0, \a} [\sigma_n,\bb{O}]
	;\a=\a_{l,m}}
	 {\ca{F}^{(n)}}^{\bar{\sigma}_n \sigma_n }_{\bb{O}\dg{\bb{O}}}(n h_{\a_{l,m}}|z)  \\
&+\int_{\fr{Q}{2}+0}^{\fr{Q}{2}+i \infty} \dd \a \
	{{\bold M}^{(n)}}_{0, \a} [\sigma_n,\bb{O}]
{\ca{F}^{(n)}}^{\bar{\sigma}_n \sigma_n }_{\bb{O}\dg{\bb{O}}}(n h_{\a}|z),
\end{aligned}
\end{equation}
where $\a_{l,m}\equiv 2\a_O+lb+mb^{-1}$. We would like to emphasize that the discrete/continuum transition occurs as the Liouville momentum $\a_O$ (Not $\a_\bb{O}$!) crosses the line $\a_O=\fr{Q}{4}$ in the same way as in the seed pure CFT.

The above expectation comes from the relation (\ref{eq:replicatwist}), which makes sense only if $n\geq2$. Therefore, we do not have any evidence of the validity for (\ref{eq:ZnRegge}) at $n<2$. If anything, it is naturally expected that the replica transition occurs at $n<2$ as mentioned in Section \ref{subsec:2ndREE}. From these observations,  we conjecture
\footnote{If the dominant contribution is given by not a single-pole but a double-pole residue, the $z$-dependence cannot be consistent with the relation  (\ref{eq:replicatwist}).}
\begin{equation}\label{eq:ZnRegge2}
\begin{aligned}
 {\ca{F}^{(n)}}^{\bar{\sigma}_n \sigma_n }_{\bb{O}\dg{\bb{O}}}(0|z)
\ar{(1-z) \to \ex{-2\pi i}(1-z)}
 &\sum_{\substack{\a_{l,m}<\fr{Q}{2} \\ l,m \in \bb{Z}_{\geq0}}}\ \text{Res}\pa{    -2\pi i 
	{{\bold M}^{(n)}}_{0, \a} [\sigma_n,\bb{O}]
	;\a=\a_{l,m}}
	 {\ca{F}^{(n)}}^{\bar{\sigma}_n \sigma_n }_{\bb{O}\dg{\bb{O}}}(n h_{\a_{l,m}}|z)  \\
+& \sum_{\substack{\b_{l,m}<\fr{Q}{2} \\ l,m \in \bb{Z}_{\geq0}}}\ \text{Res}\pa{    -2\pi i 
	{{\bold M}^{(n)}}_{0, \a} [\sigma_n,\bb{O}]
	;\a=\b_{l,m}}
	 {\ca{F}^{(n)}}^{\bar{\sigma}_n \sigma_n }_{\bb{O}\dg{\bb{O}}}(n h_{\b_{l,m}}|z)  \\
+&\int_{\fr{Q}{2}+0}^{\fr{Q}{2}+i \infty} \dd \a \
	{{\bold M}^{(n)}}_{0, \a} [\sigma_n,\bb{O}]
{\ca{F}^{(n)}}^{\bar{\sigma}_n \sigma_n }_{\bb{O}\dg{\bb{O}}}(n h_{\a}|z),
\end{aligned}
\end{equation}
where $\a_{l,m}\equiv 2\a_O+lb+mb^{-1}$ and $\b_{l,m}\equiv 2\a_{\sigma_n}+lb+mb^{-1}$. 
Here we might abuse the notation of the replica monodromy matrix because they are precisely defined by 
some representative of an orbit of $\bb{Z}_n$ and the Drinfeld double of some stabilizer in  $\bb{Z}_n$ \cite{Bantay1998} (see also Appendix in \cite{Caputa2017}), which are too complicated to express as a simple form. Therefore, we abbreviate it  just by  ${{\bold M}^{(n)}}_{0, \a} [\sigma_n,\bb{O}]$. In fact, we expect that the explicit form of  ${{\bold M}^{(n)}}_{0, \a} [\sigma_n,\bb{O}]$ does not matter in the calculation of the entanglement entropy. We mean that  $\text{Res}\pa{    -2\pi i {{\bold M}^{(n)}}_{0, \a} [\sigma_n,\bb{O}] ;\a=\b_{l,m}}$ could be approximated  by $ \text{Res}\pa{    -2\pi i {\bold M}_{0, \a} [\sigma_n,O] ;\a=\b_{l,m}}$ in the limit $n \to 1$ 
\footnote{This assumption is often used in the calculation of the entanglement entropy and it is verified in the holographic CFT (see, for example, \cite{Asplund2015}).}
and consequently the entanglement entropy is
\begin{equation}\label{eq:EEinpureCFT}
\begin{aligned}
\D S_A (t)
&\ar{\fr{t}{\e} \to \infty} \lim_{n \to 1} \fr{-nh_{2\a_{\sigma_n}}}{1-n}\log\fr{ti}{2 \e}+\lim_{n \to 1} \fr{1}{1-n} \log  \BR{  \text{Res}\pa{  -2\pi i \   {\bold M}_{0,\a}[\sigma_n,O];\a=2\a_{\sigma_n}}} \\
&=\fr{c}{6}\log\fr{ti}{2 \e}+\lim_{n \to 1} \fr{1}{1-n} \log  \BR{  \text{Res}\pa{  -2\pi i \   {\bold M}_{0,\a}[\sigma_n,O];\a=2\a_{\sigma_n}}}.
\end{aligned}
\end{equation}
What we want to emphasize is that the $t$-dependence of this result is perfectly consistent with the holographic result \cite{Nozaki2013}, which supports our conjecture (\ref{eq:ZnRegge2}).
In fact, the constant part also reproduces the holographic result, which justifies the reduction ${\bold M}^{(n)} \ar{n \to 1} {\bold M}$.  (We will explain it in the next subsection.)
In other words, this reduction is the reason why the CFT calculation \cite{Asplund2015} reproduces the holographic result \cite{Nozaki2013} even though it is calculated not by the $(\text{Vir})^n/\bb{Z}_n$ block but by just the Virasoro block.

Note that the exact form of the $n$-th ($n\geq2$) Renyi entropy after a local quench $h_O<\fr{c-1}{32}$ is
\begin{equation}\label{eq:REEheavy}
\begin{aligned}
\D S^{(n)}_A (t)
&\ar{\fr{t}{\e} \to \infty}  \fr{n h_{2\a_O}}{1-n}\log  \pa{\fr{2 \e}{t i}}    + \fr{1 }{1-n}\log \BR{ \text{Res}\pa{  -2\pi i \   {{\bold M}^{(n)}}_{0,\a}[\sigma_n,\bb{O}];\a=2\a_{O}}}.
\end{aligned}
\end{equation}
This result completely matches the result from the different way (\ref{eq:$n$-th}). This consistency is the key to obtaining the pole structure of the $(\text{Vir})^n/\bb{Z}_n$ monodromy matrix ${\bold M}^{(n)}$ (\ref{eq:ZnRegge2}).
Note that the exact replica transition point associated to $h_{\a_{\sigma_n}}=\fr{c-1}{32}$  is given by $n_*=\fr{2}{\s{1+\fr{3}{c}}} \ar{c \to \infty} 2 $. In particular, this point satisfies the inequality $1 < n_* <2$ (if $c>1$).

It is interesting to make a comparison between the constant parts of (\ref{eq:REEheavy}) and (\ref{eq:$n$-th}). We obtain the relationship of the monodromy matrix with the fusion matrix for $n\geq2$ and $h_O<\fr{c-1}{32}$,
\begin{equation}
\text{Res}\pa{    -2\pi i {{\bold M}^{(n)}}_{0, \a} [\sigma_n,\bb{O}] ;\a=2\a_O} = \pa{\fr{i}{2n\sin\fr{\pi}{n}}}^{nh_{2\a_O}}  \pa{\text{Res}\pa{  -2\pi i   {\bold F}_{0, \a} [O] ; \a=2\a_O} }^{n-1} \bb{f}(h_O) .
\end{equation}
It is naturally expected that a similar relation between the fusion and monodromy matrices would be also satisfied for $n<2$ and we could re-express (\ref{eq:EEinpureCFT}) in terms of the fusion matrix.

\subsection{Entanglement Entropy and Renyi Entropy in Large $c$ Limit}\label{subsec:largecRenyi}

\begin{figure}[t]
 \begin{center}
  \includegraphics[width=8.0cm,clip]{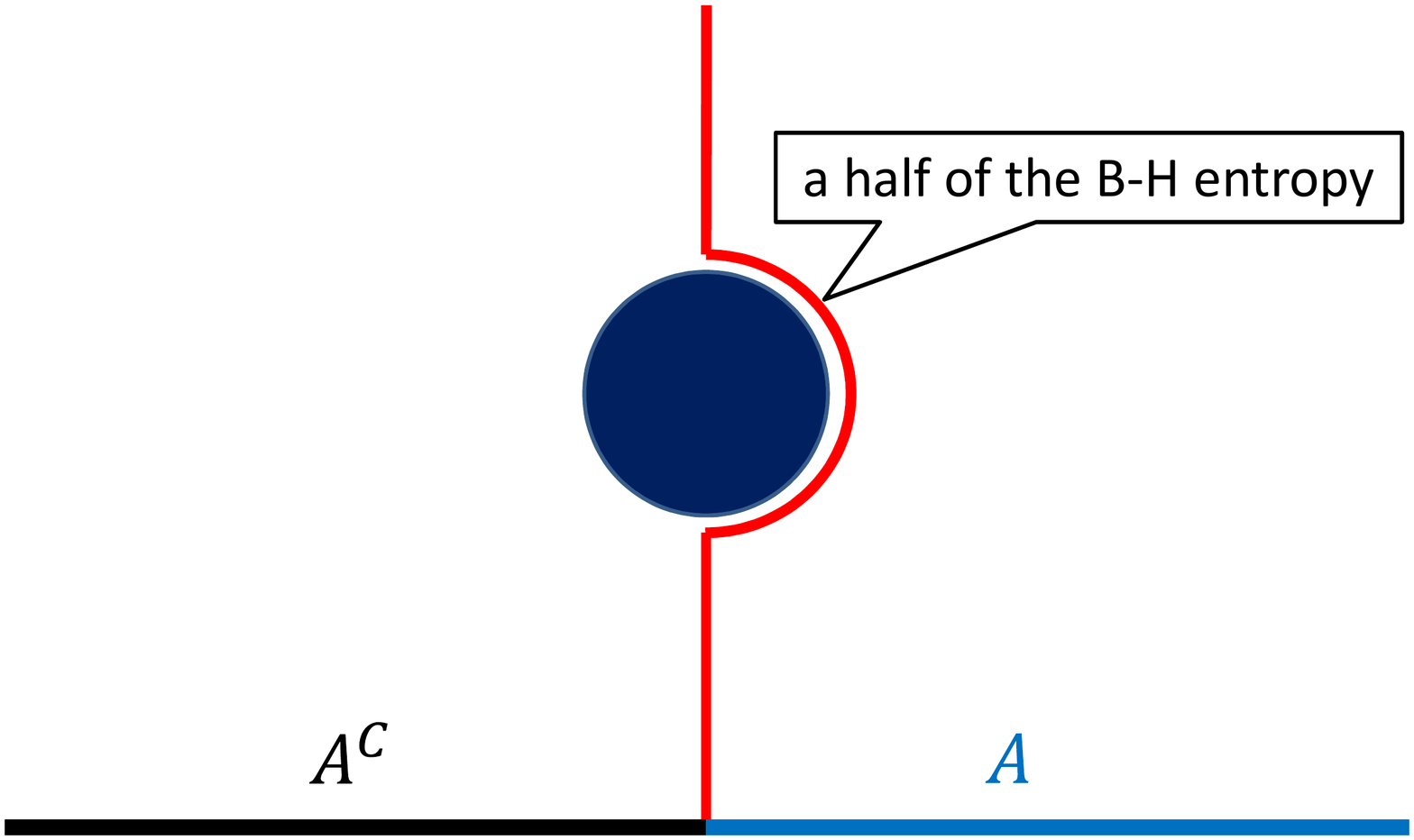}
 \end{center}
 \caption{ The interpretation of a half of the Bekenstein-Hawking entropy in the entanglement entropy (\ref{eq:EEatHHLLBH}). }
 \label{fig:halfBH}
\end{figure}

Let us consider entanglement entropy in the large $c$ limit. We first consider the case when $h_O=O(c)$ and $h_O<\fr{c-1}{24}$. Then we can apply the HHLL formula of the monodromy matrix (\ref{eq:HHLLMonodromyGeneral}) to (\ref{eq:EEinpureCFT}), and we have
\begin{equation}\D S_A (t)
=\fr{c}{6}\log\fr{t}{\e}+\fr{c}{6} \log  \pa{ \fr{\sin\pa{\pi \gamma_H}}{\gamma_H} },\label{eq:EEatHHLL}
\end{equation} 
consistent with the result given in \cite{Asplund2015}. Let us consider the other case when $h_O>\fr{c-1}{24}$. Applying the HHLL expression of the monodromy matrix (\ref{eq:HHLLMonodromy}) to (\ref{eq:EEinpureCFT}), we have
\begin{equation}\D S_A (t)
=\fr{c}{6}\log\fr{t}{\e}+\fr{c}{6} \log  \pa{ \fr{\sinh\pa{\pi \sqrt{\fr{24h_H}{c}-1}}}{\sqrt{\fr{24h_H}{c}-1}} }\label{eq:EEatHHLLBH},
\end{equation}
which is exactly the same as the analytical continuation of (\ref{eq:EEatHHLL}). Note that the constant term in (\ref{eq:EEatHHLLBH}) is identical to a chiral half of the Bekenstein Hawking entropy of the black hole, dual to the primary operator $O$ with $h_O>\fr{c-1}{24}$. Namely, we have
\begin{equation}\lim_{n \to 1} \fr{1}{1-n} \log  \BR{ -2i \text{Res}\pa{  -2\pi i \   {\bold M}_{0,\a}[\sigma_n,O];\a=2\a_{\sigma_n}}}\ar{c \to \infty} S_{\text{BH}}\pa{O},
\end{equation}
where $S_{\text{BH}}\pa{O} $ is given by the Cardy formula, 
\begin{equation}
S_{\text{BH}}\pa{O}=2\pi\s{\fr{c}{6}\pa{h_{ O}-\fr{c}{24}}}.
\end{equation}
In terms of the Ryu-Takayanagi formula, this constant contribution has a simple holographic interpretation. Following \cite{Nozaki2013}, the holographic dual of the local quench is given by a coordinate transformation of the static black hole solution 
\begin{equation}
\dd s^2=-\pa{r^2+R^2-M}\dd\tau^2+\fr{R^2\dd r^2}{R^2+r^2-M}+r^2\dd\theta^2,
\label{eq:MovingBH}\end{equation} which is asymptotically global AdS. $M$ is a mass parameter proportional to the mass of the black hole, and $R$ is the AdS radius. After an appropriate coordinate transformation (its explicit form is given in (2.15) of \cite{Nozaki2013}), this geometry is mapped to the asymptotically Poincare AdS with the moving massive particle, with the trajectory of the particle is given by
\begin{equation}\begin{aligned}
z^2-t^2&=R^2\ex{2\beta},\\&
x=0.
\end{aligned}
\end{equation}
Taking the limit $\b \to -\infty$ yields the geometry holographically dual to the local quench. We consider entanglement entropy of the half line $\left[0,L\right]$ $\pa{L\to \infty}$ on the plain after the local quench. The corresponding Ryu-Takayanagi surface at time $t$ is the geodesic connecting two boundary points $\pa{\tau,\theta}=\pa{\pi-\fr{2R\ex{\b}}{t},0}$ and $\pa{\fr{2tR\ex{\b}}{L^2},\pi-\fr{2R\ex{\b}}{L}}$ in (\ref{eq:MovingBH}). In the late time limit ($R\ex{\beta}\ll t\ll L$), these points come close to being light like separated. And the trajectory of the geodesic also approaches the black hole horizon, wrapping half of the horizon. This contributes to the holographic entanglement entropy $\D S_A (t)$ by half of the Bekenstein Hawking entropy (see Figure \ref{fig:halfBH}), which is the chiral half of the B-H entropy with $h_O=\bar{h}_O$.

Let us discuss the relation between the monodromy matrices ${{\bold M}^{(n)}}_{0, \a} [\sigma_n,\bb{O}]$ and ${{\bold M}}_{0, \a} [\sigma_n,O]$ in the large $c$ limit, which played an important role in the holographic CFT calculation. If the monodromy matrix on the orbifold CFT could be approximated  by,
\begin{equation}\label{eq:largecM}
\text{Res}\pa{    -2\pi i {{\bold M}^{(n)}}_{0, \a} [\sigma_n,\bb{O}] ;\a=2\a_O} \ar{c \to \infty} \pa{ \text{Res}\pa{    -2\pi i {\bold M}_{0, \a} [\sigma_n,O] ;\a=2\a_O}}^n  
\end{equation}
 in the large $c$ limit (instead of $n \to 1$), we can give the $n$-th $(n\geq2)$ Renyi entropy for a locally excited state $h_O<\fr{c}{32}$ as
\begin{equation}\label{eq:ReggelargecRenyi}
\begin{aligned}
\D S^{(n)}_A (t)
&\ar{\fr{t}{\e} \to \infty}  \fr{n }{1-n}\log   \BR{ \pa{\fr{2 \e}{t i}}^{h_{2\a_O}} \text{Res}\pa{  -2\pi i \   {\bold M}_{0,\a}[\sigma_n,O];\a=2\a_{O}}},
\end{aligned}
\end{equation}
where we used the Regge singularity (\ref{eq:ZnRegge2}). In particular, when $O$ is light, the HHLL limit simplifies the Renyi entropy,
\begin{equation}\label{eq:ReggeHHLLRenyi}
\begin{aligned}
\D S^{(n)}_A (t)
&\ar{\fr{t}{\e} \to \infty}   \fr{n }{1-n}\log   \BR{ \pa{\fr{2 \e}{t i}}^{h_{2\a_O}} \pa{\fr{\g_{\sigma_n} i}{2 \sin \pi \g_{\sigma_n}}}^{h_{2\a_O}} }
=  \fr{ n h_{2\a_O} }{n-1}\log \pa{\fr{ n t \sin \fr{\pi}{n}}{\e}},
\end{aligned}
\end{equation}
where $\g_H\equiv1-\fr{2}{Q}\a_h=\s{1-\fr{24}{c}h_H}$ and $\g_{\sigma_n} =\fr{1}{n}$.
On the other hand, we obtained the $n$-th Renyi  entropy in the other way as in (\ref{eq:$n$-th}),
\begin{equation}\label{eq:LCHHLLRenyi}
\begin{aligned}
\D S^{(n)}_A (t)
&\ar{\fr{t}{\e} \to \infty}   
 \fr{ n h_{2\a_O} }{n-1}\log \pa{\fr{ n t \sin \fr{\pi}{n}}{\e}},
\end{aligned}
\end{equation}
where we used
\begin{equation}
      \text{Res}\pa{    -2\pi i
  {\bold F}_{0, \a} [O]
  ; \a=2\a_O} \ar{\fr{h_O}{c} \to 0} 1,
\end{equation}
and
\begin{equation}
 \bb{f}(h_O) \ar{\fr{h_O}{c} \to 0} 1.
\end{equation}

The agreement between (\ref{eq:ReggeHHLLRenyi}) and (\ref{eq:LCHHLLRenyi}) implies the validity of the large $c$ approximation of the monodromy matrix (\ref{eq:largecM}). 
Note that we could also provide a similar relation for $h_O>\fr{c}{32}$. 
It would be interesting to generalize the relation  (\ref{eq:largecM}) by comparing between the sun-leading contributions in $\fr{\e}{t}$ to (\ref{eq:ReggelargecRenyi}) and (\ref{eq:$n$-th}).

\subsection{Renyi Entropy in Pure CFT}\label{subsec:REEinpureCFT}

There are some comments about the Renyi entropy in the pure CFT. Firstly, we would like to mention that the non-vacuum contributions to the Renyi entropy can be neglected in the limit $\e \to 0$. This is because the light cone limit and the Regge limit of the conformal block do not depend on the intermediate state. This independency had also been seen in the late time behavior of the non-vacuum block\cite{Fitzpatrick2017}. The full correlator is given by the following form,
\begin{equation}
\begin{aligned}
\braket{O_B(\infty) O_B(1) O_A(z,\bar{z}) O_A(0)}
&\ar{\substack{\text{Light cone limit} \\ \text{or} \\ \text{Regge limit} }}
z^\#\pa{C_0+C_1\bar{z}^{\bar{h}_{p_1}}+C_2\bar{z}^{\bar{h}_{p_2}}+\cdots }\\
&=\pa{\fr{t}{\e}}^\#\pa{C_0+C_1 \pa{\fr{t}{\e}}^{-\bar{h}_{p_1}}+C_2 \pa{\fr{t}{\e}}^{-\bar{h}_{p_2}}+\cdots },
\end{aligned}
\end{equation}
where $C_i$ are some constants. Therefore, we find that the limit $\e \to 0$ obviously damp the non-vacuum contributions.

Secondly, the constant part of the entanglement entropy depends only on the chiral part of the CFT. According to \cite{David2016}, this decomposition comes from the limit of large interval and late time.
It means that the entanglement growth by a local quench is controlled by a chiral CFT in the late time limit.

Thirdly, we want to emphasize that our calculation does not rely on the large $c$ approximation.
It had been expected that the logarithmic growth of the entanglement entropy would be {\it resolved} and the correct entanglement entropy would be rendered by finite if we may take non-perturbative effects in $c$ into account
\cite{Caputa2014a, David2016}. However, our result shows that the entanglement entropy even for finite $c$ also increases logarithmically. 
\subsection{Renyi Entropy in RCFT}\label{subsec:RCFT}
We first have to emphasize that when one of the external operators corresponds to a degenerate operator, the decomposition of the $s$-channel in terms of the $t$-channel is not continuum, but is discrete. That is, in such a case, the crossing kernel needs to be written as a linear combination of delta functions. This is because  $S_b$ or $\G_b$ in the denominator of the fusion matrix (see (\ref{eq:crossing2})) with the degenerate external operator dimension diverges unless $\a_t$ takes particular values, which means that the integral over $\a_t$ is replaced by the sum over the particular values. This concept is explained in greater detail in \cite{Hadasz2005,EsterlisFitzpatrickRamirez2016}. We have to mention that this never happens for unitary CFTs with $c>1$ because if the central charge is larger than one, then the conformal dimensions of degenerate operators are negative.

In the same reason as explained above, the definition of the monodromy matrix (\ref{eq:defmono}) is slightly changed by replacing the integral with the sum as 
\begin{equation}
 {\bold M}_{0, \a}
   \left[
    \begin{array}{cc}
    \a_A   & \a_A  \\
     \a_B  &   \a_B\\
    \end{array}
  \right]
=
\sum_{\b} \ 
\ex{-2\pi i (h_\b-h_A-h_B)}
 \ {\bold F}_{0, \b}
   \left[
    \begin{array}{cc}
    \a_A   & \a_A  \\
     \a_B  &   \a_B\\
    \end{array}
  \right]
 \ {\bold F}_{\b, \a}
   \left[
    \begin{array}{cc}
    \a_A   & \a_B  \\
     \a_A  &   \a_B\\
    \end{array}
  \right].
\end{equation}
For our purposes, we note the useful identities \cite{Moore1989};
\begin{description}
\item[The symmetries of the fusion matrix]
\begin{equation}\label{eq:fusionsym}
\begin{aligned}
{\bold F}_{\a_s, \a_t} 
   \left[
    \begin{array}{cc}
    \a_2   & \a_1  \\
     \a_3  &   \a_4\\
    \end{array}
  \right]
={\bold F}_{\a_s, \a_t} 
   \left[
    \begin{array}{cc}
    \a_1   & \a_2  \\
     \a_4  &   \a_3\\
    \end{array}
  \right]
={\bold F}_{\a_s, \a_t} 
   \left[
    \begin{array}{cc}
    \a_3   & \a_4  \\
     \a_2  &   \a_1\\
    \end{array}
  \right],
\end{aligned}
\end{equation}
\item[The relation between the fusion matrix and the braiding matrix]
\begin{equation}\label{eq:FtoB}
\begin{aligned}
{\bold F}_{\a_s, \a_t} 
   \left[
    \begin{array}{cc}
    \a_2   & \a_1  \\
     \a_3  &   \a_4\\
    \end{array}
  \right]
=\ex{-\e i\pi(h_1+h_3-h_{\a_s}-h_{\a_t})}
{\bold B}_{\a_s, \a_t} ^{(\e)}
   \left[
    \begin{array}{cc}
    \a_2   & \a_1  \\
     \a_4  &   \a_3\\
    \end{array}
  \right].
\end{aligned}
\end{equation}
\end{description}
Note that by making use of the hexagon identity \cite{Moore1989}, our definition of the monodromy matrix can be related to the more familiar form (which is used in a proof of the Verlinde formula \cite{Moore1989b}),
\begin{equation}\label{eq:M=BB}
 {\bold M}_{0, \a}
   \left[
    \begin{array}{cc}
    \a_A   & \a_A  \\
     \a_B  &   \a_B\\
    \end{array}
  \right]
=\sum_{\b} \ 
 \ {\bold B}_{0, \b}^{(-)}
   \left[
    \begin{array}{cc}
    \a_A   & \a_A  \\
     \a_B  &   \a_B\\
    \end{array}
  \right]
 \ {\bold B}_{\b, \a}^{(-)}
   \left[
    \begin{array}{cc}
    \a_A   & \a_B  \\
     \a_B  &   \a_A\\
    \end{array}
  \right].
\end{equation}
Utilizing the pentagon identity, the hexagon identities, and the modular invariance \cite{Moore1989}, we can show the following relation for RCFTs,
\begin{equation}
 {\bold M}_{0, 0}
   \left[
    \begin{array}{cc}
    \a_A   & \a_A  \\
     \a_B  &   \a_B\\
    \end{array}
  \right]
=\fr{S_{AB}}{S_{00}} \fr{S_{00}}{S_{0A}} \fr{S_{00}}{S_{0B}}.
\end{equation}

Now all is ready to calculate the Renyi entropy in RCFTs. Since an orbifold CFT of a RCFT is also RCFT, the Regge limit of the four point function with the twist operators can be approximated by the vacuum block as
\begin{equation}
\begin{aligned}
\braket{ \sigma_n(0)\bar{\sigma}_n(z)O^{\otimes n}(1)O^{\otimes n}(\infty)}
&\ar{\text{Regge limit}}
{{\bold M}^{(n)}}_{0, 0} 
   \left[
    \begin{array}{cc}
     \a_{\bar{\sigma}_n} & \a_{\sigma_n}  \\
     \a_{\bb{O}}  &   \a_{\dg{\bb{O}}}\\
    \end{array}
  \right]
 {\ca{F}^{(n)}}^{\bar{\sigma}_n \sigma_n }_{\bb{O}\dg{\bb{O}}}(0|z)  \overline{{\ca{F}^{(n)}}^{\bar{\sigma}_n \sigma_n }_{\bb{O}\dg{\bb{O}}}} (0|\bar{z})\\
&=
{{\bold M}^{(n)}}_{0, 0} 
   \left[
    \begin{array}{cc}
     \a_{\bar{\sigma}_n} & \a_{\sigma_n}  \\
     \a_{\bb{O}}  &   \a_{\dg{\bb{O}}}\\
    \end{array}
  \right]
\abs{z^{2h_{\sigma_n}}}^2.
\end{aligned}
\end{equation}
Note that this situation is quite different from that in the pure CFTs because in the pure CFTs, the dominant contribution of the four point function in the Regge limit is not given by the vacuum block as seen in (\ref{eq:twistRenyi}). On the other hand, in RCFTs, the sum of the monodromy transformation includes the vacuum, which is the dominant contribution.

Inserting this result into (\ref{eq:GRenyi}), we obtain
\begin{equation}\label{eq:MtoQD}
\Delta S^{(n)}_A=\frac{1}{1-n}\log 
{{\bold M}^{(n)}}_{0, 0} 
   \left[
    \begin{array}{cc}
     \a_{\bar{\sigma}_n} & \a_{\sigma_n}  \\
     \a_{\bb{O}}  &   \a_{\dg{\bb{O}}}\\
    \end{array}
  \right]
=\frac{1}{1-n}\log  \fr{{S^{(n)}}_{\dg{\sigma_n} \bb{O}}}{{S^{(n)}}_{00}} \fr{{S^{(n)}}_{00}}{{S^{(n)}}_{0 \sigma_n}} \fr{{S^{(n)}}_{00}}{{S^{(n)}}_{0\bb{O}}}.
\end{equation}
The quantum dimension of the twist operator is given in \cite{Caputa2017},
\begin{equation}
d_{\sigma_n}=\fr{{S^{(n)}}_{0 \sigma_n}}{{S^{(n)}}_{00}}=\fr{1}{\pa{S_{00}}^{n-1}}.
\end{equation}
We expect a similar relation is also satisfied (see \cite{Bantay1998}) as,
\begin{equation}
\fr{{S^{(n)}}_{\dg{\sigma_n}\bb{O}}}{{S^{(n)}}_{0\bb{O}}}=\fr{1}{\pa{S_{0O}}^{n-1}}.
\end{equation}
As a result, the growth of the Renyi entropy is given in terms of the quantum dimension as
\begin{equation}
\Delta S^{(n)}_A=\log \fr{S_{0O}}{S_{00}}=\log d_O.
\end{equation}
This  exactly matches the result from the replica method \cite{He2014}.

\subsection{Renyi Entropy in Liouville CFT}
It might be interesting to note that our monodromy matrix approach can be also used to calculate the growth of the Renyi entanglement entropy in the Liouville CFT in the same way as Section \ref{subsec:twistframe}. However, we have no knowledge about the three point function of the orbifold CFT $(\text{Liouville})^n/\bb{Z}_n$, therefore, we cannot determine the dominant pole in the $t$-channel expansion.
 Nevertheless, we naively expect that the limit $n\to 1$ of the Renyi entropy can be evaluated by the same logic as in the holographic CFT,
\footnote{The Hilbert space of the Liouville CFT does not have the vacuum state, therefore, the concept of the growth  $\Delta S^{(n)}_A(t)$  (\ref{eq:difs}) becomes subtle. Nevertheless, we can formally {\it define} the correlator (\ref{eq:GRenyi}) in the orbifold CFT  $(\text{Liouville})^n/\bb{Z}_n$.
}
\begin{equation}\label{eq:LiouvilleREE}
\fr{2h_{2\a_n}}{n-1}\log\fr{t}{\e}\ar{n \to1} \fr{c}{3}\log\fr{t}{\e},
\end{equation}
where $h_{\sigma_n}=\a_n(Q-\a_n)$. The details of the Liouville four point function in the Regge limit can be found in Section \ref{subsec:LiouvilleOTOC}.
We find that the growth of the Liouville entanglement entropy is a double of the holographic entanglement entropy. It particularly means that the Liouvile entanglement entropy also shows the logarithmic behavior like the holographic entanglement entropy. The Liouville Renyi entropy for a local quench is also investigated in \cite{He2017} by using the light cone limit. However, we have to mention that we must take account of the possibility of the Renyi transition at $n=2$ as we explained in this section.
In that, the analytic continuation $n \to 1$ of the result from the light cone analysis does not provide the correct entanglement entropy.

In the above, we formally define the entanglement growth compared to the vacuum by the correlator (\ref{eq:GRenyi}). However, there is no vacuum in the Hilbert space of the Liouville CFT, therefore, its physical meaning seems to be subtle.
To avoid this problem, it is natural to use the following definition in place of (\ref{eq:difs}),
\footnote{
We have to mention that this definition does not show the growth of the entanglement because the entanglement entropy for the vacuum also grows in a similar way. The real entanglement growth is given by (\ref{eq:LiouvilleREE}).
}
 \be
 \Delta \ti{S}^{(n)}_A(t)=S^{(n)}_A(\ket{\Psi(t)})-S^{(n)}_A(\ket{\Psi(0)}).
 \ee
In terms of the correlator, this entanglement growth is given by
\begin{equation}\label{eq:defREE2}
\Delta \ti{S}^{(n)}_A=\frac{1}{1-n}\log \frac{\lim_{\text{Regge limit}}\ave{O^{\otimes n}O^{\otimes n}\sigma_n \bar{\sigma_n} }}{\lim_{z,\bar{z} \to 0}\ave{O^{\otimes n}O^{\otimes n} \sigma_n \bar{\sigma_n}}}.
\end{equation}
In the pure CFT, this definition leads to the same conclusion as given in Section \ref{subsec:twistframe}, because the denominator $\lim_{z,\bar{z} \to 0}\ave{O^{\otimes n}O^{\otimes n} \sigma_n \bar{\sigma_n}}$ is decomposed into $\ave{O^{\otimes n}O^{\otimes n}}\ave{\sigma_n \bar{\sigma_n}}$, and therefore, the entanglement growth (\ref{eq:defREE2}) reduces to just the definition (\ref{eq:defREE}). On the other had, in the Liouville CFT, the denominator cannot be factorized even in the limit $z, \bar{z} \to 0$ (see (\ref{eq:LiouvilleNOC}) in more details). Hence, the entanglement growth is changed from (\ref{eq:LiouvilleREE}) to
\begin{equation}
\begin{aligned}
\Delta \ti{S}^{(n)}_A= &\ar{\e \to 0} \left\{
    \begin{array}{ll}
       \text{Res}\pa{    -2\pi i {{\bold M}^{(n)}}_{2\a_{O}, \a} [\sigma_n,\bb{O}] ;  \a=2\a_{O}}   ,& \text{if }   \a_{O}<\fr{Q}{4}, \a_{\sigma_n}    ,\\
       \text{Res}\pa{    -2\pi i {{\bold M}^{(n)}}_{ 2\a_{\sigma_n}, \a} [\sigma_n,\bb{O}] ;   \a=2\a_{\sigma_n}}  ,& \text{if } \a_{\sigma_n}<\fr{Q}{4},\a_{O}    ,\\  
    \fr{\s{\pi}}{8} \del_{\a}^2 \left. {{\bold M}^{(n)}}_{\fr{Q}{2}, \a} [\sigma_n,\bb{O}] \right|_{\a=\fr{Q}{2}}    ,&\text{if } \a_{\sigma_n}, \a_O >\fr{Q}{4} , \\
    \end{array}
  \right.\\
\end{aligned}
\end{equation}
where we abused the notation of the replica monodromy matrix, but what we want to emphasize is clear in this description: The entanglement growth approaches constant and there are three types of the constant, depending on the replica number $n$ and the conformal dimension $h_O$.

\section{Out-of-Time Ordered Correlator}

\subsection{Late Time Regime}

\begin{figure}[t]
 \begin{center}
  \includegraphics[width=5.0cm,clip]{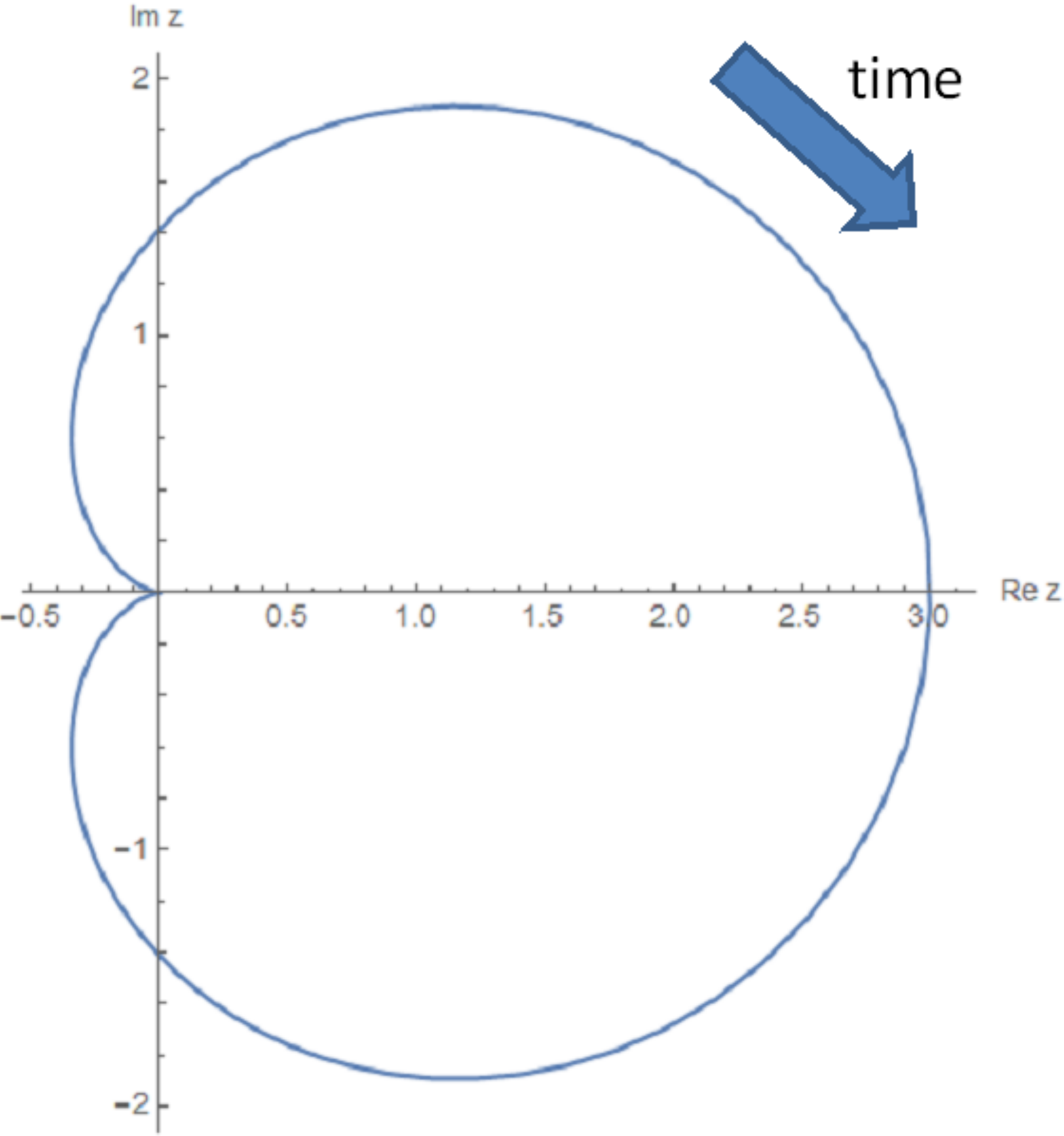}
 \end{center}
 \caption{The time dependence of the holomorphic cross ratio.}
 \label{fig:OTOCtime}
\end{figure}

In this section, we will calculate the late time behavior of the OTOC by making use of the result in Section \ref{subsec:Regge}.
As mentioned in the introduction, the OTOC in the late time is given by the Regge asymtotics of the correlator,
\begin{equation}\label{eq:Reggevacuum}
G(z,\bar{z}) \ar{\substack{\text{Regge limit} \\ (1-z) \to \ex{-2 \pi i }(1-z)}} \ca{F}^{{\dg W}W}_{V{\dg V}}(0|z)  \overline{\ca{F}^{{\dg W}W}_{V{\dg V}}}(0|\bar{z}),
\end{equation}
where the function $G(z,\bar{z})$ is defined by
\begin{equation}\label{eq:Cbeta}
C_\beta(x,t)\equiv\frac{\langle V^\dagger W^\dagger(t)V W(t)\rangle_\beta}{\langle V^\dagger V\rangle_\beta\langle W^\dagger W\rangle_\beta}
=\abs{z^{2h_W}}^2 G(z,\bar{z}).
\end{equation}
Here we also assume that we restrict ourselves to the pure CFTs.
The late time behavior of the cross ratio is expressed by
\begin{equation}\label{eq:OTOCcross}
\begin{aligned}
z\simeq -\ex{-\frac{2\pi (t-x)}{\beta}}\epsilon^*_{12}\epsilon_{34}, \ \ \ \ 
\bar{z}\simeq -\ex{-\frac{2\pi (t+x)}{\beta}}\epsilon^*_{12}\epsilon_{34},
\end{aligned}
\end{equation}
where $\epsilon_{ij}=i \left(\ex{\frac{2\pi i}{\beta}\epsilon_i}-\ex{\frac{2\pi i}{\beta}\epsilon_j}\right)$. From these expressions, we can straightforwardly find that the cross ratio goes to zero in the late time. The time evolution of the holomorphic cross ratio is shown as in Figure \ref{fig:OTOCtime}, which clearly explains why we need to pick up a monodromy around $z=1$ to investigate the OTOC in the late time limit.

The Regge limit of the conformal block is derived in Section \ref{subsec:Regge} as
\begin{equation}\label{eq:Regge2}
\begin{aligned}
\ca{F}^{AA}_{BB}(h_p|z)
&\ar{\text{Regge limit}} \left\{
    \begin{array}{ll}
    \text{Res}\pa{   -2\pi i 
  {\bold M}_{\a_p, \a} [\a_A,\a_B]
  ; \a=2\a_A} z^{h_{2\a_A}-2h_A}   ,& \text{if } \a_A<\fr{Q}{4},\a_B   ,\\
    \text{Res}\pa{   -2\pi i 
  {\bold M}_{\a_p, \a} [\a_A,\a_B]
  ; \a=2\a_B} z^{h_{2\a_B}-2h_A}   ,& \text{if } \a_B<\fr{Q}{4},\a_A   ,\\
    -i\fr{\s{\pi}}{8} \del_{\a}^2 \left. {\bold M}_{\a_p,\a}[\a_A,\a_B] \right|_{\a=\fr{Q}{2}} z^{\fr{Q^2}{4}-2h_A}\pa{-\log z}^{-\fr{3}{2}}   ,& \text{if } \a_A,\a_B>\fr{Q}{4}  .\\
    \end{array}
  \right.\\
\end{aligned}
\end{equation}
Substituting these into (\ref{eq:Cbeta}), we obtain the late time behavior of the OTOC as
\begin{equation}\label{eq:OTOCresult}
\begin{aligned}
C_\b(x,t) &\ar{\text{late time}} \left\{
    \begin{array}{ll}
      \ex{-h_{2\a_W} \fr{2 \pi t}{\b}}  ,& \text{if } \a_W<\fr{Q}{4},\a_V     ,\\
     \ex{-h_{2\a_V} \fr{2 \pi t}{\b}}   ,& \text{if } \a_V<\fr{Q}{4},\a_W    ,\\
    \ex{-\fr{Q^2}{4}\fr{2 \pi t}{\b}}t^{-\fr{3}{2}}   ,&\text{if } \a_W, \a_V>\fr{Q}{4}    .\\
    \end{array}
  \right.\\
\end{aligned}
\end{equation}
In the test mass limit $h_W \ll c$, the conformal dimension $h_{2\a_W}$ is approximated by $2h_W$, and then we obtain
\begin{equation}
C_\b(x,t) \ar{h_W\ll c}  \ex{-2h_W \fr{2 \pi t}{\b}},
\end{equation}
which perfectly matches the result from the HHLL Virasoro block \cite{Perlmutter2016}.
It is interesting to point out that the late time behavior of the OTOC has the transition as the conformal dimensions $h_W, h_V$ increase, which is first suggested in \cite{Kusuki2018b, Kusuki2018} from the numerical calculation.
This transition should be caused by a creation of a block hole in gravity side, therefore, it is a very interesting future work to discover this transition from the gravitation calculation \cite{Maldacena2016}.

In RCFTs, the late time behavior of the OTOC approaches the constant \cite{Caputa2016, Gu2016, Fan2018}. Moreover, this constant is  equal to the $(0,0)$ element of the monodromy matrix,
\begin{equation}\label{eq:RCFTOTOC}
C_\b(x,t) \ar{\text{late time}}   {\bold M}_{0, 0} [\a_W,\a_V].
\end{equation}
Note that the way to obtain this result is exactly the same as in Section \ref{subsec:RCFT}, in that, making use of the monodromy matrix approach.
Now we find that the contribution from the monodromy matrix also appears in the pure CFTs, from the expression of the Regge asymptotics (\ref{eq:Regge2}).
In that, the counterpart of the constant $ {\bold M}_{0, 0} [\a_W,\a_V]$ is given by the coefficient of the exponential decay in the pure CFTs. From this viewpoint, we may interpret the contribution from the monodromy matrix element more physically from the gravity side. It would be interesting to proceed in this direction.

We would like to mention that we only assume our CFT to be a unitary compact CFT with $c>1$ (and without chiral Virasoro primaries). In particular, we do not impose the large $c$ on our CFT. In \cite{Caputa2016}, it is discussed there which the OTOC or the entanglement entropy dynamics is better to characterize the chaos nature of a given quantum field theory and it is concluded that the OTOC is better because the large $c$ limit of the $SU(N)_k$ WZW model does not break down the constant behavior of the OTOC, whereas it spoils the constant behavior of the $2$nd Renyi entropy and leads to the logarithmic growth. However, we now find that the time-dependent behavior is not caused by the large $c$ limit but by assumption on our CFT to be a pure CFT.
Therefore, we should rather conclude that
\begin{enumerate}
\item
The late time behavior of the OTOC is characterized by whether a given CFT is a pure CFT or not.
\item
The logarithmic growth of the entanglement entropy is not caused by the large $c$ but by the properties of the pure CFT.
(The non-perturbative correction in $c$ is not be able to recover the constant behavior of the entanglement growth.)
\end{enumerate}

We will comment on a case where there is an extra current besides the Virasoro current.
In such a case, the OTOC is approximated by
\begin{equation}\label{eq:Reggevacuum2}
G(z,\bar{z}) \ar{\substack{\text{Regge limit} \\ (1-z) \to \ex{-2 \pi i }(1-z)}} {\ca{F}^{(J)}}^{{\dg W}W}_{V{\dg V}}(0|z)  \overline{{\ca{F}^{(J)}}^{{\dg W}W}_{V{\dg V}}}(0|\bar{z}),
\end{equation}
where $\ca{F}^{(J)}$ is conformal blocks associated to the chiral algebra $J$. The logic to estimate the Regge singularity of $\ca{F}^{(J)}$ is the same as the Virasoro conformal blocks, in that, 
\begin{equation}\label{eq:Reggecurrent}
\begin{aligned}
 {\ca{F}^{(J)}}^{{\dg W}W}_{V{\dg V}}(h_p|z) 
&\overset{?}{=}
\int \dd \a \ {{\bold M}^{(J)}}_{\a_p, \a}[\a_W, \a_V] {\ca{F}^{(J)}}^{{\dg W}W}_{V{\dg V}}(h_\a|z) \\
&\ar{\text{Regge limit}} z^{h_{\a_{\text{min}}}-2h_W}.
\end{aligned}
\end{equation}
We have the monodromy symmetry of the correlator,
\begin{equation}
G(1-z,1-\bar{z})=G\pa{\ex{-2 \pi i}(1-z),\ex{2 \pi i}(1-\bar{z})}.
\end{equation}
The left hand side has the asymptotic behavior $(1-z)^{-2h_{W}}$ in the limit $z \to 1$. On the other hand, the right hand side is given by the sum of the asymptotics (\ref{eq:Reggecurrent}). From this viewpoint, we find that $h_{\a_{\text{min}}}$ is not allowed to be smaller than zero. In RCFTs, the right hand side is given by a finite sum. Therefore, $h_{\a_{\text{min}}}$ must be zero to be consistent with the monodromy symmetry. As a result, we can conclude that the late time behavior of the OTOC is given by
\begin{equation}\label{eq:OTOCresult2}
C_\b(x,t) \ar{\text{late time}}  \ex{-2h_{\a_{\text{min}}} \fr{2 \pi t}{\b}}.
\end{equation}
\begin{enumerate}
\item
$h_{\a_{\text{min}}}=0$ in RCFT.
\item
If $h_{\a_{\text{min}}}>0$, it is naturally expected to have similar transitions as (\ref{eq:OTOCresult}).
\item
We can expect that if taking a large enough central charge with a number of chiral primaries fixed, the OTOC shows the exponential decay in the late time. It results that the OTOC in the holographic CFTs (with a large central charge) decays exponentially in general.
\end{enumerate}

It is interesting to note that the OTOC in a permutation orbifold at large $c$ does not decay in late time \cite{Belin2017a}.
More precisely, a permutation orbifold CFT is defined by $\ca{M}^n/G_N$ for $G_N \subseteq S_N$, giving an orbifold CFT with central charge $Nc$, therefore, the large $c$ limit is realized by taking the limit $N \to \infty$. We can see that the OTOC in such a theory approaches constant. This is one of examples of a CFT with $h_{\a_{\text{min}}}=0$ other than RCFTs.
\footnote{The limit $N \to \infty$ does not fix a number of chiral primaries, therefore, this phenomena does not contradict the expectation (3) below (\ref{eq:OTOCresult2}). A similar limit is the large $c$ limit of $SU(N)_k$ WZW model, studied in \cite{Caputa2016}.}

\subsection{Non-Vacuum Contribution in Late Time}\label{subsec:nonOTOC}

Let us introduce the non-vacuum contribution $C_\b(h_p|x,t)$ as
\begin{equation}
C_\b(h_p|x,t)\equiv\abs{z^{2h_W}}^2
\pa{\int_{\bb{S}} \dd \a \ {\bold M}_{\a_p, \a}[\a_W, \a_V] \ca{F}^{{\dg W}W}_{V{\dg V}}(h_\a|z)}
\overline{\ca{F}^{{\dg W}W}_{V{\dg V}}}(\bar{h}_p|\bar{z}),
\end{equation}
where the cross ratio is related to $(x,t)$ by (\ref{eq:OTOCcross}).
Then the full OTOC has the following expansion,
\begin{equation}
C_\b(x,t)=\sum_{p} C_{\dg{W} W p} C_{\dg{V}V p} C_\b(h_p|x,t).
\end{equation}
The Regge asymptotics of the non-vacuum contributions is also obtained from the pole structure of the fusion matrix with the non-vacuum intermediate state (\ref{eq:bootstrapBAAB}) in the same as the vacuum case. The result is
\begin{equation}
\begin{aligned}
C_\b(h_p|x,t) &\ar{\text{late time}} \left\{
    \begin{array}{ll}
      \ex{-\pa{h_{2\a_W}+\bar{h}_p} \fr{2 \pi t}{\b}}  ,& \text{if } \a_W<\fr{Q}{4},\a_V     ,\\
     \ex{-\pa{h_{2\a_V}+\bar{h}_p} \fr{2 \pi t}{\b}}   ,& \text{if } \a_V<\fr{Q}{4},\a_W    ,\\
    \ex{-\pa{\fr{Q^2}{4}+\bar{h}_p}  \fr{2 \pi t}{\b}}t^{-\fr{3}{2}}   ,&\text{if } \a_W, \a_V>\fr{Q}{4}    .\\
    \end{array}
  \right.\\
\end{aligned}
\end{equation}
This means that the non-vacuum contributions are suppressed more strongly than that of the vacuum. Therefore, we can conclude that the vacuum approximation (\ref{eq:Reggevacuum}) is justified in the Regge limit.

In the test mass limit $h_W \ll c$, the non-vacuum contributions reduce to
\begin{equation}
C_\b(h_p|x,t) \ar{\text{late time}}    \ex{-\pa{2h_W+\bar{h}_p} \fr{2 \pi t}{\b}}.
\end{equation}
This late time behavior of the non-vacuum contributions $C_\b(h_p|x,t) $ in the test mass limit has already derived in \cite{Hampapura2018}, which is consistent with our result from the monodromy matrix approach.
Moreover, we can generalize the results in \cite{Hampapura2018} beyond the test mass limit. We now assume that the smearing length scale of the operators $V, W$ is close to the thermal length scale and the OPE coefficients $C_{\dg{W} W p}, C_{\dg{V}V p}$ with the light intermediate state are of order $\ca{O}(1)$. Let $\bb{M}_{\a_p, \a_{\text{min}}}[\a_W,\a_V]$ denote the coefficient of the Regge asymptotics (\ref{eq:Regge2}). Then the {\it decay time} when the OTOC becomes much smaller than one is given by
\footnote{We neglect the polynomial decay, which only gives negligible contributions to ${t_{\text{decay}}}_p$ in the large $c$ limit. }
\begin{equation}
\begin{aligned}
{t_{\text{decay}}}_p=
 \left\{
    \begin{array}{ll}
      \fr{\b}{2\pi}\fr{1}{h_{2\a_W}}  \log\pa{\bb{M}_{\a_p, 2\a_W}[\a_W,\a_V]} ,& \text{if } \a_W<\fr{Q}{4},\a_V     ,\\
    \fr{\b}{2\pi}\fr{1}{h_{2\a_V}}  \log\pa{\bb{M}_{\a_p, 2\a_V}[\a_W,\a_V]},& \text{if } \a_V<\fr{Q}{4},\a_W    ,\\
   \fr{\b}{2\pi}\fr{4}{Q^2} \log\pa{\bb{M}_{\a_p, \fr{Q}{2}}[\a_W,\a_V]} ,&\text{if } \a_W, \a_V>\fr{Q}{4}    .\\
    \end{array}
  \right.\\
\end{aligned}
\end{equation}
In the test mass limit $h_W \ll c$, the decay time ${t_{\text{decay}}}_p$ reduces to
\begin{equation}
{t_{\text{decay}}}_p \ar{h_W \ll c} \fr{\b}{2\pi} \pa{\fr{2h_W+h_p}{2h_W+\bar{h}_p}}\log\pa{\fr{c}{h_W}}. 
\end{equation}
In order to the decay time is given by $t_{\text{decay}}={t_{\text{decay}}}_0$, the inequality $t_{\text{decay}}\geq {t_{\text{decay}}}_p$ should be satisfied.
More details can be found in \cite{Hampapura2018}.

\subsection{OTOC in Liouville CFT}\label{subsec:LiouvilleOTOC}

Besides pure CFTs, there is another interesting irrational CFT, {\it Liouville CFT}. Compared to other irrational CFTs, we have some useful tools to determine the correlator in the Liouville CFT \cite{Teschner2001a}. With this backdrop, we will also try to investigate the Liouville OTOC in this paper. Our physical motivation of this challenge is to clarify whether the Liouville OTOC has  something in common with the OTOC in a RCFT or a pure CFT (including the holographic CFT).
 In particular, it is interesting to bring out whether the Liouville OTOC reduces the OTOC in the holographic CFT in the large $c$ limit, which could be true because the Liouville CFT also has the gravitational interpretation \cite{McGough2013,Jackson2015} (but which is not the AdS/CFT correspondence). We will address these questions by using our monodromy matrix approach.

Our monodromy matrix approach can be straightforwardly generalized to the Liouville CFT.
The Liouville four point function has the following form,
\begin{equation}\label{eq:Liouvillecorr}
\begin{aligned}
&\braket{\ca{V}_{\a_B}(\infty)\ca{V}_{\a_B}(1)\ca{V}_{\a_A}(z)\ca{V}_{\a_A}(0)}\\
=&
-2\pi i  \sum_{\substack{\a_{n,m}^{(A)}<\fr{Q}{2} \\ n,m \in \bb{Z}_{\geq0}}}\ 
	\text{Res}\pa{   
		C^{\text{DOZZ}}(\a_A,\a_A,\a) C^{\text{DOZZ}}(Q-\a,\a_B,\a_B)
		  \ca{F}^{AA}_{BB}(h_{\a}|z) \overline{\ca{F}^{AA}_{BB}}(h_{\a}|\bar{z})
	;\a=\a_{n,m}^{(A)}}\\
&-2\pi i  \sum_{\substack{\a_{n,m}^{(B)}<\fr{Q}{2} \\ n,m \in \bb{Z}_{\geq0}}}\ 
	\text{Res}\pa{   
		C^{\text{DOZZ}}(\a_B,\a_B,\a) C^{\text{DOZZ}}(Q-\a,\a_A,\a_A)
		  \ca{F}^{AA}_{BB}(h_{\a}|z) \overline{\ca{F}^{AA}_{BB}}(h_{\a}|\bar{z})
	;\a=\a_{n,m}^{(B)}}\\
&+
\int_{\fr{Q}{2}+0}^{\fr{Q}{2}+i \infty} \dd \a
	C^{\text{DOZZ}}(\a_A,\a_A,\a) C^{\text{DOZZ}}(Q-\a,\a_B,\a_B)
  \ca{F}^{AA}_{BB}(h_{\a}|z) \overline{\ca{F}^{AA}_{BB}}(h_{\a}|\bar{z}) ,
\end{aligned}
\end{equation}
where $\a_{n,m}^{(I)}\equiv 2\a_I+mb+nb^{-1}$ ($I=A,B$)  and $C^{\text{DOZZ}}$ is the Liouville three point function, which is the so-called DOZZ formula \cite{Dorn1994a, Zamolodchikov1996a}. The reason for the appearance of the discrete terms is caused by the fusion rule of the Liovulle CFT (\ref{eq:fusion rule}), which is nicely reviewed in \cite{Ribault2014}.
We first want to mention that the Liouville four point function in the trivial OPE limit $\abs{z} \to 0$ shows a quite different behavior from the behavior of a four point function in a compact unitary CFT.
That is, this is the exception of the factorization rule of the four point function as mentioned around (\ref{eq:introCb}).
 As explained in \cite{Ribault2014,Ribault2015}, the leading and the first sub-leading contribution of the Liouville four point function are given by
\begin{equation}\label{eq:LiouvilleNOC}
\begin{aligned}
&\braket{\ca{V}_{\a_B}(\infty)\ca{V}_{\a_B}(1)\ca{V}_{\a_A}(z)\ca{V}_{\a_A}(0)}\\
&\ar{\abs{z} \to 0} \left\{
    \begin{array}{ll}
 \text{Res}\pa{   -2\pi i 
	  {\bb C}_{\a} [\a_A,\a_B]
	  ; \a=2\a_A} 
	\abs{z^{h_{2\a_A}-2h_A}  }^2
 	,& \text{if } \a_A<\fr{Q}{4},\a_B   ,\\
\text{Res}\pa{     -2\pi i 
 	 {\bb C}_{\a} [\a_B,\a_A]
 	 ; \a=2\a_B} 
	 \abs{z^{h_{2\a_B}-2h_A}}^2 
  	,& \text{if } \a_B<\fr{Q}{4},\a_A   ,\\
\fr{\s{\pi}}{8} \del_{\a}^2 \left. {\bb C}_{\a}[\a_A,\a_B] \right|_{\a=\fr{Q}{2}}
\abs{ z^{\fr{Q^2}{4}-2h_A}\pa{\log z}^{-\fr{3}{2}}}^2   ,& \text{if } \a_A,\a_B>\fr{Q}{4}  ,\\
    \end{array}
  \right.\\
\end{aligned}
\end{equation}
where we defined $ {\bb C}_{\a} [\a_A,\a_B]\equiv C^{\text{DOZZ}}(\a_A,\a_A,\a) C^{\text{DOZZ}}(Q-\a,\a_B,\a_B)$.
We find that the asymptotic behavior drastically changes as the Liouville momentum $\a_A, \a_B$ cross the line $\a_A,\a_B=\fr{Q}{4}$.
This is because of the appearance of the discrete terms in the same way as the transition of the light cone and the Regge singularities.
Note that the appearance of the logarithmic correction $\pa{\log z}^{-\fr{3}{2}}$ has a physical interpretation in random energy models  \cite{Cao2017}.
It is also interesting to comment that this power law appears in the late time OTOC of the Sachdev-Ye-Kitaev (SYK) model \cite{Bagrets2016, Bagrets2017}. 
In more precisely, the SYK OTOC has the crossover from the Lyapunov exponential decay to the polynomial decay in the late time regime. The polynomial decay is also related to the Liouville quantum mechanics as explained in \cite{Bagrets2016, Bagrets2017} (see also \cite{Mertens2017, Lam2018}, which discuss the OTOC in the Schwarzian theory (as the low energy limit of the SYK model) by making use of the connection to the Liouville CFT).

From now, we will consider the Regge limit of the Liouville four point function. Since the Regge limit of the conformal block (\ref{eq:Regge2}) is independent of the internal dimension, the dominant contribution of the Liouville four point function (\ref{eq:Liouvillecorr}) in the Regge limit is given by the lowest intermediate dimension as
\footnote{Note that the monodromy matrix $ \text{Res}\pa{ -2\pi i {\bold M}_{\b, \a} [\a_A,\a_B] ; 2\a_A} $ does not diverge or vanish at $\b=2\a_A$ and $\b=\fr{Q}{2}$, which justifies the Regge limit (\ref{eq:LiouvilleRegge}). }
\begin{equation}\label{eq:LiouvilleRegge}
\begin{aligned}
&\braket{\ca{V}_{\a_B}(\infty)\ca{V}_{\a_B}(1)\ca{V}_{\a_A}(z)\ca{V}_{\a_A}(0)}\\
&\ar{\text{Regge limit}} \left\{
    \begin{array}{ll}
 \text{Res}\pa{   -2\pi i 
	  {\bb C}_{\a} [\a_A,\a_B]
	  ; \a=2\a_A} 
	   \text{Res}\pa{    -2\pi i
 	 {\bold M}_{2\a_A, \a} [\a_A,\a_B]
  	; \a=2\a_A} 
	\abs{z^{h_{2\a_A}-2h_A}  }^2
 	,& \text{if } \a_A<\fr{Q}{4},\a_B   ,\\
\text{Res}\pa{     -2\pi i 
 	 {\bb C}_{\a} [\a_B,\a_A]
 	 ; \a=2\a_B} 
 	  \text{Res}\pa{   -2\pi i 
 	 {\bold M}_{2\a_B, \a} [\a_A,\a_B]
  	; \a=2\a_B} 
	\abs{z^{h_{2\a_B}-2h_A}}^2 
  	,& \text{if } \a_B<\fr{Q}{4},\a_A   ,\\
\fr{\s{\pi}}{8} \del_{\a}^2 \left. {\bb C}_{\a}[\a_A,\a_B] \right|_{\a=\fr{Q}{2}}
  \fr{\s{\pi}}{8} \del_{\a}^2 \left. {\bold M}_{\fr{Q}{2},\a}[\a_A,\a_B] \right|_{\a=\fr{Q}{2}} 
\abs{ z^{\fr{Q^2}{4}-2h_A}\pa{\log z}^{-\fr{3}{2}}}^2   ,& \text{if } \a_A,\a_B>\fr{Q}{4} .\\
    \end{array}
  \right.\\
\end{aligned}
\end{equation}

As a result, we can conclude that the Liouville OTOC behaves as
\begin{equation}\label{eq:LiouvilleOTOC}
\begin{aligned}
C_\b(x,t) &\ar{\text{late time}} \left\{
    \begin{array}{ll}
      \text{Res}\pa{    -2\pi i {\bold M}_{2\a_A, \a} [\a_A,\a_B] ; \a=2\a_A}  ,& \text{if } \a_A<\fr{Q}{4},\a_B     ,\\
       \text{Res}\pa{   -2\pi i {\bold M}_{2\a_B, \a} [\a_A,\a_B]; \a=2\a_B}   ,& \text{if } \a_B<\fr{Q}{4},\a_A    ,\\
    \fr{\s{\pi}}{8} \del_{\a}^2 \left. {\bold M}_{\fr{Q}{2},\a}[\a_A,\a_B] \right|_{\a=\fr{Q}{2}}    ,&\text{if } \a_A, \a_B>\fr{Q}{4}   ,\\
    \end{array}
  \right.\\
\end{aligned}
\end{equation}
where we have to divide (\ref{eq:LiouvilleRegge}) not by the factorization $\braket{\ca{V}_{\a_B}\ca{V}_{\a_B}} \braket{\ca{V}_{\a_A}\ca{V}_{\a_A}}$ but by (\ref{eq:LiouvilleNOC}), according to the definition (\ref{eq:introCb}).
This particularly shows that the Liouville OTOC approaches constant in the late time limit, which is a common feature of the OTOC in RCFTs. In this sense, we expect that the Liouville CFT is not chaotic even if it is classified as irrational CFTs.

There was a naive expectation \cite{Caputa2016} that the Liouville OTOC would approach the constant associated to the $(0,0)$ element of the monodromy matrix (\ref{eq:RCFTOTOC}). However, as seen in (\ref{eq:LiouvilleOTOC}), the correct formula of the Liouville OTOC is given by the more complicated element of the monodromy matrix. Moreover, the Liouville OTOC has a richer structure with three phases in a similar way as the late time OTOC in the pure CFT.

Unfortunately due to a lack of the knowledge about the block with the non-zero vacuum intermediate state $h_p \sim c$, we do not have any information to determine the Liouville Lyapunov exponent in the scrambling time. Nevertheless, we believe that further investigations of our monodromy matrix  approach would reveal the Liouville OTOC beyond the late time regime. We leave it as a future work.

\subsection{Comments on Scrambling Regime}\label{subsec:Scrambling}

We are also interested in the {\it scrambling time} when the OTOC shows the Lyapunov decay as
\begin{equation}\label{eq:Lyapunov}
C_\b(x,t) \sim 1- \fr{\#}{c} \ex{\lambda_L(t-t_*)}+\cdots,
\end{equation}
where $\lambda_L$ is the so-called Lyapunov exponent. It is shown that the exponent has an upper bound as
\begin{equation}
\l_L \leq \fr{2\pi}{\b},
\end{equation}
and this bound is saturated by the holographic CFTs \cite{Maldacena2016}. This saturation can be seen from the HHLL Virasoro block \cite{Roberts2015},
\begin{equation}
\begin{aligned}
\ca{F}^{LL}_{HH}(0|z)&=(1-z)^{h_L(\g-1)}\pa{\fr{1-(1-z)^\g}{\g}}^{-2h_L} \\
&\ar{\text{Reege limit}}	
z^{-2h_L}\pa{\fr{1}{1-\fr{24h_W \pi i}{c z}}}^{2h_L}.
\end{aligned}
\end{equation}
where we assume $h_H \ll c$. Inserting this conformal block into (\ref{eq:Cbeta}), we obtain the OTOC as
\begin{equation}
C_\b (x,t)\ar{h_W \ll h_V \ll c}\pa{\fr{1}{1-\fr{24h_W \pi i}{c z}}}^{2h_W}
=\pa{\fr{1}{1+C\ex{\fr{2\pi}{\b}\pa{t-x-t_*}}}}^{2h_W},
\end{equation}
where $C=\fr{24\pi i h_W}{\e_{12} \e_{34}^*}$ and $t_*=\fr{\b}{2\pi}\log c$.  At the scrambling time $t-x \sim t_*$, this OTOC behaves as (\ref{eq:Lyapunov}) with the Lyapunov exponent $\l_L=\fr{2\pi}{\b}$.

However, as pointed out in \cite{Chang2018}, the vacuum block approximation is not justified at the scrambling time because the non-vacuum contributions may be larger than the vacuum contribution in the scrambling time regime. This might suggest that we have to impose some other assumptions on the CFT data of the holographic CFTs. To go further in this direction, we need the Regge asymptotics of the Virasoro blocks with general internal dimensions. In \cite{Chang2018}, it is discussed there by using the HHLL Virasoro block, the semiclassical block (with $h_p\gg c$) and the Zamolodchikov recursion relation. Here we would like to propose the other approach to address the issue by using the {\it monodromy matrix approach}.

We have the exact form of the non-vacuum contribution,
\begin{equation}
C_\b(h_p|x,t)\equiv\abs{z^{2h_W}}^2
\pa{\int_{\bb{S}} \dd \a \ {\bold M}_{\a_p, \a}[\a_W, \a_V] \ca{F}^{{\dg W}W}_{V{\dg V}}(h_\a|z) }
\overline{\ca{F}^{{\dg W}W}_{V{\dg V}}}(\bar{h}_p|\bar{z}),
\end{equation}
and the monodromy matrix is simplified in the large $c$ limit, therefore, we expect that the monodromy matrix approach is very useful to investigate the scrambling time of the non-vacuum contributions both in an analytic and a numerical way (because we can use the simple asymptotics of the block $\ca{F}^{{\dg W}W}_{V{\dg V}}(h_\a|z) \sim z^{h_{\a}-2h_W}$ in this expression).
In more details, the scrambling time for $C_\b(h_p|x,t)$ could be evaluated in the following way,
\begin{equation}
\begin{aligned}
&\abs{z^{2h_W}}^2  C_{\dg{W} W p} C_{\dg{V}V p} 
\int \dd \a \ {\bold M}_{\a_p, \a}[\a_W, \a_V] \ca{F}^{{\dg W}W}_{V{\dg V}}(h_\a|z)\overline{\ca{F}^{{\dg W}W}_{V{\dg V}}}(\bar{h}_p|\bar{z})\\
&\ar{\substack{\text{saddle point}\\ \text{approximation}}}
 C_{\dg{W} W p} C_{\dg{V}V p} {\bold M}_{\a_p, \a_{\text{saddle}}}[\a_W, \a_V] \ex{-\pa{h_{\a_{\text{saddle}}}+\bar{h}_p}\fr{2\pi t}{\b} }\\
&\ar{c \gg 1}
\ex{D-\pa{h_{\a_{\text{saddle}}}+\bar{h}_p}\fr{2\pi t}{\b} },
\end{aligned}
\end{equation}
where $\ex{D} \sim C_{\dg{W} W p} C_{\dg{V}V p} {\bold M}_{\a_p, \a_{\text{saddle}}}[\a_W, \a_V] $. Therefore, the scrambling time could be determined by ${t_*}_p=\fr{\b}{2\pi} \fr{D}{h_{\a_{\text{saddle}}}+\bar{h}_p}$.
We shall discuss this issue in a future work.

\section{Discussion}

In this paper, we studied the Renyi entropy and the OTOC from the light cone singularity and the Regge singularity.
The key point is that the conformal block can be simplified in the light cone and the Regge limit by means of the fusion and the monodromy matrices. We believe that this idea utilizing the fusion and the monodromy matrices will have many applications in further investigations of the AdS${}_3$/CFT${}_2$. Apart from the holography, we have a wealth of interesting questions
in 2D CFT that can now be attacked.

We will propose some remaining questions and interesting future works at the end of this paper:
\begin{description}
\item[Renyi Entropy]\
\\
We would like to understand the physical meaning of the replica transition. It is expected that this transition can be interpreted as the instability of the bulk. A similar transition of the Renyi entanglement entropy is found in \cite{Metlitski2009,Belin2013,Belin2015,Belin2017,Dong2018}, which might be related to our replica transition. It would be interesting to explore this issue.

It would be also interested to consider the Renyi entropy after a bilocal quench. In the CFT side, there are already several works about the Renyi entropy for a multiple locally excited state \cite{Nozaki2014a, Numasawa2016, Guo2018}, which argues that the Renyi entropy after a multiple local quench just obeys the sum rule. On the other hand, it does not seem that in the gravity side,  the entanglement entropy for a bilocal quench is given by such a simple rule \cite{Ageev2016, Arefeva2017}. 
In our upcoming publication, we will present the results for the bilocal quench from the viewpoint of the fusion matrix approach.

\item[OTOC]\
\\
We found that the late time OTOC had three phases as in (\ref{eq:OTOCresult}). This predicts that we would find the relevant transition in the bulk side as the mass of particle is increased. Therefore, it would be interesting to investigate the late time OTOC from the bulk calculation and provide the bulk interpretation of the phase transition.

An important future work is to generalize our works to the scrambling time regime more clearly. We expect that a further investigation of the monodromy matrix would make clear the OTOC at the scrambling time. It is also interesting to reveal the validity of the vacuum block approximation \cite{Liu2018, Hampapura2018, Chang2018} more explicitly as in Section \ref{subsec:nonOTOC}. We would like to carry out such investigations in the future.

We also studied the OTOC in the Liouville CFT and discovered that the Liouville OTOC approaches constant in the late time as in RCFTs, however, its constant was not given by the $(0,0)$ element of the monodromy matrix.
This constant behavior may imply that the Liouvile CFT is not chaotic in a sense. It would be interesting to study the Liouville OTOC beyond the late time limit.

\item[Virasoro Block]\
\\
We had studied the light cone singularity of the Virasoro block for the motivation of applying the conformal bootstrap and understanding the AdS/CFT correspondence in \cite{Kusuki2018a, Kusuki2018c, Kusuki2018} and succeeded in solving the bootstrap equation in \cite{Kusuki2018c, Collier2018}. In this article, we provided the Regge singularity of the Virasoro block to study the Renyi entropy and the OTOC.
Hopefully, the monodromy matrix would also make a new analytic approach to solve the conformal bootstrap in a certain limit, as suggested in \cite{Chen2017}.

Another question is how to interpret the transition of the light cone limit and Regge limit singularities in the bulk side. Further, we would like to understand our results more directly from the relation between the Virasoro block and the geodesics \cite{Fitzpatrick2017,Fitzpatrick2015}.
We will provide more clear understanding of the relation between the Virasoro block and the bulk objects in terms of the fusion matrix in the future

\end{description}

\section*{Acknowledgments}
We thank Chi-Ming Chang,  Mukund Rangamani, Pawel Caputa, Song He, Ryo Sato, and Tadashi Takayanagi for fruitful discussions and comments.
YK and MM are supported by the JSPS fellowship. 
In this research work, we used the supercomputer of ACCMS, Kyoto University.

\appendix
\section{Explicit Form of Fusion Matrix} \label{app:FM}

In the following, we introduce the notations usually found in Liouville CFTs.
\begin{equation}
c=1+6Q^2, \ \ \ \ \ Q=b+\fr{1}{b}, \ \ \ \ \ h_i=\a_i(Q-\a_i).
\end{equation}
Note that we can relate the parameter $\eta_i$ appearing in \cite{Kusuki2018a} to $\a_i$ as $\a_i=Q\eta_i$.

The fusion matrix is defined by the  invertible fusion transformations between $s$ and $t$- channel conformal blocks \cite{Teschner2001a} as follows:
\begin{equation}\label{eq:fusiontrans}
\begin{aligned}
\ca{F}^{21}_{34}(h_{\a_s}|z)=\int_{\bb{S}} \dd \a_t {\bold F}_{\a_s, \a_t} 
   \left[
    \begin{array}{cc}
    \a_2   & \a_1  \\
     \a_3  &   \a_4\\
    \end{array}
  \right]
  \ca{F}^{23}_{14}(h_{\a_t}|1-z),
\end{aligned}
\end{equation}
where the contour $\bb{S}$ runs from $\fr{Q}{2}$ to $\fr{Q}{2}+ i\infty$, and also runs anti-clockwise around $\a_t=\a_1+\a_4+mb+nb^{-1}<\fr{Q}{2}$ and $\a_t=\a_2+\a_3+mb+nb^{-1}<\fr{Q}{2}$ for $m, n\in \mathbb{Z}_{\geq 0}$. The kernel $ {\bold F}_{\a_s, \a_t} $ is called the {\it crossing matrix} or {\it fusion matrix}. The explicit form of the fusion matrix is given in \cite{Ponsot1999,Teschner2001a}as follows:
\begin{equation}\label{eq:crossing}
\begin{aligned}
{\bold F}_{\a_s, \a_t} 
   \left[
    \begin{array}{cc}
    \a_2   & \a_1  \\
     \a_3  &   \a_4\\
    \end{array}
  \right]
=\fr{N(\a_4,\a_3,\a_s)N(\a_s,\a_2,\a_1)}{N(\a_4,\a_t,\a_1)N(\a_t,\a_3,\a_2)}
   \left\{
    \begin{array}{cc|c}
    \a_1   & \a_2    &  \a_s  \\
     \a_3  & \a_4    &  \a_t   \\
    \end{array}
  \right\}_b,
\end{aligned}
\end{equation}
where the function $N(\a_3,\a_2,\a_1)$ is
\begin{equation}
N(\a_3,\a_2,\a_1)=\fr{\G_b(2\a_1)\G_b(2\a_2)\G_b(2Q-2\a_3)}{\G_b(2Q-\a_1-\a_2-\a_3)\G_b(Q-\a_1-\a_2+\a_3)\G_b(\a_1+\a_3-\a_2)\G_b(\a_2+\a_3-\a_1)},
\end{equation}
and 
$ \left\{
    \begin{array}{cc|c}
    \a_1   & \a_2    &  \a_s  \\
     \a_3  & \a_4    &  \a_t   \\
    \end{array}
  \right\}_b$
is the Racah--Wigner coefficient for the quantum group $U_q(sl(2,\bb{R}))$, which is given by
\footnote{Ponsot--Teschner have derived a more symmetric form of the Racah--Wigner coefficient \cite{Teschner2014} than the traditional expression found in \cite{Ponsot1999,Teschner2001a}. In this study, we used the new expression derived in \cite{Teschner2014}.}
\begin{equation}\label{eq:6j}
\begin{aligned}
&\left\{
    \begin{array}{cc|c}
    \a_1   & \a_2    &  \a_s  \\
     \a_3  & \bar{\a_4}    &  \a_t   \\
    \end{array}
  \right\}_b\\
&= \fr{S_b(\a_1+\a_4+\a_t-Q)S_b(\a_2+\a_3+\a_t-Q)S_b(\a_3-\a_2-\a_t+Q)S_b(\a_2-\a_3-\a_t+Q)}{S_b(\a_1+\a_2-\a_s)S_b(\a_3+\a_s-\a_4)S_b(\a_3+\a_4-\a_s)}\\
&\times \abs{S_b(2\a_t)}^2 \int^{2Q+i \infty}_{2Q-i \infty} \dd u 
\fr{S_b(u-\a_{12s})S_b(u-\a_{s34})S_b(u-\a_{23t})S_b(u-\a_{1t4})}{S_b(u-\a_{1234}+Q)S_b(u-\a_{st13}+Q)S_b(u-\a_{st24}+Q)S_b(u+Q)},
\end{aligned}
\end{equation}
where we have used the notations $\bar{\a}=Q-\a$, $\a_{ijk}=\a_i+\a_j+\a_k$ and $\a_{ijkl}=\a_i+\a_j+\a_k+\a_l$.
The functions $\G_b(x)$ and $S_b(x)$ are defined as
\begin{equation}
\G_b(x)= \fr{\G_2(x|b,b^{-1})}{\G_2\pa{\fr{Q}{2}|b,b^{-1}}}, \ \ \ \ \ S_b(x)=\fr{\G_b(x)}{\G_b(Q-x)},
\end{equation}
$\G_2(x|\w_1,\w_2)$ is the double gamma function,
\begin{equation}
\log \G_2(x|\w_1,\w_2)=\pa{\pd{t}\sum^{\infty}_{n_1,n_2=0} \pa{x+n_1 \w_1+n_2\w_2}^{-t}}_{t=0}.
\end{equation}
Note that the function $\G_b(x)$ is introduced such that $\G_b(x)=\G_{b^{-1}}(x)$ and satisfies the following relationship:
\begin{equation}
\G_b(x+b)=\fr{\s{2 \pi}b^{bx-\fr{1}{2}}}{\G(bx)}\G_b(x).
\end{equation}
By substituting the explicit form of the Racah--Wigner coefficients (\ref{eq:6j}) into (\ref{eq:crossing}), we can simplify the expression for the fusion matrix into
\begin{equation}\label{eq:crossing2}
\begin{aligned}
&{\bold F}_{\a_s, \a_t} 
   \left[
    \begin{array}{cc}
    \a_2   & \a_1  \\
     \a_3  &   \a_4\\
    \end{array}
  \right]\\
&=\fr{\G_b(Q+\a_2-\a_3-\a_t)\G_b(Q-\a_2+\a_3-\a_t)\G_b(2Q-\a_1-\a_4-\a_t)\G_b(\a_1+\a_4-\a_t)}{\G_b(2Q-\a_1-\a_2-\a_s)\G_b(\a_1+\a_2-\a_s)\G_b(Q+\a_3-\a_4-\a_s)\G_b(Q-\a_3+\a_4-\a_s)}\\
&\times \fr{\G_b(Q-\a_2-\a_3+\a_t)\G(-Q+\a_2+\a_3+\a_t)\G_b(\a_1-\a_4+\a_t)\G_b(-\a_1+\a_4+\a_t)}{\G_b(\a_1-\a_2+\a_s)\G_b(-\a_1+\a_2+\a_s)\G_b(Q-\a_3-\a_4+\a_s)\G_b(-Q+\a_3+\a_4+\a_s)}\\
&\times \abs{S_b(2\a_t)}^2 \fr{\G_b(2Q-2\a_s)\G_b(2\a_s)}{\G_b(2Q-2\a_t)\G_b(2\a_t)} \\
&\times \int^{2Q+i \infty}_{2Q-i \infty} \dd u 
\fr{S_b(u-\a_{12s})S_b(u-\a_{s34})S_b(u-\a_{23t})S_b(u-\a_{1t4})}{S_b(u-\a_{1234}+Q)S_b(u-\a_{st13}+Q)S_b(u-\a_{st24}+Q)S_b(u+Q)}.
\end{aligned}
\end{equation}

\section{Singularity of $n$-point Block} \label{app:n-point}

In this Appendix, we will give a detailed explanation (for beginners) of how to calculate the singularity of this $n$-point conformal partial wave in terms of 3-point blocks.
To express the $n$-point conformal partial wave explicitly, we will introduce the following notations;

\hrulefill
\begin{description}

\item[Primary field ($\nu, \bar{\nu}$):]
\begin{equation}
\lim_{z,\bar{z} \to 0} O_i (z,\bar{z}) \ket{0} = \ket{\nu_i \otimes \bar{\nu}_i}.
\end{equation}
By using this notation, we will denote descendant states in the Verma module $\ca{V}_{\nu_i}$ as
\begin{equation}
\ket{\nu_{i,M}}=L_{-M}\ket{\nu_i}\equiv L_{m_j}\dots L_{-m_1}\ket{\nu_i},
\end{equation}
where $M$ is arbitrary ordered set of indices $M=\{m_1,m_2,\dots,m_j \in \bb{N} |m_1\geq m_2 \geq \dots \geq m_j   \text{ and } |M|=m_1+m_2+\cdots+m_j \}$. We will denote the operator corresponding to $\ket{\nu_{i,M}}$ as $O_{\nu_{i,M}}$ via state-operator mapping.

\item[Three point conformal block $\rho(\xi_1;\xi_2;\xi_3|z)$:]
\begin{equation}
\braket{\xi_1, \bar{\xi}_1| O_2(z,\bar{z})| \xi_3,\bar{\xi}_3 }=\rho(\xi_1;\xi_2;\xi_3|z)\rho(\bar{\xi}_1;\bar{\xi}_2;\bar{\xi}_3|\bar{z})C_{123},
\end{equation}
where $\xi_i$ is any state in the Verma module $\ca{V}_{\nu_i}$. For $L_0$-eigenstates ($L_0\xi_i=\Delta(\xi_i)\xi_i$), 
the $z$ dependence of the 3-point conformal block is fixed by the Ward identity as
\begin{equation}\label{eq:3blockz}
\rho(\xi_1;\xi_2;\xi_3|z)=z^{\D(\xi_1)-\D(\xi_2)-\D(\xi_3)}\rho(\xi_1;\xi_2;\xi_3|1).
\end{equation}
By definition, any 3-point conformal block for primary states $\nu_i$ is  given by
\begin{equation}
\rho(\nu_1;\nu_2;\nu_3|1)=1.
\end{equation}
We also define 
\begin{equation}
\braket{\xi_1, \bar{\xi}_1| O_2(z_2,\bar{z}_2) O_3(z_3,\bar{z}_3) }=\rho(\xi_1;\xi_2,\xi_3|z_2,z_3)\rho(\bar{\xi}_1;\bar{\xi}_2,\bar{\xi}_3|\bar{z}_2,\bar{z}_3)C_{123},
\end{equation}
\begin{equation}
\braket{O_2(z_1,\bar{z_1}) O_2(z_2,\bar{z_2}) O_3(z_3,\bar{z_3})}=\rho(\xi_1,\xi_2,\xi_3|z_1,z_2,z_3)\rho(\bar{\xi}_1,\bar{\xi}_2,\bar{\xi}_3|\bar{z}_1,\bar{z}_2,\bar{z}_3)C_{123}.
\end{equation}

\item[Gram matrix $G^{(n)}_{MN}:$]
\begin{equation}
\br{G^{(n)}_i}_{MN}=\braket{\nu_{i,M}|\nu_{i,N} },
\end{equation}
where  $\abs{M}=\abs{N}=n $.
\end{description}

\hrulefill \\
Following the above notations, the 4-point conformal partial wave is given by
\footnote{For the purpose of generalizing the results in Section \ref{sec:LC} to $n$-point conformal blocks,  we used a different notation than in (\ref{eq:blockdef}). The relation is given by $\ca{F}^{h_1,h_2,h_3, h_4}_{h_{p}}(z)=\ca{F}^{21}_{34}(h_p|z) $. }
\begin{equation}
\begin{aligned}
\ca{F}^{h_1,h_2,h_3, h_4}_{h_{p}}(z_1,z_2, z_3, z_4)=\sum_{\substack{l=\abs{M}=\abs{N}  }}  \rho(\nu_4, \nu_3, \nu_{p,M}|z_4,z_3,0)
\br{G_{p}^{(l)}}_{M, N}^{-1} \rho(\nu_{p,N}; \nu_{2}, \nu_1|z_2,z_1).
\end{aligned}
\end{equation}
The fusion transformation of the 4-point conformal partial wave can be expressed as
\begin{equation}
\begin{aligned}
&\ca{F}^{h_1,h_2,h_3, h_4}_{h_{p}}(z_1,z_2, z_3, z_4)\\&=\int_{\bb{S}}\dd \a_{q} \ {\bold F}_{\a_{p}, \a_{q}}
   \left[
    \begin{array}{cc}
     \a_{2}  &   \a_{1}\\
    \a_{3}   & \a_{4}  \\
    \end{array}
  \right]
 \sum_{\substack{l=\abs{M}=\abs{N} }}
    \rho(\nu_4, \nu_1,\nu_{q,M}|z_4,z_1,0)\br{G_q^{(l)}}_{M, N}^{-1} \rho(\nu_{q,N}; \nu_2, \nu_3|z_2,z_3),
\end{aligned}
 \end{equation}
where the contour $\bb{S}$ runs from $\fr{Q}{2}$ to $\fr{Q}{2}+ i\infty$, and also runs anti-clockwise around $\a_q=\a_1+\a_4+mb+nb^{-1}<\fr{Q}{2}$ and $\a=\a_2+\a_3+mb+nb^{-1}<\fr{Q}{2}$, in that, this contour is the shorthand for (\ref{eq:bootstrapAABB2}) (and its generalization). From now on, the notation $\a_i$ stands for $h_i \equiv \a_i(Q-\a_i)$ and $\a_{p_i}$ stands for $h_{p_i} \equiv \a_{p_i}(Q-\a_{p_i})$. We introduce another expression of the fusion transformation \cite{Moore1989} here for later use,
\begin{equation}\label{eq:part wave}
\begin{aligned}
&\ca{F}^{h_1,h_2,h_3, h_4}_{h_{p}}(z_1,z_2, z_3, z_4)\\&=\int_{\bb{S}}\dd \a_{q} \ {\bold F}_{\a_{p}, \a_{q}}
   \left[
    \begin{array}{cc}
     \a_{2}  &   \a_{1}\\
    \a_{3}   & \a_{4}  \\
    \end{array}
  \right]
 \sum_{\substack{l=\abs{M}=\abs{N} }}
    \rho(\nu_4,  \nu_{q,M}, \nu_1 |z_{4},z_{2},z_1)\br{G_q^{(l)}}_{M, N}^{-1} \rho(\nu_{q,N}; \nu_3; \nu_2|z_{32}),
\end{aligned}
 \end{equation}
where $z_{ij}\equiv z_i-z_j$.
By the $z$-dependence of the 3-point blocks (\ref{eq:3blockz}), we find that in the limit $z_{32} \to 0$, the dominant contribution to the sum comes from the term $l=0$. Consequently, we obtain the following singularity of the 4-point partial wave,
\begin{equation}\label{eq:partial sing}
\begin{aligned}
&\ca{F}^{h_1,h_2,h_3, h_4}_{h_{p}}(z_1,z_2, z_3, z_4)\\&\ar{z_{32} \to 0}\int_{\bb{S}}\dd \a_{q} \ {\bold F}_{\a_{p}, \a_{q}}
   \left[
    \begin{array}{cc}
     \a_{2}  &   \a_{1}\\
    \a_{3}   & \a_{4}  \\
    \end{array}
  \right]
    \rho(\nu_4, \nu_q, \nu_1  |z_4,z_2,z_1) \times (z_3-z_2)^{h_q-h_2-h_3}.
\end{aligned}
 \end{equation}
From the same discussion as that in Section \ref{sec:LC}, we can identify the light cone singularity of the conformal partial wave (in that, the conformal block with four parameters $z_{i=1,2,3,4}$).

\begin{figure}[t]
 \begin{center}
  \includegraphics[width=16.0cm,clip]{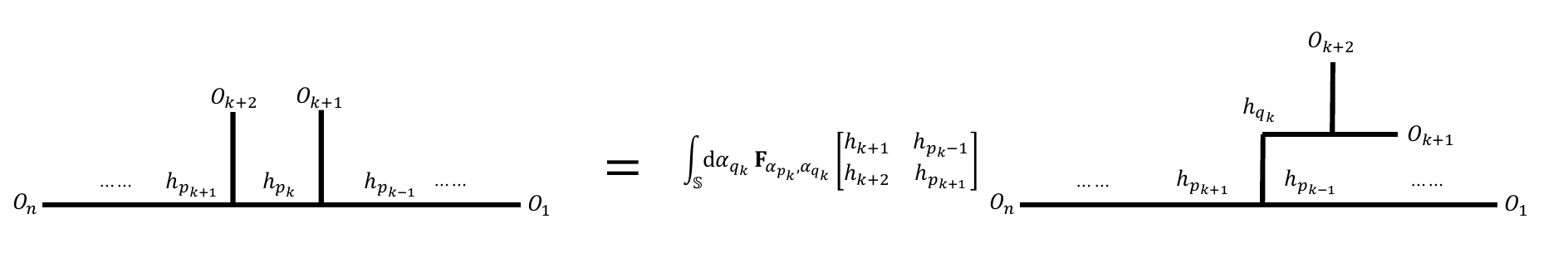}
 \end{center}
 \caption{The fusion transformation of the $n$-point conformal partial wave.}
 \label{fig:fusion}
\end{figure}

We will give the generalization of this light cone singularity (\ref{eq:partial sing}) to any $n$-point. The $n$-point conformal partial wave can be expressed in terms of the 3-point blocks as
\begin{equation}\label{eq:pants}
\begin{aligned}
&\ca{F}^{h_1,h_2,\dots, h_n}_{h_{p_1}, h_{p_2},\dots, h_{p_{n-3}}}(z_1,z_2, \dots, z_{n})\\
&=\sum_{\substack{l_i=\abs{M_i}=\abs{N_i} \\i=1,2, \dots,n-3 }}  
	\rho(\nu_n, \nu_{n-1}, \nu_{p_{n-3},M_{n-3}}|z_n,z_{n-1},0)
	 \br{G_{p_{n-3}}^{(l_{n-3})}}_{M_{n-3}, N_{n-3}}^{-1} \\
	&\times\pa{ \prod^{n-4}_{k=1}  \rho(\nu_{p_{k+1},N_{k+1}}; \nu_{k+2}; \nu_{p_{k},M_{k}}|z_{k+2})  \br{G_{p_k}^{(l_k)}}_{M_k, N_k}^{-1}  }
	\rho( \nu_{p_{1},N_{1}}; \nu_{2}, \nu_1|z_{2},z_{1}).
\end{aligned}
\end{equation}
In order to apply the fusion transformation to the $n$-point conformal partial wave, we focus on the following part of the $n$-point conformal partial wave, which can be thought of as the $4$-point conformal partial wave,
\begin{equation}
\sum_{\substack{l_{k}=\abs{M_{k}}=\abs{N_{k}} }}\rho(\nu_{p_{k+1},N_{k+1}}; \nu_{k+2}; \nu_{p_{k},M_{k}}|z_{k+2})\br{G_{p_{k}}^{(l_{k})}}_{M_{k}, N_{k}}^{-1} \rho(\nu_{p_{k},N_{k}}; \nu_{k+1}; \nu_{p_{k-1},M_{k-1}}|z_{k+1}).
 \end{equation}
We consider the fusion transformation of this part in the same way as (\ref{eq:part wave}), which can be depicted graphically in Figure \ref{fig:fusion}.
Consequently, we obtain
\footnote{Here, two arguments of the each 3-point block are given by descendants (not primaries), which is slightly different from  (\ref{eq:part wave}). Nevertheless, the fusion transformation acts on them in a similar manner. Actually one can find that this is trivial from the definition of the fusion transformation in the context of the vertex operator algebra \cite{Moore1989}.}

\begin{equation}
\begin{aligned}
&\sum_{\substack{l_{k}=\abs{M_{k}}=\abs{N_{k}} }}\rho(\nu_{p_{k+1},N_{k+1}}; \nu_{k+2}; \nu_{p_{k},M_{k}}|z_{k+2})\br{G_{p_{k}}^{(l_{k})}}_{M_{k}, N_{k}}^{-1} \rho(\nu_{p_{k},N_{k}}; \nu_{k+1}; \nu_{p_{k-1},M_{k-1}}|z_{k+1})\\
&=\int_{\bb{S}}\dd \a_{q_{k}} \ {\bold F}_{\a_{p_{k}}, \a_{q_{k}}}
   \left[
    \begin{array}{cc}
     \a_{k+1}  &   \a_{p_{k-1}}\\
    \a_{k+2}   & \a_{p_{k+1}}  \\
    \end{array}
  \right]\\
& \times\sum_{\substack{l_{k}=\abs{M_{k}}=\abs{N_{k}} }}\rho(\nu_{p_{k+1},N_{k+1}}; \nu_{q_{k},M_{k}}; \nu_{p_{k-1},M_{k-1}} |z_{k+1})\br{G_{q_{k}}^{(l_{k})}}_{M_{k}, N_{k}}^{-1} \rho(\nu_{q_{k},N_{k}}; \nu_{k+2};\nu_{k+1}|z_{k+2,k+1})\\
&\ar{z_{k+2,k+1}\to 0 }
\int_{\bb{S}}\dd \a_{q_{k}} \ {\bold F}_{\a_{p_{k}}, \a_{q_{k}}}
   \left[
    \begin{array}{cc}
     \a_{k+1}  &   \a_{p_{k-1}}\\
    \a_{k+2}   & \a_{p_{k+1}}  \\
    \end{array}
  \right] \rho(\nu_{p_{k+1},N_{k+1}}; \nu_{q_{k}}; \nu_{p_{k-1},M_{k-1}} |z_{k+1}) (z_{k+2}-z_{k+1})^{h_{q_{k}}-h_{k+1}-h_{k+2}}.
\end{aligned}
 \end{equation}
In a similar manner, we have
\begin{equation}\label{eq:edge1}
\begin{aligned}
&\sum_{\substack{l_1=\abs{M_1}=\abs{N_1}}}  \rho(\nu_{p_{2},N_{2}}; \nu_3; \nu_{p_1,M_1}|z_3)
 \br{G_{p_1}^{(l_1)}}_{M_1, N_1}^{-1} \rho(\nu_{p_1,N_1}; \nu_{2}, \nu_1  |z_2,z_1 ) \\
&=\int_{\bb{S}}\dd \a_{q_{1}} \ {\bold F}_{ \a_{p_{1}}, \a_{q_{1}}}
   \left[
    \begin{array}{cc}
     \a_{2}  &   \a_{1}\\
    \a_{3}   & \a_{p_{2}}  \\
    \end{array}
  \right]     \\
  &\times\sum_{\substack{l_{1}=\abs{M_{1}}=\abs{N_{1}} }}
	\rho(\nu_{p_2,N_2} ; \nu_{p_1,M_1}, \nu_1 |z_{2},z_1)
	\br{G_{q_{1}}^{(l_{1})}}_{M_{1}, N_{1}}^{-1}
	\rho(\nu_{p_1,N_1}; \nu_3, \nu_2 | z_{32})\\
&\ar{z_{32}\to 0}
\int_{\bb{S}}\dd \a_{q_{1}} \ {\bold F}_{ \a_{p_{1}}, \a_{q_{1}}}
   \left[
    \begin{array}{cc}
     \a_{2}  &   \a_{1}\\
    \a_{3}   & \a_{p_{2}}  \\
    \end{array}
  \right]
	\rho(\nu_{p_2,N_2} ; \nu_{p_1}, \nu_1 |z_{2},z_1) (z_3-z_2)^{h_{p_1}-h_2-h_3}, 
  \end{aligned}
  \end{equation}
 and
  \begin{equation}\label{eq:edge2}
 \begin{aligned} 
 &\sum_{\substack{l_{n-3}=\abs{M_{n-3}}=\abs{N_{n-3}}}}
	\rho(\nu_n, \nu_{n-1}, \nu_{p_{n-3},M_{n-3}}|z_n,z_{n-1},0)
	\br{G_{p_{n-3}}^{(l_{n-3})}}_{M_{n-3}, N_{n-3}}^{-1} 
	\rho( \nu_{p_{n-3},N_{n-3}}; \nu_{n-2}, \nu_{p_{n-4},M_{n-4}}  |z_{n-2})
   \\&=\int_{\bb{S}}\dd \a_{q_{n-3}} \ {\bold F}_{\a_{p_{n-3}}, \a_{q_{n-3}} }
   \left[
    \begin{array}{cc}
     \a_{n-2}  &   \a_{p_{n-4}}\\
    \a_{n-1}   & \a_{n}  \\
    \end{array}
  \right]\\
 &\times \sum_{\substack{l_{n-3}=\abs{M_{n-3}}=\abs{N_{n-3}}}}
	\rho(\nu_n, \nu_{p_{n-3},M_{n-3}},  \nu_{p_{n-4},M_{n-4}}|z_{n} , z_{n-2},0)
	\br{G_{p_{n-3}}^{(l_{n-3})}}_{M_{n-3}, N_{n-3}}^{-1} 
	\rho( \nu_{p_{n-3},N_{n-3}}; \nu_{n-1}, \nu_{n-2}  |z_{n-1,n-2})\\
&\ar{z_{n-1,n-2} \to 0}
	\int_{\bb{S}}\dd \a_{q_{n-3}} \ {\bold F}_{\a_{p_{n-3}}, \a_{q_{n-3}} }
   \left[
    \begin{array}{cc}
     \a_{n-2}  &   \a_{p_{n-4}}\\
    \a_{n-1}   & \a_{n}  \\
    \end{array}
  \right]
	\rho(\nu_n, \nu_{p_{n-3}},  \nu_{p_{n-4},M_{n-4}}|z_{n} , z_{n-2},0)
	(z_{n-1}-z_{n-2})^{h_{p_{n-3}}-h_{n-2}-h_{n-1}}.
\end{aligned}   
\end{equation}
Substituting these asymptotic behaviors into (\ref{eq:pants}), we find 
\begin{equation}
\begin{aligned}
& \ca{F}^{h_1,h_2,\dots,h_{k},h_{k+1},h_{k+2},h_{k+3}\dots h_n}_{h_{p_1}, h_{p_2},\dots,h_{p_{k-1}}, h_{p_{k}}, h_{p_{k+1}}, \dots   , h_{p_{n-3}}}(z_1,z_2, \dots,z_k, z_{k+1}, z_{k+2}, z_{k+3}, \dots , z_{n})\\
&\ar{z_{k+2, k+1}\to 0}
\int_{\bb{S}}\dd \a_{q_{k}} \ {\bold F}_{\a_{p_{k}}, \a_{q_{k}}}
   \left[
    \begin{array}{cc}
     \a_{k+1}  &   \a_{p_{k-1}}\\
    \a_{k+2}   & \a_{p_{k+1}}  \\
    \end{array}
  \right]   
	(z_{k+2}-z_{k+1})^{h_{q_{k}}-h_{k+1}-h_{k+2}} \\
& \hspace{2cm} \times  
	\ca{F}^{h_1,h_2,\dots,h_{k},h_{q_k},h_{k+3}\dots h_n}_{h_{p_1}, h_{p_2},\dots,h_{p_{k-1}}, h_{p_{k+1}}, \dots   , h_{p_{n-3}}}(z_1,z_2, \dots,z_k, z_{k+1}, z_{k+3}, \dots , z_{n}).
\end{aligned}
\end{equation}
Moreover we have shown that the dominant contribution to the integral can be given by (\ref{eq:LC}) and (\ref{eq:intsingularity}).
As a result, we can conclude that the light cone singularity of the $n$-point conformal partial wave is give by
\begin{equation}\label{eq:n-point singularity}
\begin{aligned}
&\ca{F}^{h_1,h_2,\dots,h_{k},h_{k+1},h_{k+2},h_{k+3}\dots h_n}_{h_{p_1}, h_{p_2},\dots,h_{p_{k-1}}, h_{p_{k}}, h_{p_{k+1}}, \dots   , h_{p_{n-3}}}(z_1,z_2, \dots,z_k, z_{k+1}, z_{k+2}, z_{k+3}, \dots , z_{n})\\
&\ar{z_{k+2, k+1}\to 0}\left\{
    \begin{array}{ll}
\text{Res}\pa{   -2\pi i  \ 
	{\bold F}_{\a_{p_{k}}, \a_{q_{k}}} \left[
	    \begin{array}{cc}
	     \a_{k+1}  &   \a_{p_{k-1}}\\
	    \a_{k+2}   & \a_{p_{k+1}}  \\
	    \end{array}
	  \right] 
	  ; \a_{q_k}=\a_{p_{k-1},p_{k+1}}} 
	(z_{k+2}-z_{k+1})^{h_{p_{k-1},p_{k+1}}-h_{k+1}-h_{k+2}}  \\
	 \times  
	\ca{F}^{h_1,h_2,\dots,h_{k}, h_{p_{k-1},p_{k+1}}  ,h_{k+3}\dots h_n}_{h_{p_1}, h_{p_2},\dots,h_{p_{k-1}}, h_{p_{k+1}}, \dots   , h_{p_{n-3}}}(z_1,z_2, \dots,z_k, z_{k+1}, z_{k+3}, \dots , z_{n}), 
	\hspace{1cm} \text{if } \a_{p_{k-1},p_{k+1}}<\fr{Q}{2},\a_{k+1, k+2}   ,\\ \\ 
\text{Res}\pa{   -2\pi i  \ 
	{\bold F}_{\a_{p_{k}}, \a_{q_{k}}} \left[
	    \begin{array}{cc}
	     \a_{k+1}  &   \a_{p_{k-1}}\\
	    \a_{k+2}   & \a_{p_{k+1}}  \\
	    \end{array}
	  \right] 
	  ; \a_{q_k}=\a_{k+1,k+2}} 
	(z_{k+2}-z_{k+1})^{h_{k+1,k+2}-h_{k+1}-h_{k+2}}  \\
	 \times  
	\ca{F}^{h_1,h_2,\dots,h_{k}, h_{k+1,k+2}  ,h_{k+3}\dots h_n}_{h_{p_1}, h_{p_2},\dots,h_{p_{k-1}}, h_{p_{k+1}}, \dots   , h_{p_{n-3}}}(z_1,z_2, \dots,z_k, z_{k+1}, z_{k+3}, \dots , z_{n}), 
	\hspace{1cm} \text{if } \a_{k+1, k+2} <\fr{Q}{2}, \a_{p_{k-1},p_{k+1}}  ,\\ \\ 
-i\del_{\a_{q_k}}^2 \left. {\bold F}_{\a_{p_{k}}, \a_{q_{k}}} \left[
	    \begin{array}{cc}
	     \a_{k+1}  &   \a_{p_{k-1}}\\
	    \a_{k+2}   & \a_{p_{k+1}}  \\
	    \end{array}
	  \right]  \right|_{\a_{q_k}=\fr{Q}{2}}   \fr{\s{\pi}}{8} (z_{k+2}-z_{k+1})^{\fr{Q^2}{4}-h_{k+1}-h_{k+2}}\pa{-\log (z_{k+2}-z_{k+1})}^{-\fr{3}{2}} \\	 		\times  
	\ca{F}^{h_1,h_2,\dots,h_{k}, \fr{Q^2}{4}  ,h_{k+3}\dots h_n}_{h_{p_1}, h_{p_2},\dots,h_{p_{k-1}}, h_{p_{k+1}}, \dots   , h_{p_{n-3}}}(z_1,z_2, \dots,z_k, z_{k+1}, z_{k+3}, \dots , z_{n}), 
	\hspace{1cm} \text{if } \a_{k+1, k+2}, \a_{p_{k-1},p_{k+1}}  >\fr{Q}{2} ,\\ \\ 
    \end{array}
  \right.\\
&(k=2,3,\cdots, n-4)
\end{aligned}
\end{equation}
where $\a_{i,j} \equiv \a_i+\a_j$ and $h_{i,j} \equiv \a_{i,j}(Q-\a_{i,j})$.
The above procedure can be also implemented in evaluating the singularity in the limit $z_{32}\to 0$ and $z_{n-1,n-2} \to 0$ by using the asymptotic behaviors (\ref{eq:edge1}) and (\ref{eq:edge2}).
Note that the singularity in the limit $z_{21} \to 0$ can be obtained in terms of 3-point blocks as follows,
\begin{equation}
\begin{aligned}
&\sum_{\substack{l_1=\abs{M_1}=\abs{N_1}}}  
	\rho(\nu_{p_{2},N_{2}}; \nu_3; \nu_{p_1,M_1}|z_3) \br{G_{p_1}^{(l_1)}}_{M_1, N_1}^{-1}    \rho( \nu_{p_{1},N_{1}}; \nu_{2}, \nu_1|z_{2},z_{1})\\
&=\sum_{\substack{l_1=\abs{M_1}=\abs{N_1}}}  
	\rho(\nu_{p_{2},N_{2}}; \nu_3; \nu_{p_1,M_1}|z_3) \br{G_{p_1}^{(l_1)}}_{M_1, N_1}^{-1}    \rho( \nu_{p_{1},N_{1}}; \nu_{2}, \nu_1|z_{21})\\
& \ar{z_{21} \to 0}  \rho(\nu_{p_{2},N_{2}}; \nu_3, \nu_{p_1}|z_3, z_1) (z_{2}-z_1)^{h_{p_1}-h_2-h_1}.
\end{aligned}
\end{equation}
We can also obtain the singularity in the limit  $z_{n,n-1} \to 0$ in the same way.

By using the braiding matrix, we can also give the singularity in more general limits $z_{i,j} \to 0$.
The braiding matrix is defined as (\ref{eq:Bdef}), and it acts on a $n$-point partial wave as
\begin{equation}
\begin{aligned}
& \ca{F}^{h_1,h_2,\dots,h_{k},h_{k+1},h_{k+2},h_{k+3}\dots h_n}_{h_{p_1}, h_{p_2} ,\dots,h_{p_{k-1}}, h_{p_{k}}, h_{p_{k+1}} ,  \dots   , h_{p_{n-3}}}(z_1,z_2  , \dots,z_k, z_{k+1}, z_{k+2}, z_{k+3}, \dots , z_{n})\\
&=\int_{\bb{S}}\dd \a_{q_{k}} \ {\bold B}^{(\e)}_{\a_{p_{k}}, \a_{q_{k}}}
   \left[
    \begin{array}{cc}
     \a_{k+1}  &   \a_{p_{k-1}}\\
    \a_{k+2}   & \a_{p_{k+1}}  \\
    \end{array}
  \right]   
 \ca{F}^{h_1,h_2,\dots,h_{k},h_{k+2},h_{k+1},h_{k+3}\dots h_n}_{h_{p_1}, h_{p_2} ,\dots,h_{p_{k-1}}, h_{q_{k}}, h_{p_{k+1}} , \dots   , h_{p_{n-3}}}(z_1,z_2, \dots,z_k, z_{k+2}, z_{k+1}, z_{k+3}, \dots , z_{n}).
\end{aligned}
\end{equation}
Therefore, the singularity in the limit  $z_{i,j} \to 0$ can be obtained in the following way,

\begin{equation}
\begin{aligned}
& \ca{F}^{h_1,h_2,\dots,h_{i} , \dots  ,h_{j}\dots h_n}_{h_{p_1}, h_{p_2}, \dots ,h_{p_{i-1}}, \dots , h_{p_{j-1}}, \dots  , h_{p_{n-3}}}(z_1,z_2, \dots,z_i,  \dots ,  z_{j}, \dots , z_{n})\\
& \hspace{2cm} \downarrow (\text{braiding})\\
& \ca{F}^{h_1,h_2,\dots,h_{i} ,h_{j}\dots h_n}_{h_{p_1}, h_{p_2},  \dots ,h_{q_{i-1}}, \dots , h_{q_{j-2}}, h_{p_{j-1}}, \dots   , h_{p_{n-3}}}(z_1,z_2, \dots,z_i,  z_{j}, \dots , z_{n})\\
& \hspace{2cm} \downarrow (\text{fusion})\\
& \hspace{1cm}    \text{singularity described by (\ref{eq:n-point singularity})}
\end{aligned}
\end{equation}
The explicit form is
 \begin{equation}\label{eq:nbraid}
  \begin{aligned} 
& \ca{F}^{h_1,h_2,\dots,h_{i} , \dots  ,h_{j}\dots h_n}_{h_{p_1}, h_{p_2}, \dots ,h_{p_{i-1}}, \dots , h_{p_{j-2}}, \dots  , h_{p_{n-3}}}(z_1,z_2, \dots,z_i,  \dots ,  z_{j}, \dots , z_{n})\\
    &=\pa{\prod_{k=i-1}^{j-3}\int_{\bb{S}}\dd \a_{q_{k}} \ {\bold B}_{\a_{p_{k}}, \a_{q_{k}}}
   \left[
    \begin{array}{cc}
           \a_i   &     \a_{q_{k-1}}\\
           \a_{k+2}  &\a_{p_{k+1}} \\
    \end{array}
  \right]}
\ca{F}^{h_1,h_2,\dots,h_{i} ,h_{j}\dots h_n}_{h_{p_1}, h_{p_2},  \dots ,h_{q_{i-1}}, \dots , h_{q_{j-3}}, h_{p_{j-2}}, \dots   , h_{p_{n-3}}}(z_1,z_2, \dots,z_i,  z_{j}, \dots , z_{n})\\
&\ar{z_{j, i}\to 0}
\pa{\prod_{k=i-1}^{j-3}\int_{\bb{S}}\dd \a_{q_{k}} \ {\bold B}_{\a_{p_{k}}, \a_{q_{k}}}
   \left[
    \begin{array}{cc}
           \a_i   &     \a_{q_{k-1}}\\
           \a_{k+2}  &\a_{p_{k+1}} \\
    \end{array}
  \right]}
\int_{\bb{S}}\dd \a_{q_{j-2}} \ {\bold F}_{\a_{p_{j-2}}, \a_{q_{j-2}}}
   \left[
    \begin{array}{cc}
     \a_{i}  &   \a_{q_{j-3}}\\
    \a_{j}   & \a_{p_{j-1}}  \\
    \end{array}
  \right]   
	(z_{j}-z_{i})^{h_{q_{j-2}}-h_{i}-h_{j}} \\
& \hspace{2cm} \times  
\ca{F}^{h_1,h_2,\dots,h_{j-1}, h_{q_{j-2}} ,h_{j+1}\dots h_n}_{h_{p_1}, h_{p_2},  \dots ,h_{q_{i-1}}, \dots , h_{q_{j-3}}, \hat{h}_{p_{j-2}}, \dots   , h_{p_{n-3}}}(z_1,z_2, \dots,z_i,  \hat{z_{j}}, \dots , z_{n}),
\end{aligned}
\end{equation}
where the hat means that the entry is omitted and $q_{i-2} \equiv p_{i-2}$.
Note that the dominant contribution to the integral over $\a_{q_{j-2}}$ is given by
\begin{equation}
\a_{q_{j-2}}=\min \left\{ \a^{\text{min}}, \fr{Q}{2} \right\},
\end{equation}
where we defined
\begin{equation}
\a^{\text{min}} \equiv \min
\left\{ \a_{p_{i-2}} + \sum_{l=i+1}^{j-1} \a_{l} ,\ \ \   \a_i+\a_{p_{i+m-1}}+\sum_{l=i+m+1}^{j-1} \a_{l}  \ \ \  (m=1, \cdots , j-i-1)    \right\} .
\end{equation}

As a concrete example, let us consider a special singularity of the following conformal partial wave,
\begin{equation}
\begin{aligned}
\lim_{\e \to 0}\ca{F}^{h_{\a},h_{\b}, \cdots , h_{\b},h_{\a}}_{h_{\a},\cdots,h_{\a}}\pa{z_1, z_2, z_3, \cdots, z_{n-1}, z_1+\e},
\hspace{1cm}
\b=\left\{
    \begin{array}{ll}
   2\a    ,& \text{if }  \a<\fr{Q}{4}   ,\\
      \fr{Q}{2}   ,& \text{if }   \a>\fr{Q}{4}   ,\\
    \end{array}
  \right.\\
\end{aligned}
\end{equation}
From the asymptotics (\ref{eq:nbraid}), this limit can be expressed by
 \begin{equation}
  \begin{aligned} 
& \ca{F}^{h_{\a},h_{\b}, \cdots , h_{\b},h_{\a}}_{h_{\a},\cdots,h_{\a}}\pa{z_1, z_2, z_3, \cdots, z_{n-1}, z_1+\e}\\
&\ar{\e  \to 0}
\pa{\prod_{k=1}^{n-3}\int_{\bb{S}}\dd \a_{q_{k}} \ {\bold B}_{\a, \a_{q_{k}}}
   \left[
    \begin{array}{cc}
           \a   &     \a_{q_{k-1}}\\
           \b  &  \a \\
    \end{array}
  \right]}
	\e^{h_{q_{n-3}}-2h_{\a}} \\
& \hspace{2cm} \times  
\ca{F}^{h_{\b},h_{\b} \cdots , h_{\b},h_{q_{n-3}}}_{h_{q_1},h_{q_2}, \cdots,h_{q_{n-4}}  }   \pa{z_2, z_3, \cdots, z_{n-1}, z_1},
\end{aligned}
\end{equation}
where $\a_{q_0}=0$.
In this special case, the dominant contribution to the integral over $h_{q_{n-3}}$ is just given by
\begin{equation}
h_{q_{n-3}}= h_\b  .
\end{equation}
As a result, the singularity of this partial wave can be expressed by
\begin{equation}\label{eq:edge limit}
\begin{aligned}
& \ca{F}^{h_{\a},h_{\b}, \cdots , h_{\b},h_{\a}}_{h_{\a},\cdots,h_{\a}}\pa{z_1, z_2, z_3, \cdots, z_{n-1}, z_1+\e}\\
&\ar{\e  \to 0}\left\{
    \begin{array}{ll}
	\pa{\prod_{k=1}^{n-4}\int_{\bb{S}}\dd \a_{q_{k}} \ {\bold B}_{\a, \a_{q_{k}}}
	   \left[
   	 \begin{array}{cc}
     	      \a   &     \a_{q_{k-1}}\\
     	      2\a  &  \a \\
   	 \end{array}
  	\right]}
\text{Res}\pa{   -2\pi i  \ 
{\bold B}_{\a, \a_{q_{n-3}}}
	   \left[
   	 \begin{array}{cc}
     	      \a   &     \a_{q_{n-4}}\\
     	      2\a  &  \a \\
   	 \end{array}
  	\right]
	  ; \a_{q_{n-3}}=2\a}  \\
 \hspace{1cm} \times  
	\ca{F}^{h_{2\a},h_{2\a} \cdots , h_{2\a},h_{2\a}}_{h_{q_1},h_{q_2}, \cdots,h_{q_{n-4}}  }   \pa{z_2, z_3, \cdots, z_{n-1}, z_1}
	\e^{h_{2\a}-2h_{\a}}, 
	\hspace{4cm} \text{if } \a<\fr{Q}{4}  ,\\ \\ 
	\pa{\prod_{k=1}^{n-4}\int_{\bb{S}}\dd \a_{q_{k}} \ {\bold B}_{\a, \a_{q_{k}}}
	   \left[
   	 \begin{array}{cc}
     	      \a   &     \a_{q_{k-1}}\\
     	      \fr{Q}{2}  &  \a \\
   	 \end{array}
  	\right]}
\pa{-i  \fr{\s{\pi}}{8} \del_{\a_{q_{n-3}}}^2 \left. 
	{\bold B}_{\a, \a_{q_{n-3}}}
	   \left[
   	 \begin{array}{cc}
     	      \a   &     \a_{q_{n-4}}\\
     	      \fr{Q}{2}  &  \a \\
   	 \end{array}
  	\right]  \right|_{\a_{q_{n-3}}=\fr{Q}{2}}   }\\
 \hspace{1cm} \times  
	\ca{F}^{\fr{Q^2}{4} ,\fr{Q^2}{4} \cdots , \fr{Q^2}{4}, \fr{Q^2}{4} }_{h_{q_1},h_{q_2}, \cdots,h_{q_{n-4}}  }   \pa{z_2, z_3, \cdots, z_{n-1}, z_1} 
	\e^{\fr{Q^2}{4}-2h_a}\pa{-\log \e}^{-\fr{3}{2}} , 
	\hspace{2.2cm} \text{if } \a>\fr{Q}{4} .
    \end{array}
  \right.\\
\end{aligned}
\end{equation}

\section{Monodromy Matrix in Ising Model} \label{app:Ising}
We will directly and carefully confirm the relation between the monodromy matrix and the fusion matrix (\ref{eq:defmono}) in the Ising model (with the central charge $c=\fr{1}{2}$),
which has only three Virasoro primaries: the identity, the Ising spin $\sigma$ with the conformal dimensions $(h_\sigma,\bar{h}_\sigma)=(\fr{1}{16},\fr{1}{16})$ and the energy density $\e$ with  $(h_\e,\bar{h}_\e)=(\fr{1}{2},\fr{1}{2})$.
We first mention that in RCFTs, the integral of the monodromy matrix  (\ref{eq:defmono}) is replaced by the sum as
\begin{equation}\label{eq:monoIsing}
 {\bold M}_{0, \a}
   \left[
    \begin{array}{cc}
    \a_A   & \a_A  \\
     \a_B  &   \a_B\\
    \end{array}
  \right]
=
\sum_\b
\ex{-2\pi i (h_\b-h_A-h_B)}
 \ {\bold F}_{0, \b}
   \left[
    \begin{array}{cc}
    \a_A   & \a_A  \\
     \a_B  &   \a_B\\
    \end{array}
  \right]
 \ {\bold F}_{\b, \a}
   \left[
    \begin{array}{cc}
    \a_A   & \a_B  \\
     \a_A  &   \a_B\\
    \end{array}
  \right].
\end{equation}
For convenience, we list the CFT data and the conformal block of the Ising model \cite{Francesco2012}.
\footnote{We do not display the elements with zero values.}

\hrulefill
\begin{description}
\item[conformal dimension]
\begin{equation}
(h_\sigma,\bar{h}_\sigma)=\pa{\fr{1}{16},\fr{1}{16}}, \ \ \ 
(h_\e,\bar{h}_\e)=\pa{\fr{1}{2},\fr{1}{2}}. \ \ \ 
\end{equation}
\item[OPE coefficient]
\begin{equation}
C_{\sigma \sigma 0} =C_{\e \e 0} =1, \ \ \ 
C_{\sigma \sigma \e} =\fr{1}{2}.
\end{equation}
\item[quantum dimension]
\begin{equation}
d_0=d_\e=1, \ \ \ 
d_\sigma=\s{2}.
\end{equation}
\item[fusion matrix] (+permutation (\ref{eq:fusionsym}))
\begin{equation}\label{eq:fusionIsing}
\begin{aligned}
 {\bold F}_{0, 0}
   \left[
    \begin{array}{cc}
    \sigma   &  \sigma  \\
      \sigma  &    \sigma\\
    \end{array}
  \right]
&=2 {\bold F}_{0, \e}
   \left[
    \begin{array}{cc}
    \sigma   &  \sigma  \\
      \sigma  &    \sigma\\
    \end{array}
  \right]
=\fr{1}{2} {\bold F}_{\e, 0}
   \left[
    \begin{array}{cc}
    \sigma   &  \sigma  \\
      \sigma  &    \sigma\\
    \end{array}
  \right]
=- {\bold F}_{\e, \e}
   \left[
    \begin{array}{cc}
    \sigma   &  \sigma  \\
      \sigma  &    \sigma\\
    \end{array}
  \right]
=\fr{1}{\s{2}}, \\
 {\bold F}_{0, 0}
   \left[
    \begin{array}{cc}
    \e   &  \e \\
      \e  &    \e \\
    \end{array}
  \right]
&=
 2{\bold F}_{0, \sigma}
   \left[
    \begin{array}{cc}
    \sigma   &  \sigma \\
      \e  &    \e \\
    \end{array}
  \right]
=
 \fr{1}{2}{\bold F}_{\sigma, 0}
   \left[
    \begin{array}{cc}
    \e   &  \sigma \\
      \e  &    \sigma \\
    \end{array}
  \right]
=- {\bold F}_{\sigma, \sigma}
   \left[
    \begin{array}{cc}
    \e   &  \sigma \\
      \sigma  &    \e \\
    \end{array}
  \right]
=1.
\end{aligned}
\end{equation}
\item[braiding matrix] (general form)
\begin{equation}
\begin{aligned}
{\bold F}_{\a_s, \a_t} 
   \left[
    \begin{array}{cc}
    \a_2   & \a_1  \\
     \a_3  &   \a_4\\
    \end{array}
  \right]
=\ex{-\e i\pi(h_1+h_3-h_{\a_s}-h_{\a_t})}
{\bold B}_{\a_s, \a_t} ^{(\e)}
   \left[
    \begin{array}{cc}
    \a_2   & \a_1  \\
     \a_4  &   \a_3\\
    \end{array}
  \right].
\end{aligned}
\end{equation}
\item[monodromy matrix] (We will derive these elements in the following.)
\begin{equation}\label{eq:monomatIsing}
2{\bold M}_{0, \e}
   \left[
    \begin{array}{cc}
    \sigma   &  \sigma  \\
      \sigma  &    \sigma\\
    \end{array}
  \right]
=\fr{1}{2} {\bold M}_{\e, 0}
   \left[
    \begin{array}{cc}
    \sigma   &  \sigma  \\
      \sigma  &    \sigma\\
    \end{array}
  \right]
=\ex{\fr{\pi i}{4}}, \ \ \ 
{\bold M}_{0, 0}
   \left[
    \begin{array}{cc}
    \e   &  \e \\
      \e  &    \e \\
    \end{array}
  \right]
=1.
\end{equation}
\item[conformal block]
\begin{equation}\label{eq:blockIsing}
\ca{F}^{\sigma \sigma}_{\sigma \sigma}(0|z) =\fr{1}{\s{2}}\fr{\s{1+\s{1-z}}}{\pa{z(1-z)}^{\fr{1}{8}}}, \ \ \ 
\ca{F}^{\sigma \sigma}_{\sigma \sigma}(\e|z) =\fr{2}{\s{2}}\fr{\s{1-\s{1-z}}}{\pa{z(1-z)}^{\fr{1}{8}}}, \ \ \ 
\ca{F}^{\e \e}_{\e \e}(0|z) =\fr{1-z+z^2}{z(1-z)}.
\end{equation}
\end{description}
\hrulefill

The fusion matrix elements can be given by using the following identities,
\begin{equation}
\begin{aligned}
\s{1+\s{1-z}}&=\fr{1}{\s{2}}\pa{\s{1+\s{z}}+\s{1-\s{z}}},\\
\s{1-\s{1-z}}&=\fr{1}{\s{2}}\pa{\s{1+\s{z}}-\s{1-\s{z}}}.
\end{aligned}
\end{equation}
These identities relate the t-channel block to the s-channel block as
\begin{equation}
\begin{aligned}
\ca{F}^{\sigma \sigma}_{\sigma \sigma}(0|z)=\fr{1}{\s{2}}\ca{F}^{\sigma \sigma}_{\sigma \sigma}(0|1-z)+\fr{1}{2\s{2}}\ca{F}^{\sigma \sigma}_{\sigma \sigma}(\e|1-z), \\
\ca{F}^{\sigma \sigma}_{\sigma \sigma}(\e|z)=\fr{2}{\s{2}}\ca{F}^{\sigma \sigma}_{\sigma \sigma}(0|1-z)-\fr{1}{\s{2}}\ca{F}^{\sigma \sigma}_{\sigma \sigma}(\e|1-z).
\end{aligned}
\end{equation}
From these relations, we find that the fusion matrix elements read respectively the equations  (\ref{eq:fusionIsing}).
As mentioned in Section \ref{subsec:Regge}, the monodromy matrix element around $z=0$ is trivially given by the phase factor $\ex{-2\pi i \pa{h_p-h_A-h_B}}$, which can be directly found from the conformal blocks (\ref{eq:blockIsing}) as follows;
\begin{equation}
\begin{aligned}
\ca{F}^{\sigma \sigma}_{\sigma \sigma}(0|z)
=\fr{1}{2}\fr{\s{1+\s{z}}+\s{1-\s{z}}}{\pa{z(1-z)}^{\fr{1}{8}}}
\ar{z \to \ex{-2\pi i}z}
\ex{\fr{\pi i}{4}}\fr{1}{2}\fr{\s{1-\s{z}}+\s{1+\s{z}}}{\pa{z(1-z)}^{\fr{1}{8}}}=\ex{\fr{\pi i}{4}}\ca{F}^{\sigma \sigma}_{\sigma \sigma}(0|z),\\
\ca{F}^{\sigma \sigma}_{\sigma \sigma}(\e|z)
=\fr{\s{1+\s{z}}-\s{1-\s{z}}}{\pa{z(1-z)}^{\fr{1}{8}}}
\ar{z \to \ex{-2\pi i}z}
\ex{\fr{\pi i}{4}}\fr{\s{1-\s{z}}-\s{1+\s{z}}}{\pa{z(1-z)}^{\fr{1}{8}}}=-\ex{\fr{\pi i}{4}}\ca{F}^{\sigma \sigma}_{\sigma \sigma}(\e|z).
\end{aligned}
\end{equation}
Therefore, the definition of the monodromy matrix (\ref{eq:monoIsing}) provides the elements as
\begin{equation}
\begin{aligned}
{\bold M}_{0, \e}
   \left[
    \begin{array}{cc}
    \sigma   &  \sigma  \\
      \sigma  &    \sigma\\
    \end{array}
  \right]
=
\ex{\fr{\pi i}{4}}
 {\bold F}_{0, 0}
   \left[
    \begin{array}{cc}
    \sigma   &  \sigma  \\
      \sigma  &    \sigma\\
    \end{array}
  \right]
 {\bold F}_{0, \e}
   \left[
    \begin{array}{cc}
    \sigma   &  \sigma  \\
      \sigma  &    \sigma\\
    \end{array}
  \right]
-
\ex{\fr{\pi i}{4}}
 {\bold F}_{0, \e}
   \left[
    \begin{array}{cc}
    \sigma   &  \sigma  \\
      \sigma  &    \sigma\\
    \end{array}
  \right]
 {\bold F}_{\e, \e}
   \left[
    \begin{array}{cc}
    \sigma   &  \sigma  \\
      \sigma  &    \sigma\\
    \end{array}
  \right]
=\fr{1}{2}\ex{\fr{\pi i}{4}}, \\
{\bold M}_{\e, 0}
   \left[
    \begin{array}{cc}
    \sigma   &  \sigma  \\
      \sigma  &    \sigma\\
    \end{array}
  \right]
=\ex{\fr{\pi i}{4}}
 {\bold F}_{\e, 0}
   \left[
    \begin{array}{cc}
    \sigma   &  \sigma  \\
      \sigma  &    \sigma\\
    \end{array}
  \right]
 {\bold F}_{0, 0}
   \left[
    \begin{array}{cc}
    \sigma   &  \sigma  \\
      \sigma  &    \sigma\\
    \end{array}
  \right]
-
\ex{\fr{\pi i}{4}}
 {\bold F}_{\e, \e}
   \left[
    \begin{array}{cc}
    \sigma   &  \sigma  \\
      \sigma  &    \sigma\\
    \end{array}
  \right]
 {\bold F}_{\e, 0}
   \left[
    \begin{array}{cc}
    \sigma   &  \sigma  \\
      \sigma  &    \sigma\\
    \end{array}
  \right]
=2\ex{\fr{\pi i}{4}},
\end{aligned}
\end{equation}
which exhibit the monodromy matrix elements (\ref{eq:monomatIsing}). The monodromy matrix is also derived from the conformal blocks (\ref{eq:blockIsing}) directly as follows;
\begin{equation}
\begin{aligned}
\ca{F}^{\sigma \sigma}_{\sigma \sigma}(0|z) =\fr{1}{\s{2}}\fr{\s{1+\s{1-z}}}{\pa{z(1-z)}^{\fr{1}{8}}}
\ar{(1-z) \to \ex{-2 \pi i}(1-z)}
\ex{\fr{\pi i}{4}}\fr{1}{\s{2}}\fr{\s{1-\s{1-z}}}{\pa{z(1-z)}^{\fr{1}{8}}}=\fr{1}{2}\ex{\fr{\pi i}{4}}\ca{F}^{\sigma \sigma}_{\sigma \sigma}(\e|z), \\
\ca{F}^{\sigma \sigma}_{\sigma \sigma}(\e|z) =\fr{2}{\s{2}}\fr{\s{1-\s{1-z}}}{\pa{z(1-z)}^{\fr{1}{8}}}
\ar{(1-z) \to \ex{-2 \pi i}(1-z)}
\ex{\fr{\pi i}{4}}\fr{2}{\s{2}}\fr{\s{1+\s{1-z}}}{\pa{z(1-z)}^{\fr{1}{8}}}=2\ex{\fr{\pi i}{4}}\ca{F}^{\sigma \sigma}_{\sigma \sigma}(0|z). 
\end{aligned}
\end{equation}
Hence, we can confirm that the definition of the monodromy matrix (\ref{eq:defmono}) definitely represents the monodromy around $z=1$.
Note that the monodromy matrix of the block $\ca{F}^{\e \e}_{\e \e}$ is trivial because there is only one fusion  matrix element with a non-zero value.

\clearpage
\bibliographystyle{JHEP}
\bibliography{renyi}

\end{document}